\documentclass[british]{scrartcl}
\usepackage[T1]{fontenc}
\usepackage[latin9]{inputenc}
\usepackage{float}
\usepackage{bm}
\usepackage{amsmath}
\usepackage{accents}
\usepackage{amssymb}
\usepackage{graphicx}
\usepackage{esint}
\usepackage[authoryear]{natbib}
\usepackage{amsthm}
\usepackage{url}
\usepackage{color}
\usepackage{enumitem}
\usepackage{pdfpages}

\usepackage{tikz}
\usepackage{ifthen}

\usepackage[position=top,labelformat=empty]{subfig}
\newdimen\HilbertLastX
\newdimen\HilbertLastY
\newcounter{HilbertOrder}

\def\DrawToNext#1#2{%
   \advance \HilbertLastX by #1
   \advance \HilbertLastY by #2
   \pgfpathlineto{\pgfqpoint{\HilbertLastX}{\HilbertLastY}}
}

\def\Hilbert[#1,#2,#3,#4,#5,#6,#7,#8] {
  \ifnum\value{HilbertOrder} > 0%
     \addtocounter{HilbertOrder}{-1}
     \Hilbert[#5,#6,#7,#8,#1,#2,#3,#4]
     \DrawToNext {#1} {#2}
     \Hilbert[#1,#2,#3,#4,#5,#6,#7,#8]
     \DrawToNext {#5} {#6}
     \Hilbert[#1,#2,#3,#4,#5,#6,#7,#8]
     \DrawToNext {#3} {#4}
     \Hilbert[#7,#8,#5,#6,#3,#4,#1,#2]
     \addtocounter{HilbertOrder}{1}
  \fi
}
\def\hilbert((#1,#2),#3){%
   \advance \HilbertLastX by #1
   \advance \HilbertLastY by #2
   \pgfpathmoveto{\pgfqpoint{\HilbertLastX}{\HilbertLastY}}
   \setcounter{HilbertOrder}{#3}
   \Hilbert[0mm,1mm,0mm,-1mm,1mm,0mm,-1mm,0mm]
   \pgfusepath{stroke}%
}

\newcommand{\starN}{\|_\star}
\newcommand{\infN}{\|_\infty}
\newcommand{\Dst}{D^\star} 
\usepackage{framed}
\makeatletter

\floatstyle{ruled}
\newfloat{algorithm}{tbp}{loa}
\providecommand{\algorithmname}{Algorithm}
\floatname{algorithm}{\protect\algorithmname}

\usepackage{dsfont}\usepackage{algpseudocode}
\usepackage{mathtools}

\newcounter{rmq}[section]
\newcommand{\R}{\mathbb{R}}
\newcommand{\Rp}{\R_+}

\newcommand{\contb}{\mathcal{C}_b}

\newcommand{\setX}{\mathcal{X}}
\newcommand{\setY}{\mathcal{Y}}

\renewcommand{\P}{\mathbb{P}}
\newcommand{\mP}{\mathcal{P}}
\newcommand{\Pf}{\mP_f(\setX)} 

\newcommand{\E}{\mathbb{E}}
\newcommand{\var}{\mathrm{Var}}
\newcommand{\cov}{\mathrm{Cov}}

\newcommand{\as}{\mbox{a.s.}}

\newcommand{\Q}{\mathbb{Q}} 

\newcommand{\N}{\mathcal{N}} 

\newcommand{\bx}{x}
\newcommand{\bu}{u}

\newcommand{\ind}{\mathds{1}}
\newcommand{\dd}{\mathrm{d}}
\newcommand{\dx}{\dd \bx}

\newcommand{\eqdef}{:=} 
\newcommand{\bigO}{\mathcal{O}} 
\newcommand{\smallo}{{\scriptscriptstyle\mathcal{O}}} 

\newcommand{\FK}{Feynman-Kac}

\newcommand{\comment}[1]{ \ifthenelse{ \equal{\showcomment}{true} }{ {\bf #1} }{} }
\newcommand{\showcomment}{true}
\newcommand{\subscript}[2]{$#1 _ #2$}
\newcommand{\multi}{\mathrm{multi}}

\newcommand{\syst}{\mathrm{syst}}
\newcommand{\strat}{\mathrm{strat}}

\newcommand{\res}{\mathrm{res}}
\newcommand{\ssp}{\mathrm{ssp}}

\newcommand{\wc}{\xRightarrow{\text{w}}}

\newtheorem{thm}{Theorem}
\newtheorem{prop}{Proposition}
\newtheorem{corollary}{Corollary}
\newtheorem{lemma}{Lemma}
\newtheorem{definition}{Definition}

\makeatother

\usepackage{babel}
\usepackage{hyperref}

\title{Negative association, ordering and convergence of resampling methods}
\date{}

\newcommand*\samethanks[1][\value{footnote}]{\footnotemark[#1]}
\author{Mathieu Gerber\thanks{
  School of Mathematics, University of Bristol, UK.}
  \and Nicolas Chopin\thanks{CREST-ENSAE, France.}
  \and Nick Whiteley\samethanks[1]
  } 
  
\begin{document}

\maketitle


\begin{abstract}
We study convergence and convergence rates for resampling schemes. Our first main
result is a general consistency theorem based on the notion of negative
association, which is applied to establish the almost sure weak convergence of
measures output from Kitagawa's \citeyearpar{Kitagawa1996} stratified
resampling method. Carpenter et al's \citeyearpar{CarClifFearn} systematic
resampling method is similar in structure but can fail to converge depending on
the order of the input samples. We introduce a new resampling algorithm based
on a stochastic rounding technique of \cite{Srinivasan2001}, which shares some
attractive properties of systematic resampling, but which exhibits negative
association and therefore converges irrespective of the order of the input
samples. We confirm a conjecture made by \cite{Kitagawa1996} that ordering
input samples by their states in $\mathbb{R}$ yields a faster
rate of convergence; we establish that when particles are ordered using the
Hilbert curve in $\mathbb{R}^d$, the variance of the resampling error is
$\smallo(N^{-(1+1/d)})$ under mild conditions, where $N$ is the number of
particles.
We use these results to establish asymptotic properties of particle algorithms
based on resampling schemes that differ from multinomial resampling.



\end{abstract}



\section{Introduction}

A resampling scheme is a randomized procedure that takes
as input random samples $X^n$ with nonnegative weights $W^n\geq 0$,
$n=1,\ldots,N$, such that $\sum_{n=1}^N W^n =1$,
and returns as an output resampled variables $X^{A^n}$, where $A^n$ is a random
index in $\{1,\ldots,N\}$, such that, in some sense,
\begin{equation}\label{eq:empirical_measures}
\frac 1 N \sum_{n=1}^N \delta(X^{A^n})
\approx
\sum_{n=1}^N W^n \delta({X^n}).
\end{equation}
Here $\delta(x)$ denotes the Dirac measure at point $x$ (this slightly unconventional notation
will make our equations more readable).



Resampling appears in various statistical procedures. The present work is
primarily motivated by resampling within Sequential Monte Carlo methods, also
known as particle filters  \citep{DouFreiGor}. Particle filters approximate
recursively a sequence of probability distributions by propagating N
`particles' through weighting,
resampling and mutation steps.  The resampling steps play a crucial role
in stabilizing the Monte Carlo error over time \citep{Gordon}.
In particular, without resampling, the largest normalised
weight of the particle sample converges quickly to one as the
number of iterations increases \citep{delmoraldoucet2003}. This means that
most of the computational effort is wasted on particles that contribute
little to the end results.

Resampling also appears in survey sampling  under the name of `unequal
probability sampling'  \citep{MR2225036}, but in a context  slightly
different from the one we consider in this paper. In survey sampling
only $M<N$ `units' are selected  and the object of interest after the
(re)sampling operation, the  Horvitz-Thompson empirical process \citep[HTEP,
see e.g.][]{Bertail2017}  is another un-normalized  weighted sum of Dirac
measures. Adapting the statement and the assumptions
of our first main result, Theorem \ref{thm:NA_resampling} 
in Section \ref{sec:preliminaries}, in order
to study the asymptotic behaviour of the HTEP is possible  but beyond the scope of this paper. Yet another statistical procedure where resampling appears is the
the weighted bootstrap \citep{MR2195545}.


There are various existing resampling methods.  Multinomial resampling is
perhaps the simplest technique, where given the weights, the indices $A^n$ are
generated conditionally independently from the finite distribution that assigns
probability $W^n$ to outcome $n$.  In particle filtering it is common practice
to replace multinomial resampling with techniques which are computationally
faster and empirically more accurate. However, these advanced resampling
techniques are generally not straightforward to analyse because they induce
complicated dependence between output samples, and various aspects of their
behaviour are still not understood.

Following definitions and an account of what is known about existing resampling
techniques, our first main result, Theorem \ref{thm:NA_resampling} in Section
\ref{sec:preliminaries}, is a general consistency result for resampling based
on the notion of \emph{negative association} \citep{joag1983negative}. An
application of this theorem gives, to our knowledge, the first proof of
almost sure weak convergence of the random probability measures output from the
stratified resampling method of \cite{Kitagawa1996}. A notable feature of
Theorem \ref{thm:NA_resampling} is that, although its assumptions do not require the input particles to be algorithmically ordered in a particular way, its
proof involves establishing a necessary and sufficient condition for
almost sure weak convergence involving ordering using the Hilbert space-filling curve. Here
we build on \citet{MR3351446}, who used the Hilbert curve to derive and
analyse a quasi-Monte Carlo version of sequential Monte Carlo samplers.

The systematic resampling method of \cite{CarClifFearn},  which involves a sampling technique first proposed by \citet{Madow1944},
 is a very popular
and computationally cheap resampling technique, with the property that the
number of offspring of any sample with weight $W$ in a population of size $N$ is
with probability $1$ either $\left\lfloor NW\right\rfloor$  or $\left\lfloor
NW\right\rfloor +1$. However, depending on the order of the input particles,
the error variance for systematic resampling can fail to converge to zero as
$N\to+\infty$, see \citet{Douc2005} and \cite{LEcuyer2000}. We
complement this insight by providing a counter-example to almost sure weak
convergence.
We then introduce a new resampling method, called Srinivasan Sampling Process
(SSP) resampling, which corrects this deficiency: it also has the property that
offspring numbers are of the form either $\left\lfloor NW\right\rfloor$  or
$\left\lfloor NW\right\rfloor +1$, but it provably converges irrespective of
the order of input particles, by another application of our Theorem
\ref{thm:NA_resampling}.

\cite{Kitagawa1996} conjectured that in the case that the state-space is
$\mathbb{R}$, ordering the particles input to stratified resampling according
to their states leads to faster convergence. In particular, he suggested that
the integrated square error between empirical cdf's before and after resampling
behaves as $\bigO(N^{-2})$, compared to the standard Monte Carlo rate
$\bigO(N^{-1})$ in the un-ordered case. We confirm this conjecture by proving,
under mild conditions, that for stratified resampling on state-space
$\mathbb{R}^{d}$ with input particles ordered by their states using the Hilbert
curve, the variance of the resampling error is $\smallo(N^{-(1+1/d)})$.
Kitagawa also examined the behaviour of a deterministic resampling scheme; we
identify the variant of it which is optimal in terms of the Kolmogorov metric
when the state-space is $\mathbb{R}$. We also prove the almost sure weak
consistency of stratified and systematic when the particles are
Hilbert-ordered.

Finally, we discuss the implications of our results on particle filtering.
In particular, we show that particle estimates are consistent when resampling schemes
such as e.g. SSP or stratified resampling are used. In addition, we show that
the ordered version of stratified resampling dominates other resampling schemes
in terms of asymptotic variance of particle estimates.

All the proofs are gathered in the supplementary materials.

\section{Preliminaries\label{sec:preliminaries}}

\subsection{Notation and conventions}\label{subsec:notation}

Let $\setX$ be an open subset of $\R^d$, $\mathbb{X}$ its Borel $\sigma$-algebra,
$\mP(\setX)$ the set of  probability measures on $(\setX,\mathbb{X})$,
$\mP_b(\setX)\subset\mP(\setX)$ the subset of measures in $\mP(\setX)$ which admit
a continuous and bounded density with respect to $\lambda_d$,
the Lebesgue measure on $\setX$, and
$\mP_f(\setX)\subset\mP(\setX)$ the subset of measures in $\mP(\setX)$
whose support is a finite set.





For integers $1\leq a\leq b$, we will often use the index shorthands $z_{a:b}=(z_a,\dots,z_b)$ and $z^{a:b}=(z^a,\dots,z^b)$, and let $1:N = \{1,\ldots,N\}$.

For any measurable mapping $\varphi$ from $(\setX,\mathbb{X})$ to some measurable space $(\setY,\mathbb{Y})$  and a probability measure $\pi\in\mP(\setX)$, we write $\pi_{\varphi}$ for the pushforward of $\pi$ by $\varphi$. The set of continuous and bounded functions on $\setX$ is denoted by $\mathcal{C}_b(\setX)$ and we use the symbol ``$\wc$'' to denote  weak convergence; that is, for sequence $(\pi^N)_{N\geq 1}$ in $\mathcal{P}(\setX)$ and $\pi\in\mathcal{P}(\setX)$,
$$
\pi^N\wc\pi\quad\Leftrightarrow\quad \lim_{N\rightarrow+\infty}\pi^N(\varphi)=\pi(\varphi),\quad\forall\varphi\in \mathcal{C}_b(\setX).
$$

Throughout the paper we consider a fixed probability space $(\Omega,\mathcal{F},\mathbb{P})$ on which all random variables are defined. With $\mathcal{B}([0,1]^{\mathbb{N}})$ denoting the Borel $\sigma$-algebra on $[0,1]^{\mathbb{N}}$,   let $U=(U_{1},U_{2},\ldots)$ be a $([0,1]^{\mathbb{N}},\mathcal{B}([0,1]^{\mathbb{N}}))$-valued random variable on $(\Omega,\mathcal{F},\mathbb{P})$,
such that $\mathbb{P}$ makes $(U_{1},U_{2},\ldots)$ independent
of each other and all other random variables, and such that each $U_{i}$
is distributed uniformly on $[0,1]$.

We note that one can choose a countable subset of $\mathcal{C}_b(\setX)$ that completely determines weak convergence, hence for random measures $(\pi^N)_{N\geq1}$, the event $\{\pi^N\wc\pi\}$ is measurable.

For $\pi\in \mathcal{P}(\setX)$, we denote by
$\pi(\varphi)$ the expectation $\int_\setX \varphi(x)\pi(\dx)$, and for a random variable $Z=(Z_1,\ldots,Z_d)$ whose distribution is $\pi$ we denote by
$F_{\pi}(a)=\P(Z_1\leq a_1,\ldots,Z_d \leq a_d)$, $a=(a_1,\ldots,a_d)$, its CDF (cumulative distribution function)   and, when $d=1$,  by $F^{-}_{\pi}$ its generalized inverse:
$F^{-}_{\pi}(u)=\inf\{x:\, F_\pi(x)\geq u\} $.


For each $N\geq1$ we consider a distinguished collection of random variables $\zeta^{N}=(X^{n,N},W^{n,N})_{n=1}^{N}$,
with each $X^{n,N}$ valued in $\setX$, each $(W^{n,N})_{n=1}^{N}$ valued in $\mathbb{R}_+$, and such that $\mathbb{P}$-a.s.,   $\sum_{n=1}^{N}W^{n,N}=1$.
When no confusion may arise, we suppress dependence on $N$ and write $\zeta^{N}=(X^{n},W^{n})_{n=1}^{N}$. We associate with $\zeta^{N}$ the random measure $\pi^N=\sum_{n=1}^N W^n \delta(X^n)$, the (random) CDF
$$
F_N(n)=\sum_{m=1}^N W^m \ind(m\leq n),\quad n\in 1:N,
$$
and its inverse is denoted $F_N^-$.

To lighten notation we shall write $\P_{\zeta^N}(\cdot)$, $\E_{\zeta^N}[\cdot]$, $\var_{\zeta^N}[\cdot]$, $\cov_{\zeta^N}[\cdot,\cdot]$ for conditional probability, expectation, variance and covariance given $\zeta^N$.

Let $\mathcal{Z}^{N}= \{ (x,w)\in\setX^N\times \R_+^N:\,\sum_{n=1}^N w_n=1\}$
and define the disjoint union $\mathcal{Z}\coloneqq\bigcup_{N=1}^{+\infty}\mathcal{Z}^{N}$.
So we may think of $\zeta^{N}$ as a random point in $\mathcal{Z}^{N}$, and hence $\mathcal{Z}$.


\begin{definition}\label{def:X_Set}
$\setX\subseteq\mathbb{R}^d$ is said to be cubifiable
if there exist measurable sets $\setX_i\subseteq\mathbb{R}$, $i=1,\dots,d$, such that
\begin{enumerate}
\item  $\setX=\times_{i=1}^d\setX_i$;
\item For any $i\in 1:d$, there exists a   $C^1$-diffeomorphism $\psi_i:\setX_i\rightarrow (0,1)$ which is strictly increasing on $\setX_i$.
\end{enumerate}
We shall write $\psi(x)=(\psi_1(x_1),\ldots,\psi_d(x_d))$, $x=x_{1:d}\in\setX$, the resulting $C^1$-diffeomorphism from $\setX$ into $(0,1)^d$.
\end{definition}

We recall the reader that function $\psi:\setX\rightarrow(0,1)^d$ is a $C^1$-diffeomorphism if it is a bijection and its inverse $\psi^{-1}:(0,1)^d\rightarrow\setX$ is continuously differentiable. In what follows, for a cubifiable set $\setX$ we denote by $\mathcal{D}(\setX)$  the set of all $C^1$-diffeomorphisms from $\setX$ into $(0,1)^d$ that verify the conditions of Definition \ref{def:X_Set}.

Cubifiable sets are   sets that   can be written as $\setX=\times_{i=1}^d(a_i,b_i)$ for some $a_i,b_i\in\R\cup\{-\infty,+\infty\}$. The point of these sets  is to be able to work `as if' $\setX=(0,1)^d$. The hypercube $(0,1)^d$  will   play a key role below because the Hilbert space-filling curve, which is  essential in this work, is defined on this hypercube.

Most of the results presented below   assume that the limiting distribution $\pi$ admits a continuous and bounded density. Consequently, to work `as if'  $\setX=(0,1)^d$ we will often assume that  $\pi$ belongs to
$$
\widetilde{\mP}_b(\setX)=\big\{\pi\in\mP_b(\setX):\,\exists \psi\in\mathcal{D}(\setX)\text{ s.t. } \pi_\psi\in\mP_b((0,1)^d)\Big\}.
$$

The following result provides a sufficient condition to have $\pi\in \widetilde{\mP}_b(\setX)$. We denote by $p_\pi$ the density (w.r.t. $\lambda_d$) of $\pi\in\mP_b(\setX)$ and, for $I\subset 1:d$, we write $x_I=(x_i,\,i\in I)$ and $x_{\setminus I}=(x_i,\,i\not\in I)$.

\begin{lemma}\label{lem:tails}
Let $\setX$ be a cubifiable set, $\delta>0$ and
$\pi \in\mP_b(\setX)$ such that   $\forall I\subseteq 1:d$ and $\forall x_{\setminus I}\in\times_{i\not\in I}\setX_i$ we have
$\sup_{ x_I\in \times_{i\in I}\setX_i} p_\pi(x)\prod_{i\in I} |x_i|^{1+\delta}\leq  C$ for some $C<+\infty$.
  Then $\pi \in \widetilde{\mP}_b(\setX)$.
\end{lemma}

Recall that  $\sup_{x\in\R} p_\pi(x)|x|<+\infty$ for any $\pi\in\mP_b(\R)$.
Therefore, as $\delta>0$ is arbitrary in the lemma, very few extra conditions
on the tails of $\pi\in\mP_b(\R)$ are needed in order to have
$\pi\in\widetilde{\mP}_b(\R)$ when $d=1$. When $d>1$, assuming that $\pi\in
\widetilde{\mP}_b(\R^d)$ is more restrictive  since the lemma requires some
uniformity in the behaviour of  tails. However, we note   that members of
$\widetilde{\mP}_b(\R^d)$ may not have a first moment and therefore the
sufficient  condition  of Lemma \ref{lem:tails} appears to be quite weak.

\subsection{Resampling schemes: definitions and properties}
\label{sub:resampling_schemes}



\begin{definition}\label{def:resamplingSheme}
A resampling scheme is a mapping $\rho:[0,1]^\mathbb{N}\times  \mathcal{Z}\rightarrow \Pf$ such that, for any $N\geq1$ and $z=(x^{n},w^{n})_{n=1}^{N}\in\mathcal{Z}^{N}$,
\[
\rho(u,z)=\frac{1}{N}\sum_{n=1}^{N}\delta(x^{a_{N}^{n}(u,z)}),
\]
where for each $n$, $a_{N}^{n}:[0,1]^{\mathbb{N}}\times\mathcal{Z}^{N}\to1:N$
is a measurable function.
\end{definition}

Given $u\in[0,1]^{\mathbb{N}}$, the mapping  $\rho(u,\cdot)$ therefore takes as input a weighted point set $z=(x^{n},w^{n})_{n=1}^{N}$, selects $N$ indices $(a_{N}^{n}(u,z))_{n=1}^N$ in the set $1:N$  and returns a  probability measure on $(x^{n})_{n=1}^N$ with the property that each  $x^{a_{N}^{n}(u,z)}$ has   weight  $N^{-1}$.

Instances of the function $a_{N}^{n}$ are given below. We shall use the shorthands $\rho(z)$ for the random measure
$\rho(U,z)$, $z\in{\mathcal{Z}^N}$, and $A^n$ for the random indices $ a_{N}^{n}(U,\zeta^N)$. Introducing the quantities,
\begin{equation}
\#^n(u,z)=\mathrm{card}\{i\in1:N\;\;\mathrm{s.t.}\;\;a_N^i(u,z)=n\},\,\Delta_{\rho,z}^n=\#^n(U,z)-Nw^n, \label{eq:def_Psi}
\end{equation}
a resampling scheme $\rho$ is said to be \emph{unbiased} if, for any $N\geq1$, $n\in 1:N$ and $z\in\mathcal{Z}^{N}$,
\begin{equation*}
\E[\Delta_{\rho,z}^n]=0.
\end{equation*}

We now define the resampling schemes of primary interest in this work.

\begin{itemize}
\item \textbf{Multinomial resampling:}  $\rho_{\mathrm{multi}}$ such that
\[a_{N}^{n}(u,\zeta^{N})=F_{N}^{-}(u_{n}).\]
In this case the
$a^n_N(U,\zeta^N)$ are i.i.d. (independent and identically distributed) draws
from the distribution which assigns probability $W^n$ to outcome $n$.

    \item  \textbf{Stratified resampling:} $\rho_{\mathrm{strat}}$ such that
    $$
    a_{N}^{n}(u,\zeta^{N})=F_{N}^{-}\left(\frac{n-1+u_{n}}{N}\right).
    $$

    \item  \textbf{Systematic resampling:} $\rho_{\mathrm{syst}}$ such that
    $$
    a_{N}^{n}(u,\zeta^{N})=F_{N}^{-}\left(\frac{n-1+u_{1}}{N}\right).
    $$



\end{itemize}


The following definition captures the notion of almost sure weak convergence of the random measures $(\pi^N)_{N\geq1}$ which we shall study and is similar to condition (9) in  \citep{crisan2002survey}.

\begin{definition}\label{def:resampling}
Let $\mathcal{P}_0\subseteq\mathcal{P}(\setX)$. Then, we say that a resampling scheme  $\rho:[0,1]^{\mathbb{N}}\times \mathcal{Z}\rightarrow \mP_f(\setX)$ is $\mathcal{P}_0$-consistent if, for any $\pi\in\mathcal{P}_0$ and  $(\zeta^N)_{N\geq 1}$ such that $\pi^N\wc \pi$, $\P$-a.s., one has
\[\rho(\zeta^N)\wc \pi,\quad \P-\mbox{a.s.}
\]
\end{definition}


It is well known that multinomial, stratified and systematic resampling are
unbiased. An account of various properties of these methods can be found in
\citet{Douc2005}.

\citet[][Lemma 2]{crisan2002survey} shows that multinomial resampling is
$\mathcal{P}(\setX)$-consistent for any measurable set
$\setX\subseteq\mathbb{R}^d$.


It is easy to show  \citep{MR887702,Douc2005}  that
stratified resampling dominates multinomial resampling in terms of variance, i.e.,
\[\var\left[\rho_{ \mathrm{strat}}(z)(\varphi)\right]
\leq
\var \left[\rho_{\mathrm{multi}}
(z)(\varphi)\right],\quad\forall z\in\mathcal{Z}
\]
for any measurable $\varphi:\setX\rightarrow\R$.
Similar results are harder to derive for systematic resampling,
owing to the strong dependencies between the resampled indices. 
However, it is known \citep{Douc2005} that the variance of $\rho_{\mathrm{syst}}(\zeta^N)(\varphi)$ may not converge to $0$ as $N\rightarrow+\infty$  \citep[see also][for an explanation of this phenomenon]{LEcuyer2000}.

\section{Convergence of resampling schemes based on negative association\label{sec:no_order}}

\subsection{A general consistency result}

Before stating the main result of this section we  recall the definition of negatively associated (NA) random  variables \citep{joag1983negative}.

\begin{definition}\label{def:NA}
A collection of random variables $(Z^n)_{n=1}^N$ are negatively associated if, for every pair of disjoint subsets $I_1$ and $I_2$ of $\{1,\dots N\}$,
$$
\cov\Big(\varphi_1\big(Z^n, n\in I_1\big),\,\varphi_2\big(Z^n, n\in I_2\big)\Big)\leq  0
$$
for all  coordinatewise non-decreasing functions $\varphi_1$ and $\varphi_2$  such that for $k\in \{1,2\}$, $\varphi_k:\mathbb{R}^{|I_k|}\rightarrow\mathbb{R}$,  and  such that the covariance is well-defined.
\end{definition}



\begin{thm}\label{thm:NA_resampling}
Let $\setX$ be a cubifiable set and   $\rho$ be an unbiased resampling scheme such that the following  conditions hold:
\begin{enumerate}[label=(\subscript{H}{\arabic*})]
\item\label{H1} For any $N\geq 1$ and $z\in\mathcal{Z}^N$, the random variables  $(\#^n(U,z))_{n=1}^{N}$ are negatively associated;
\item\label{H2} There exists a sequence $(r_N)_{N\geq 1}$ of non-negative real numbers such that
$ r_N = \smallo( N/\log N)$ and, for $N$ large enough,
\[
	\sup_{z\in\mathcal{Z}^N}\sum_{n=1}^N\E\big[(\Delta_{\rho,z}^n)^2\big]  \leq r_N\,N, \quad
	\sum_{N=1}^\infty \sup_{z\in\mathcal{Z}^N}\P\Big(\max_{n\in 1:N}\big|\Delta_{\rho,z}^n\big|> r_N \Big) <+\infty.
\]
\end{enumerate}
Then, $\rho$ is $\widetilde{\mP}_b(\setX)$-consistent.
\end{thm}

The strategy of the proof is the following. In a first step, we show that when $\sigma_N^*$ is a permutation of $1:N$  which corresponds to ordering input particles using the Hilbert space filling curve (details of which we postpone to Section \ref{sec:Hilbert}),
the resampling scheme $\rho$ is $\widetilde{\mP}_b(\setX)$-consistent \textit{if and only if}
\begin{align}\label{eq:sigma_star}
\lim_{N\rightarrow+\infty}\max_{m\in 1:N}\big|\sum_{n=1}^m\Delta_{\rho,z_N}^{\sigma_N^*(n)}\big|=0,\quad\P-a.s.
\end{align}
for any sequence $(z_N)_{N\geq 1}$ with $z_N\in\mathcal{Z}^N$.
In a second step, we show that the hypotheses \ref{H1} and \ref{H2} are sufficient
to establish \eqref{eq:sigma_star}, via a maximal inequality for negatively
associated random variables due to \citet{MR1777538}. We stress here that the permutation $\sigma_N^*$ is introduced solely as a device in the proof; there is no assumption in Theorem \ref{thm:NA_resampling} that the input particles are \emph{algorithmically} sorted in any particular way. The reader should note, in fact, that \ref{H1} must
hold for all $z$, and \ref{H2} is uniform in $z$, and hence all permutations of
the input particles. 


\subsection{Discussion of \ref{H1} and \ref{H2}}

From the definition of $\#^n(U,z)$ given in \eqref{eq:def_Psi} it follows  that 
$\sum_{n=1}^N \#^n(U,z)=N$, $\P$-as. Intuitively, this constraint suggests that
at least some random variables in the set $(\#^n(U,z))_{n=1}^N$ are negatively
correlated.  \ref{H1} may be understood as imposing that \emph{all} these
random variables are negatively correlated.

\ref{H2} alone is not sufficient to guarantee the consistency of an unbiased
resampling scheme.  If a resampling  scheme $\rho$ violates \ref{H1} then it is
indeed  possible to find examples where the offspring numbers are positively correlated in a way that, with positive
probability, prevents  the limit in \eqref{eq:sigma_star} from being zero. The next
result formalizes this assertion in the context of systematic resampling. Its proof involves a somewhat technical construction of a counter-example.


\begin{prop}\label{prop:counterExample}
The systematic resampling scheme $\rho_{\mathrm{syst}}$ is unbiased,   satisfies \ref{H2} with $r_N=1$ but is not  $\widetilde{\mP}_b(\setX)$-consistent.
\end{prop}

On the other hand,  \ref{H1}  alone is not enough to  guarantee consistency. If we consider the resampling scheme $\rho$   such that $\#^n(U,z)=N$ with probability $w^n$, it is easily checked $\rho$ is unbiased and \ref{H1} holds, but this resampling scheme is obviously
not $\widetilde{\mP}_b(\setX)$-consistent. \ref{H2} rules out this kind of situation via constraints on the second moments and negligibility of the deviations of the offspring numbers $(\#^n(U,z))_{n=1}^N$ from their respective means $(Nw^n)_{n=1}^N$.

\subsection{Some comments about systematic resampling}

Systematic resampling has the property that $\#^n(U,z)$ is either $\lfloor N
w^n\rfloor$ or $\lfloor N w^n\rfloor + 1$, $\P$-a.s., hence
$|\Delta_{\rho_\syst,z^N}^n|\leq 1$, $\P$-a.s., so that \ref{H2} holds with
$r_N=1$ as stated in Proposition \ref{prop:counterExample}.

A corollary of this latter is that systematic resampling violates \ref{H1}. A simple way to establish this result is to take a $z\in\mathcal{Z}^N$ such that we have $Nw^n-\lfloor N w^n\rfloor=1/2$ for $n=1,\dots,3$. Then,
$$
\P\big(\#^1(U,z)=\#^3(U,z)=1\big)=\frac{1}{2}>\P\big(\#^1(U,z)=1) \P(\#^3(U,z)=1\big)=\frac{1}{4}
$$
showing that the collection of random variables $(\#^n(U,z))_{n=1}^N$ is not NA.

To overcome the lack of consistency (in the sense of Definition \ref{def:resampling}) of systematic resampling we introduce  below (Section \ref{subsec:SSP}) a new resampling scheme, named  SSP (for Srinivasan Sampling Process) resampling, which both satisfies the NA condition \ref{H1} and shares the property of systematic resampling  that $|\Delta_{\rho_\syst, z^N}^n|\leq 1$ for all $n\in 1:N$, $\P$-a.s., so that \ref{H2} also holds with $r_N=1$ for this new resampling scheme.

\subsection{Applications of Theorem \ref{thm:NA_resampling}}

\subsubsection{Multinomial resampling}\label{subsec:Multinomal}

As already mentioned, it is a known result  that multinomial resampling is
$\mP(\setX)$-consistent for any measurable $\setX\subseteq\mathbb{R}^d$
\citep[][Lemma 2]{crisan2002survey}. Theorem
\ref{thm:NA_resampling} may be applied to obtain a similar result.

\begin{corollary}\label{cor:multi}
Let $\setX$ be a cubifiable set. Then, the resampling scheme $\rho_\multi$ verifies conditions \ref{H1} and \ref{H2} of Theorem \ref{thm:NA_resampling} and is therefore  $\widetilde{\mP}_b(\setX)$-consistent.
\end{corollary}

Condition \ref{H1} holds for multinomial resampling as shown by \cite{joag1983negative} while \ref{H2} is verified using properties of the binomial distribution and Hoeffding's inequality.


For similar reasons,
the conditions of Theorem \ref{thm:NA_resampling} are also satisfied by
the residual resampling scheme of \citet{LiuChen}.

\subsubsection{Stratified resampling}\label{subsec:Stratified}

To the best of our knowledge the following corollary of Theorem \ref{thm:NA_resampling} is the first  almost sure weak convergence
result  for Kitagawa's \citeyearpar{Kitagawa1996} stratified resampling scheme.

\begin{corollary}\label{cor:stratified}
Let $\setX$ be a  cubifiable set. Then  the resampling scheme $\rho_{\strat}$
verifies conditions \ref{H1} and \ref{H2} of Theorem \ref{thm:NA_resampling} and is
therefore  $\widetilde{\mP}_b(\setX)$-consistent.
\end{corollary}

Verifying \ref{H1} in this situation involves the observation that stratified resampling is a ``Balls and Bins'' experiment  \citep{MR1642566}  in which  $N$ balls are independently thrown into $N$ bins, the total number of balls occupying the $n$th bin is $\#^n(U,z)$, and  where the probability of falling in a given bin varies across balls, due to the stratified nature of the sampling. The fact that \ref{H1} holds is then a direct consequence  of Theorem 14 in  \cite{MR1642566}, which establishes the NA of occupancy numbers in a slightly  more general balls and bins problem  where the number of balls is not necessarily equal to the number of bins. \ref{H2} holds because $|\Delta_{\rho_{\strat}, z}|\leq 2$, $\P$-a.s.



It is worth noting that the conditions of Theorem \ref{thm:NA_resampling} are also satisfied by the stratified version of the residual resampling scheme of \citet{LiuChen}, where the multinomial resampling part is replaced by a stratified resampling step. Denoting these two resampling schemes by $\rho_{\mathrm{res/multi}}$ and $\rho_{\mathrm{res/strat}}$ respectively, the stratified version of residual resampling has the interesting property that, for any  measurable  $\varphi:\setX\rightarrow\mathbb{R}$  we have \citep[see][for the second inequality]{Douc2005}
$$
\var\big[\rho_{\mathrm{res/strat}}(z)(\varphi)\big]\leq  \var\big[\rho_\mathrm{res/multi}(z)(\varphi)\big]\leq \var\big[\rho_\multi(z)(\varphi)\big],\quad\forall z\in\mathcal{Z}.
$$
In addition, $\rho_{\mathrm{res/strat}}$ has the advantage to be easier and slightly cheaper to implement than $\rho_{\mathrm{res/multi}}$.

\subsubsection{SSP resampling}\label{subsec:SSP}



The underlying idea of SSP resampling is to see the  resampling scheme  as a
rounding operation, where the vector of `weights' $(N w^1,\dots, N w^N)$ is
$\P$-a.s. transformed into a point $(Y_1(U),\dots, Y_N(U))$ in $\mathbb{N}^N$
satisfying the constraint $\sum_{n=1}^N Y_n(U)=N$.

Before proceeding further we recall the terminology   that, for $\xi\in\mathbb{R}_+$, a random variable $Y:\Omega\rightarrow \mathbb{N}$ is called a randomized rounding of $\xi$ if
$$
\P\big(Y=\lfloor \xi\rfloor +1\big)= \xi-\lfloor \xi\rfloor ,\quad  \P\big(Y=\lfloor \xi\rfloor \big)=1-\big(\xi-\lfloor \xi\rfloor\big).
$$
Hence, any algorithmic technique for constructing randomized roundings that takes as input a vector $(\xi_1,\dots,\xi_N)\in  \mathbb{R}_+^N$ and returns $\P$-a.s. as output a vector $(Y_1(U),\dots, Y_N(U))\in\mathbb{N}^N$
verifying
$\sum_{n=1}^N Y_n(U)=\sum_{n=1}^N\xi_n
$
may be used to construct an unbiased resampling mechanism; systematic
resampling can be viewed as being constructed in this way.

The SSP resampling  scheme $\rho_{\ssp} :[0,1]^\mathbb{N}\times
\mathcal{Z}\rightarrow \Pf$ is based  on the Srinivasan's
\citeyearpar{Srinivasan2001} randomized rounding technique \citep[also known as pivotal sampling in the sampling survey literature, see e.g.][]{Deville1998} and is presented in
Algorithm \ref{alg:SSP}. To see that this latter  indeed defines a randomized
rounding process it suffices to note that step (2) leaves unchanged the
expectation of the vector $(Y_\ssp^n(U))_{n=1}^N$ while, by construction, each
iteration of the algorithm leaves the quantity  $\sum_{n=1}^N Y_\ssp^n(U)$
unchanged with $\P$-probability one. By  \citet[][Theorem 5.1; see also
\citealp{Kramer2011}]{Dubhashi2007}  the collection of random variables
$(Y_\ssp^n(U))_{n=1}^N$ produced by the SSP described in Algorithm
\ref{alg:SSP} is NA. Together with Theorem \ref{thm:NA_resampling}, this
result allows to readily show the consistency of  $\rho_{\ssp}$.



\begin{corollary}\label{cor:ssp}
Let $\setX$ be a  cubifiable set. Then,  the resampling scheme $\rho_{\ssp}$ verifies conditions \ref{H1} and \ref{H2} of Theorem \ref{thm:NA_resampling} and is therefore  $\widetilde{\mP}_b(\setX)$-consistent.
\end{corollary}

Algorithm \ref{alg:SSP} has complexity $\bigO(N)$, like other standard resampling schemes.
An open question is whether or not SSP resampling
dominates multinomial resampling in terms of variance. See Section
\ref{sec:numerics} for a numerical comparison.

Lastly in this section, we note that a resampling scheme proposed in \citet{Crisan} may also be
interpreted as a randomized rounding technique. However, to the best of our
knowledge, there are no convergence results for this resampling scheme.

\begin{algorithm}
\caption{\label{alg:SSP}  SSP resampling }
\begin{description}
\item\textbf{Inputs:} $u\in [0,1]^{\mathbb{N}}$ and $(\xi_1,\dots,\xi_N)\in\mathbb{R}_+^N$ such that $\sum_{n=1}^N\xi_n\in\mathbb{N}$.

\item\textbf{Output:} $\big(Y_\ssp^1(u),\dots,Y_\ssp^N(u)\big)\in\mathbb{N}^N$ such that $\sum_{n=1}^N Y_\ssp^n(u)=\sum_{n=1}^N\xi_n$.
\vspace{0.1cm}

\item[]Initialization:  $\big(Y_\ssp^1(u),\dots,Y_\ssp^N(u)\big)\gets (\xi_1,\dots,\xi_N)$, $(n,m,k)\gets (1,2,1)$
\vspace{0.1cm}

\item []Iterate the following steps until $\big(Y_\ssp^1(u),\dots,Y_\ssp^N(u)\big)\in\mathbb{N}^N$:
\item[]\textbf{(1)} Let $\delta$ be the smallest number in $(0,1)$ such that at least one of $Y_\ssp^n(u)+\delta$ or $Y_\ssp^m(u)-\delta$ is an integer, and let $\epsilon$ be the smallest number in $(0,1)$ such that at least one of $Y_\ssp^n(u)-\epsilon$ or $Y_\ssp^m(u)+\epsilon$ is an integer.
\item[]\textbf{(2)}  If $u_k\leq \epsilon/(\delta+\epsilon)$ set $(Y_\ssp^n(u),Y_\ssp^m(u))\gets (Y_\ssp^n(u)+\delta,Y_\ssp^m(u)-\delta)$; otherwise set  $(Y_\ssp^n(u),Y_\ssp^m(u))\gets (Y_\ssp^n(u)-\epsilon,Y_\ssp^m(u)+\epsilon)$.
\item[]\textbf{(3)} Update $n$ and $m$ as follows:
\begin{enumerate}
\item If $\big(Y_\ssp^n(u), Y_\ssp^m(u)\big)\in\mathbb{N}^2$, $(n,m)\gets (m+1,m+2)$;
\item If $Y_\ssp^n(u)\in\mathbb{N}$ and $Y_\ssp^m(u)\not\in\mathbb{N}$ set $(n,m)\gets(m,m+1)$;
\item if $Y_\ssp^n(u)\not\in\mathbb{N}$ and $Y_\ssp^m(u) \in\mathbb{N}$ set $(n,m)\gets(n,m+1)$.
\end{enumerate}
\item\textbf{(4)} $k\gets k+1$
\end{description}
\end{algorithm}


\section{Convergence of  ordered resampling schemes\label{sec:Hilbert}}

\citet[Appendix A]{Kitagawa1996} provided numerical results about the behaviour
of stratified resampling in the case that $d=1$ and the input particles are
ordered according to their states. He conjectured that in this situation, the
error of stratified resampling is of size $\bigO(N^{-2})$, compared to
$\bigO(N^{-1})$ without the ordering. He also considered a deterministic
resampling scheme, and found that in same $d=1$ case and with ordered
particles, it also exhibited $\bigO(N^{-2})$ convergence.

The purpose of this section is to provide a rigorous investigation of this
topic. While \citet{Kitagawa1996} measured the error introduced by a resampling
scheme by the integrated square error between empirical CDF's before and after
resampling, we compare below the probability measures before and after
resampling by comparing their expectations for some  test functions. Notably,
we present in this section results on the convergence rate of the variance of
stratified resampling when applied on ordered input particles. We first
consider the case $d=1$ and then the general $d\geq1$ case in which particles
input to resampling are ordered using the Hilbert space filling curve.




\subsection{Ordered resampling schemes on univariate sets}

In this subsection we present results for a univariate set  $\setX$, which is
the set-up considered by \citet{Kitagawa1996}. The existence of a  natural order in this context
greatly facilitates the presentation and allows to derive more precise
convergence results than in multivariate settings.

 We denote below by $\rho^*_\strat$ the ordered stratified resampling scheme; that is, $\rho^*_\strat:[0,1]^{\mathbb{N}}\times\mathcal{Z}\rightarrow\mP_f(\setX)$ is defined by
$$
\rho^*_\strat(u,z)=\rho_\strat\big(u,(z_{\sigma^*_N(n)})_{n=1}^N\big),\quad (u,z)\in [0,1]^{\mathbb{N}}\times\mathcal{Z}^N
$$
with $\sigma^*_N$   a permutation of $1:N$ such that $z_{\sigma^*_N(1)}\leq\dots\leq z_{\sigma^*_N(N)}$. In words, $\rho^*_\strat(\zeta^N)$ simply  amounts to apply the stratified resampling scheme $\rho_{\strat}$ on the ordered input point set $\zeta^{N,\sigma_N^*}:=(X^{\sigma_N^*(n)}, W^{\sigma_N^*(n)})_{n=1}^N$. Notice that  $\rho^*_\strat(\zeta^N)$ is such that
\begin{align}\label{eq:inv_cdf}
X^{A^n}=F^-_{\pi^N}\bigg(\frac{n-1+U_n}{N}\bigg),\quad n\in 1:N;
\end{align}
that is, the resampled particles $(X^{A^n})_{n=1}^N$ are obtained by sampling from the empirical distribution $\pi^N$ using the stratified point set $((n-1+U_n)/N)_{n=1}^N$.

The following theorem shows that under mild conditions  the variance induced by  ordered  stratified resampling converges faster than $N^{-1}$. In addition, it also provides conditions under which one has a non-asymptotic bound of size $N^{-2}$ for this resampling method.

\begin{thm}\label{thm:univariate_var}
Let $\setX\subseteq \R$ be a cubifiable set. Then, the following results hold:
\begin{enumerate}
\item  Let $\pi\in\widetilde{\mP}_b(\setX)$ have a strictly positive density  and
$(\zeta^N)_{N\geq 1}$ be such that $\pi^N\wc\pi$, $\P$-a.s., and such that, $\lim_{N\rightarrow+\infty}\big(\max_{n\in 1:N} W^{n,N}\big)=0$, $\P$-a.s.
Then, for any $\varphi\in\mathcal{C}_b(\setX)$, $\var_{\zeta^N}\left[\rho^*_{\mathrm{strat}}(\zeta^N)(\varphi)\right]=\smallo(1/N)$, $\P$-a.s.

\item Let $\varphi:\setX\rightarrow\R$ be a continuously differentiable function  such that, for a $\delta>0$, we have $\sup_{x\in\setX} \frac{\dd \varphi}{\dd x}(x)|x|^{1+\delta}<+\infty$. Then, there exists a constant $C_\varphi<+\infty$ such that, for all $N\geq 1$,
$$
\var\left[\rho^*_{\mathrm{strat}}(z)(\varphi)\right]\leq C_\varphi\, N^{-2},\quad \quad\forall z\in\mathcal{Z}^N.
$$
\end{enumerate}
\end{thm}

The second observation of \citet[p.23]{Kitagawa1996} is that deterministic resamplimg mechanisms may be used when applied to  the ordered input particles $\zeta^{N,\sigma_N^*}$. In particular, he considered a resampling scheme  defined by \eqref{eq:inv_cdf} but with the random variables $(U_n)_{n=1}^N$ replaced by a deterministic point in $\alpha\in (0,1)$. In the notation of this work, for $\alpha\in (0,1)$ \citet{Kitagawa1996} considered the resampling scheme $\rho^*_\alpha$ defined by $\rho^*_\alpha(u,z)=\rho^*_\strat(\alpha_{\infty},z)$ with $\alpha_{\infty}$ the vector in $(0,1)^{\mathbb{N}}$ having $\alpha$ in all its entries. The consistency of this deterministic resampling mechanism trivially follows  from Corollary \ref{cor:Hilbert_Res} (see below) and the fact that \citep[][Theorem 2.6, p.15]{Niederreiter1992}
\begin{align}\label{eq:bound_dstar}
\|F_{\rho^*_\alpha(\zeta^N)} -F_{\pi^N}\|_{\infty}\leq \frac{1}{2N}+\Big|\frac{\alpha-1/2}{N}\Big|.
\end{align}
Notice that the right-hand side of this expression is minimized for $\alpha=0.5$. In fact, it is not difficult to check that the resampling scheme $\rho^*_{1/2}$  is optimal  in the sense that it minimises  $\|F_{\rho(\zeta^N)}-F_{\pi^N}\|_\infty$ among all resampling schemes $\rho$. One rationale for trying to minimize this quantity when considering deterministic resampling schemes   is given by the generalized Koksma-Hlawka  \citep[][Theorem 1]{Aistleitner2014} which implies that
\begin{align}\label{eq:KL}
\big|\rho(\zeta^N)(\varphi)-\pi^N(\varphi)\big|\leq  V(\varphi) \|F_{\rho(\zeta^N)}-F_{\pi^N}\|_\infty
\end{align}
with $V(\varphi)$ the variation of $\varphi$ in $\setX$.

We end this subsection  by noting that  inequality \eqref{eq:bound_dstar} shows that  systematic resampling is consistent when applied on the ordered  input particles $\zeta^{N,\sigma_N^*}$.

\subsection{Hilbert-ordered resampling schemes}

In this subsection we generalize the results presented above   to   any dimension $d\geq 1$. The main challenge when $d>1$ is to find an  ordering  of particles $\zeta^N$ which allows to improve upon the un-ordered version of the resampling scheme. Below we consider an ordering based on the Hilbert space filling curve.


\subsubsection{Hilbert space filling curve and related definitions}\label{sub:hilbert_curve}

For $\pi,\pi'\in\mathcal{P}(\setX)$, we use below the shorthand $\|\pi-\pi'\starN=\|F_{\pi}-F_{\pi'}\infN$; note that the `star' metric $\|\cdot\|_\star$ is the multivariate generalization of the Kolmogorov metric. The   star discrepancy of the point set $u_{1:N}$    in $[0,1]^d$ is defined by
$$
\Dst_N(\bu_{1:N})=\big\|N^{-1}\sum_{i=1}^N\delta_{u_i}-\lambda_d\big\|_\star.
$$

\begin{figure}
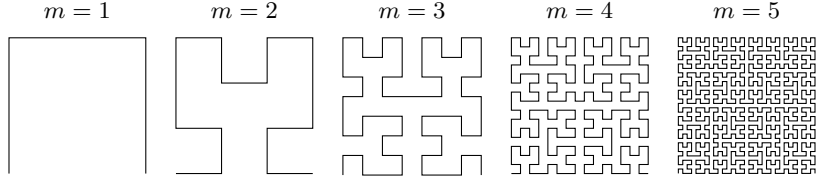
%
    \centering
    \subfloat[$m=1$]{\tikz[scale=18] \hilbert((0mm,0mm),1);}~~
    \subfloat[$m=2$]{\tikz[scale=6] \hilbert((0mm,0mm),2);}~~
    \subfloat[$m=3$]{\tikz[scale=2.6] \hilbert((0mm,0mm),3);}~~
    \subfloat[$m=4$]{\tikz[scale=1.2] \hilbert((0mm,0mm),4);}~~
    \subfloat[$m=5$]{\tikz[scale=0.58] \hilbert((0mm,0mm),5);}%
\caption{The Hilbert curve in dimension $d=2$ is defined as the limit of this sequence. (source: Marc van Dongen) \label{fig:Hilbert}}
\end{figure}

The Hilbert curve is a space-filling curve, that is a continuous surjective function $H:[0,1]\rightarrow [0,1]^d$. It is defined as the limit of the sequence of functions depicted (for $d=2$) in Figure \ref{fig:Hilbert}. Precise details of the construction and some important properties  of the Hilbert curve are given in Section S1.2 of the supplementary materials. In particular, the function $H:[0,1]\rightarrow [0,1]^d$ is H\"older continuous with exponent $1/d$ and is   measure-preserving  in the sense that $\lambda_d(H(I))=\lambda_1(I)$ for any measurable set $I\in[0,1]$. This last property plays a crucial role in the derivation  of the consistency results presented in the next subsection while the H\"older continuity of the Hilbert curve is central in our analysis of  the variance of Hilbert-ordered stratified resampling (Section \ref{sub:variance}).

In the construction of the Hilbert curve one is free to choose the value of $H(0)$, and we shall take it to be $(0,\ldots,0)$. The Hilbert curve admits a one-to-one Borel measurable pseudo-inverse $h:[0,1]^d\rightarrow [0,1]$  such that $H(h(x))=x$ for all $x\in[0,1]^d$, as shown in the next proposition.

\begin{prop}\label{prop:h_measurability}
There exists a one-to-one Borel measurable function $h:[0,1]^d\rightarrow [0,1]$ such that $H(h(x))=x$ for all $x\in [0,1]^d$.
\end{prop}

For $d=1$, we simply take $H(x)=h(x)=x$ for
$x\in[0,1]$.


For a cubifiable set $\setX$ and diffeomorphism $\psi\in\mathcal{D}(\setX)$, we denote by $h_{\setX,\psi}$ the one-to-one mapping $x\mapsto h\circ \psi (x)$.  Remark that  $h_{\setX,\psi}(\setX)=(0,1)$  under the convention $H(0)=(0,\ldots,0)$. To simplify the notation in what follows, we associate to a cubifiable set $\setX$ a   diffeomorphism $\psi_\setX\in\mathcal{D}(\setX)$ and use the shorthand $h_\setX=h_{\setX,\psi_\setX}$. In particular, when $\setX=(0,1)^d$ we assume henceforth that $\psi_\setX(x)=x$ for all $x\in\setX$.


We now define $\sigma^*_N$  as a permutation of $1:N$ such that
$$
h_{\setX}( z_{\sigma_N^*(1)})\leq \ldots\leq h_{\setX}(z_{\sigma_N^*(N)})
$$
 and use it to extend the definition  of the ordered stratified resampling scheme $\rho^*_\strat$ introduced in the previous subsection to any $d\geq 1$; that is, for any $d\geq 1$ we define $\rho^*_\strat:[0,1]^{\mathbb{N}}\times\mathcal{Z}\rightarrow\mP_f(\setX)$ by
$$
\rho^*_\strat(u,z)=\rho_\strat\big(u,(z_{\sigma^*_N(n)})_{n=1}^N\big),\quad (u,z)\in [0,1]^{\mathbb{N}}\times\mathcal{Z}^N.
$$
The resampling scheme  $\rho^*_\strat(\zeta^N)$ is such that
\begin{align}\label{eq:inv_cdf_2}
X^{A^n}=\psi_\setX^{-1}\circ H\Bigg(F^-_{\pi_{h_\setX}^N}\bigg(\frac{n-1+U_n}{N}\bigg)\Bigg),\quad n\in 1:N
\end{align}
and thus $\rho^*_\strat$ amounts to first  sample from the empirical distribution $\pi^N_{h_{\setX}}$ using the stratified point set $( (n-1+U_n)/N)_{n=1}^N$ and then to `project' the resulting sample in the original set $\setX$ using the mapping $\psi_\setX^{-1}\circ H$. Note that representation \eqref{eq:inv_cdf_2} of $\rho^*_\strat$ extends the one given in \eqref{eq:inv_cdf} for $d=1$ to any $d\geq 1$.


The  ordered systematic resampling scheme $\rho^*_\syst$ is defined in a similar way.

Although this is not apparent from the notation, when $d>1$ the resampling schemes  $\rho^*_\strat$ and $\rho^*_\syst$  depend  on $\psi_\setX$ through $\sigma^*_N$, and therefore different choices for $\psi_\setX$ lead to  different resampling mechanisms. Consequently, convergence results for these two resampling schemes will assume that the limiting distribution $\pi$ on $\setX$ belongs to the subset $\mP^*_{b}(\setX)$ of $\widetilde{\mP}_{b}(\setX)$ defined by $\mP^*_{b}(\setX)= \{\pi\in \mP_b(\setX):\,\pi_{\psi_{\setX}}\in\mP_b((0,1)^d)\}$.

 To fix the ideas, when $\setX=\R^d$ one can take for $\psi_\setX$ the diffeomorphism  $\psi(x)=(\tilde{\psi}(x_1),\dots, \tilde{\psi}(x_d))$, with $\tilde{\psi}\in\mathcal{D}(\R)$ defined by
$$
\tilde{\psi}(x)=\frac{1}{2}+\frac{\sqrt{4+x^2}-2}{2x}\ind_{\R\setminus \{0\}}(x),\quad x\in\R.
$$
In this case, following Lemma \ref{lem:tails}, it is easily checked that  $\pi \in \mP^*_b(\setX)$ when $\pi \in\mP_b(\setX)$ is such that  $\forall I\subseteq 1:d$ and  $\forall x_{\setminus I}\in\times_{i\not\in I}\setX_i$ we have, for some $C<+\infty$,
$\sup_{ x_I\in \times_{i\in I}\setX_i} p_\pi(x)\prod_{i\in I} |x_i|^{2}\leq  C$.

\subsubsection{Consistency\label{sub:consistency_H}}

The following theorem provides a necessary and sufficient condition for the consistency of a
resampling scheme.

\begin{thm}\label{thm:Hilbert_Res_Main} Let $\setX$ be a cubifiable set. Then, a  resampling
	scheme $\rho$ is $\widetilde{\mP}_b(\setX)$-consistent if and only if, for any
	$\pi\in\widetilde{\mP}_b(\setX)$ and sequence $(\zeta^N)_{N\geq 1}$ such that
 $\pi^N\wc \pi$, $\P$-a.s., we have
 \begin{equation}
\lim_{N\rightarrow\infty}\|\rho(\zeta^N)_{h_{\setX,\psi}}-\pi^N_{h_{\setX,\psi}}\starN=0,\quad\P-a.s.\label{eq:hilbert_sorted_iff}
\end{equation}
for a $\psi\in\mathcal{D}(\setX)$ such that $\pi_\psi\in\mP_b((0,1)^d)$.
\end{thm}
This result is a consequence of Theorem \ref{thm:Main_Hilbert} (see Appendix
\ref{app:hilbert}) that establishes  the equivalence between the weak
convergence and the convergence in the sense of star metric, and shows that  the
Hilbert curve $H$ and its pseudo-inverse $h$ preserve these two modes of
convergence.




A direct corollary of Theorem \ref{thm:Hilbert_Res_Main}
is that any Hilbert-ordered resampling scheme satisfying the discrepancy condition in \eqref{eq:Star} below is consistent, and in particular the Hilbert-ordered versions of stratified and systematic
resampling are consistent.

\begin{corollary}\label{cor:Hilbert_Res}
Let $\setX$ be a cubifiable set. For each $N\geq1$  and $n\in 1:N$, let $\phi_{N}^n:[0,1]^{\mathbb{N}}\to[0,1]$ be a measurable function and consider a resampling scheme of the form
\begin{equation}\label{eq:aN}
a^n_N(u,\zeta_N)=F_N^{\sigma_N^*,-}(\phi_{N}^n(u))
\end{equation}
with  $F_N^{\sigma_N^*,-}$ the inverse of the CDF $F_N^{\sigma_N^*}(n)=\sum_{m=1}^N W^{\sigma_N(m)} \ind(m\leq n)$, $n\in 1:N$.
Then, a sufficient condition for such a resampling scheme
to be $\mP^*_{b}(\setX)$-consistent is that
\begin{align}\label{eq:Star}
	\lim_{N\rightarrow+\infty}\Dst_{N}\big(\phi_{N}^{1:N}(U)\big)=0,\quad \P-\as
\end{align}
In particular, $\rho^*_\strat$ and $\rho^*_\syst$, which correspond respectively to $\phi_N^n(u)=(n-1+u_n)/N$ and $\phi_N^n(u)=(n-1+u_1)/N$, are $\mP^*_{b}(\setX)$-consistent.
\end{corollary}

\subsubsection{Variance behaviour of Hilbert-ordered  resampling}\label{sub:variance}

The main goal of this subsection is to study in detail the convergence rate of the error variance for Hilbert-ordered stratified resampling.





The next result   generalizes the first part of Theorem \ref{thm:univariate_var} to any $d\geq 1$.
\begin{thm}\label{thm:variance_Stratified}
Let $\setX$ be a cubifiable set,    $\pi\in \mP^*_{b}(\setX)$ have a strictly positive density, and let
$(\zeta^N)_{N\geq 1}$ be such that $\pi^N\wc\pi$, $\P$-a.s., and such that,
$$
\lim_{N\rightarrow+\infty}\big(\max_{n\in 1:N} W^{n,N}\big)=0,\quad\P-a.s.
$$
Then, for any $\varphi\in\mathcal{C}_b(\setX)$,
$$
\var_{\zeta^N}\left[\rho^*_\strat(\zeta^N)(\varphi)\right]=\smallo(1/N),\quad\P-a.s.
$$
\end{thm}

Theorem \ref{thm:variance_Stratified} shows that under mild conditions
Hilbert-ordered stratified resampling outperforms multinomial resampling asymptotically.
The following result establishes its
non-asymptotic behaviour under stronger assumptions on the test function
$\varphi$.

\begin{thm}\label{thm:rate_var}
Let $\setX$ be a cubifiable set and     $\varphi:\setX\rightarrow\mathbb{R}$ be  a measurable function such that there exist  constants $C_{\varphi,\psi}<+\infty$ and   $\gamma\in(0,1]$ verifying
\begin{align*}
\big|\varphi\circ\psi_{\setX}^{-1}(x)-\varphi\circ\psi_\setX^{-1}(y)\big|\leq C_{\varphi,\psi_\setX}\|x-y\|^\gamma_2,\quad\forall (x,y)\in(0,1)^d.
\end{align*}
Then, for any $N\geq 1$ we have
$$
\var\left[\rho^*_\strat(z)(\varphi)\right]\leq\big(2\sqrt{d+3}\big)^{2\gamma}\frac{C_{\varphi,\psi_\setX}^2}{ N^{1+\frac{\gamma}{d}}},\quad\forall z\in\mathcal{Z}^N.
$$
\end{thm}
The key tool to establish this result is the generalized Koksma-Hlawka inequality of   \citet[][Theorem 1]{Aistleitner2014} that we already used in \eqref{eq:KL}.

Note that, because of the use of the Hilbert curve in the resampling mechanism, the rate given in Theorem \ref{thm:rate_var} cannot be improved by assuming differentiability on $\varphi$. This is true  because  the Hilbert curve is nowhere differentiable \citep[see e.g.][Lemma 4.3, p.96]{Zumbusch2003}.
We also note  that the rate reported in Theorem \ref{thm:rate_var} for $\gamma=1$ is in line with the one reported in \citet{He2015}, where for a random  quadrature based on the Hilbert curve a variance of size $\bigO(N^{-1-1/d})$  is found for a class of discontinuous functions having a Lipschitz component.

It should also be clear that the power $1/d$ appearing in the upper bound of Theorem \ref{thm:rate_var} arises because the Hilbert curve is H\"older continuous with exponent $1/d$. This latter is `optimal' in the sense that   $1/d$ is the best  possible H\"older exponent for measure-preserving mappings  from $[0,1]$ onto $[0,1]^d$ \citep[][Lemma 6]{Jaffard2009}. For this reason it   seems hard to improve the upper bound of Theorem \ref{thm:rate_var} by considering an alternative ordering of the particles.


An interesting property of Theorem \ref{thm:rate_var} is that it holds for any $N\geq 1$ and requires no conditions on the weights and on the existence of a  $\pi\in\mP(\setX)$ such that $\pi^N\wc \pi$. At the same time, this  suggests that the rate of $N^{1+\gamma /d}$ is not optimal when a limiting distribution $\pi$ exists. Indeed, Theorem \ref{thm:rate_var} does not take into account  that, in the definition of $\rho^*_\strat(\pi^N)$ given in \eqref{eq:inv_cdf_2}, the CDF  $F_{\pi_{h_\setX}^N}$ may converge to $F_{\pi_{h_\setX}}$, the CDF of $\pi_{h_\setX}$, which is potentially a `smooth' function. This point is corrected in the next result.

\begin{thm}\label{thm:rate_var2}
Consider the set-up of Theorem \ref{thm:rate_var},  let $(\zeta^N)_{N\geq 1}$ and $\pi\in \mP^*_{b}(\setX)$  be as in Theorem \ref{thm:variance_Stratified} and assume that
\begin{align}\label{eq:condition:variance}
\var_{\zeta^N}\bigg[\frac{1}{N}\sum_{n=1}^N F_{\pi_{h_{\setX}}}^-\Big(\frac{n-1+U_n}{N}\Big)\bigg]=\smallo(N^{-2}),\quad \P-a.s.
\end{align}

Then, for any measurable function $\varphi:\setX\rightarrow\mathbb{R}$   satisfying the condition  of Theorem \ref{thm:rate_var}, we have
\begin{align}\label{eq:rate_strat}
\var_{\zeta^N}\left[\rho^*_\strat(\zeta^N)(\varphi)\right]=\smallo\big(N^{-(1+\frac{\gamma}{d})}\big),\quad\P-a.s.
\end{align}
When there exists a constant $c>0$ such that $c^{-1}\,\lambda_d(A)\leq \pi(A)\leq c\,\lambda_d(A)$ for all measurable set $A\subseteq\setX$ condition \eqref{eq:condition:variance} is   verified.
\end{thm}

We  note that   the rate in \eqref{eq:rate_strat} does not only  depend on the underlying rate in \eqref{eq:condition:variance} but also on the speed at which $\pi^N$ converges (in some sense) to $\pi$. More precisely, the rate in \eqref{eq:rate_strat} depends on the rate at which the quantity $v_N:=\|F_{\pi^N_{h_\setX}}^-(u)-F_{\pi_{h_\setX}}^-(u)\|_\infty$ converges to 0 as $N\rightarrow+\infty$. In particular, under the extra assumptions of the second part of the theorem, the rate in \eqref{eq:rate_strat}  becomes $\bigO\big(N^{-(1+\frac{2\gamma}{d})}\big)$  when  $v_N=\bigO(1/N)$.

\section{Implications for particle algorithms}\label{sec:SMC}

We apply in this section our previous results to the study of
particle algorithms.

\subsection{Set-up}

We consider a generic \FK{} model, consisting of (a)
a Markov chain, with initial distribution $\mu(\dx_{0})$,
Markov kernels $M_{t}:\setX \rightarrow \mathcal{P}(\setX)$, $t\geq 1$, acting from $\setX$ to itself;
and (b) a sequence of measurable functions,
$G_0:\setX \rightarrow \Rp$, $G_t:\setX\times\setX \rightarrow \Rp$
for $t\geq 1$. The corresponding \FK{} distributions are defined as:

\[
\mathbb{Q}_{t}(\dd x_{0:t})=\frac{1}{L_{t}}\mu(\dx_{0})G_{0}(x_{0})\prod_{s=1}^{t}M_{t}(x_{t-1},\dx_{t})G_{s}(x_{s-1},x_{s})
\]
where
$$
L_t=\int_{\setX^{t+1} }\mu(\dx_{0})G_{0}(x_{0})\prod_{s=1}^{t}M_{t}(x_{t-1},\dx_{t})G_{s}(x_{s-1},x_{s}),
$$
assuming $L_t>0$. In practice,
we are usually interested in approximating the so-called filtering distributions,
i.e. the marginal distributions $\pi_t(\dx_t)=\int_{x_{0:t-1}\in\setX^t} \Q_t(\dx_{0:t})$.
We also define
$ \ell_{t}=L_{t}/L_{t-1} = \left(\Q_{t-1}M_{t}\right)(G_{t}) $
and the operators, $V_0(\varphi)=\eta( \{\varphi-\eta(\varphi)\}^2)$, and for $t\geq 1$,
\[
V_{t}(x_{t-1},\varphi)=M_{t}\left(x_{t-1},\left\{ \varphi-M_{t}(\varphi)\right\} ^2\right),
\]
where $M_t(x_{t-1},\varphi) \eqdef \int_\setX \varphi(x_t) M_t(x_{t-1},\dx_t)$, and
$M_t(\varphi)$ is the function $x_{t-1}\rightarrow M_t(x_{t-1},\varphi)$.

The subsequent results will rely on the following assumptions.
\begin{description}
	\item [{(G)}] Functions $G_{t}$ are continuous and upper bounded.
\item [{(M)}] The Markov kernels $M_t$ define a Feller process; i.e. $M_t(\varphi)\in\contb (\setX)$ for
	all $\varphi\in\contb(\setX)$.
\end{description}

A standard particle filter (Algorithm \ref{alg:Generic-SMC-algorithm})
generates at iteration $t$ a weighted sample, $(X_t^{n},W_t^{n})_{n=1}^N$, which approximates
$\pi_t$ through the random measure 
$\pi_t^N(\dx_t) = \sum_{n=1}^N W_t^n \delta(X_t^n)$.

\begin{algorithm}[H]
\caption{Standard particle filter\label{alg:Generic-SMC-algorithm}}
\begin{description}
\item At time 0:
\begin{description}
\item\textbf{(a)}  Generate (for $n\in 1:N$) $X_{0}^{n}\sim \mu(\dx_{0})$.
\item\textbf{(b)} Compute  (for $n\in 1:N$) $w_{0}^{n}=G_{0}(X_{0}^{n})$ and  $W_{0}^{n}=w_{0}^{n}/\sum_{m=1}^{N}w_{0}^{m}$.
\end{description}

\item Recursively, for times $t=1,\ldots,T$:
\begin{description}
\item \textbf{(a)} Resample: for a given resampling scheme $\rho$,
	generate ancestor variables $A_{t}^{1:N}$, where $A_t^n = a_N^n(U_t,\zeta_{t-1}^N)$,
	$U_t\sim \P$, and
	$\zeta_{t-1}^N=(X_{t-1}^{n},W_{t-1}^{n})_{n=1}^N$
	(as in Definition \ref{def:resampling}).
\item\textbf{(b)} Generate (for $n\in 1:N$) $X_{t}^{n}\sim M_{t}(X_{t-1}^{A_{t}^{n}},\dx_{t})$.
\item\textbf{(c)} Compute (for $n\in 1:N$)  $w_{t}^{n}=G_{t}(X_{t-1}^{A_{t}^{n}},X_{t}^{n})$
and $W_{t}^{n}=w_{t}^{n}/\sum_{m=1}^{N}w_{t}^{m}$.
\end{description}
\end{description}
\end{algorithm}

\subsection{Consistency}

We first state an almost sure   weak convergence result for Algorithm
\ref{alg:Generic-SMC-algorithm} under the condition that $\rho$ is consistent
for a suitable class of distributions \citep[see][Theorem 2.3.2, p.23, for a proof]{Crisan}.

\begin{prop}\label{prop:consistency}
	Let $\mP_0\subseteq\mP(\setX)$ and assume that the \FK{} model defined by $(G_t)_{t\geq 0}$, $\mu$ and $(M_t)_{t\geq 1}$ is such that Assumptions (G) and (M) hold, and that
	$\pi_t\in \mathcal{P}_0$ for all $t\geq 0$.
	Then, for any   $\mathcal{P}_0$-consistent resampling scheme $\rho:[0,1]^{\mathbb{N}}\times \mathcal{Z}\rightarrow\mP_f(\setX)$ and $t\geq 0$, the particle approximation $\pi_t^N:=\sum_{n=1}^N W_t^n\delta(X_t^n)$ of $\pi_t$ generated by Algorithm \ref{alg:Generic-SMC-algorithm} is such that
\begin{align}\label{eq:limit_pi}
\pi_t^N\wc \pi_t,\quad  \P-a.s.
\end{align}
\end{prop}

As a corollary, when $\setX$ is a cubifiable set and the assumptions of the proposition are satisfied with $\mP_0=\widetilde{\mP}_b(\setX)$, this result shows that Algorithm
\ref{alg:Generic-SMC-algorithm} based on stratified and SSP resampling is consistent in the sense that \eqref{eq:limit_pi} holds for any $t\geq 0$.


We recall that \eqref{eq:limit_pi} implies that, for any
$\varphi\in\mathcal{C}_b(\setX)$, $\pi^N_t(\varphi)\rightarrow\pi_t(\varphi)$,
$\P$-a.s. When stratified resampling is used in  Algorithm
\ref{alg:Generic-SMC-algorithm}  we note that, because this resampling
mechanism dominates multinomial resampling in term of variance (see Section
\ref{sub:resampling_schemes}), it also holds true that
$\lim_{N\rightarrow+\infty} N
\E\big[(\pi^N_t(\varphi)-\pi_t(\varphi))^2]<+\infty $ for any
	$\varphi\in\mathcal{C}_b(\setX)$. For unbounded measurable function
	$\varphi:\setX\rightarrow\R$ such that $\pi_t(\varphi)<+\infty$, the
	results in \citet[][Chapter 9]{MR2159833} imply that
	$\pi^N_t(\varphi)\rightarrow\pi_t(\varphi)$ in $\P$-probability. 


\subsection{Central limit theorem}

As shown in  the previous section, the `noise' introduced by the Hilbert
ordered stratified resampling scheme $\rho^*_\strat$  converges to zero faster
than the usual $\bigO(N^{-1})$ Monte Carlo   rate. The next
result formalises the intuitive idea that,  when Algorithm
\ref{alg:Generic-SMC-algorithm} is based on this resampling mechanism, the
resampling step  does not contribute to the asymptotic variance of the
quantity $N^{1/2}\big(\pi_t^N(\varphi)-\pi_t(\varphi)\big)$. For sake of
completeness, Theorem \ref{thm:clt} also presents results for the
multinomial resampling ($\rho_\multi)$ and  residual reampling
($\rho_{\res/\multi}$) schemes for which a central limit theorem also exists
\citep[see][]{Chopin:CLT, Kunsch:CLT, Douc2005}.

\begin{thm}\label{thm:clt}
For Algorithm \ref{alg:Generic-SMC-algorithm}, assuming that $\setX$ is a cubifiable set, $\pi_t\in\mP^*_b(\setX)$ for all $t\geq 0$,
$\rho\in\{\rho_\multi,\rho_{\res/\multi},\rho^\star_\strat\}$
and that the \FK{} model fulfils assumptions (G) and (M), for any
test function $\varphi\in \mathcal{C}_{b}(\mathcal{X})$  we have that (for any $t\geq 0$)
\[
N^{1/2}\left\{ \sum_{n=1}^{N}W_{t}^{n}\varphi(X_{t}^{n})-\pi_{t}(\varphi)\right\}   \wc \mathcal{N}_d\left(0,\mathcal{V}_{t}\left[\varphi\right]\right)\label{eq:clt_filt}
\]
where the $\mathcal{V}_t(\varphi)$ are defined recursively as follows: $\widetilde{\mathcal{V}}_0\left[\varphi\right] = V_0(\varphi)$,
\begin{align*}
\mathcal{V}_{t}\left[\varphi\right] & =\frac{1}{\ell_{t}^{2}}\widetilde{\mathcal{V}}_{t}\left[G_{t}\left\{ \varphi-\pi_{t}(\varphi)\right\} \right]\\
\widehat{\mathcal{V}}_{t}\left[\varphi\right] & =\mathcal{V}_{t}\left[\varphi\right]+R_{t}\left(\rho, \varphi\right)\\
\widetilde{\mathcal{V}}_{t+1}\left[\varphi\right] & =\widehat{\mathcal{V}}_{t}\left[M_{t+1}(\varphi)\right]+\pi_{t}\left[V_{t+1}(\varphi)\right]
\end{align*}
and
\[
	0 = R_t(\rho^\star_{\mathrm{strat}}, \varphi)
	\leq R_t(\rho_{\res/\multi}, \varphi)
\leq R_t(\rho_\multi, \varphi).
\]

\end{thm}

The proof is a simple combination of Theorem \ref{thm:variance_Stratified} and the proofs in the aforementioned papers (see the supplementary materials).

An obvious corrolary of this theorem is that ordered stratified resampling dominates
multinomial and residual resampling, in terms of the asymptotic variance of particle estimates
generated by a particle filter. In fact, since the contribution of the resampling step
is zero when ordered stratified resampling is used, this particular scheme may be
declared as optimal (again, relative to the asymptotic variance for any test function).

\subsection{A note on the auxiliary particle filter}\label{sub:APF}

The auxiliary particle filter (APF, \citealp{PittShep}) is a variation on the standard
particle filter, where the resampling weights are `twisted' using some function
$\eta_t:\setX \rightarrow\R_{>0}$; that is, the resampling weight of ancestor $X_{t-1}^m$ is
$\widetilde{W}_{t}^m \propto W_{t-1}^m \times\eta_{t-1}(X_{t-1}^m)$; $\sum_{n=1}^N \widetilde{W}_{t-1}^n = 1$.
When a particle $X_t^n$ originates from ancestor $X_{t-1}^m$, i.e. $A_t^n = m$,
it is assigned (un-normalised) weight $w_t^n= G_t(X_{t-1}^m,X_t^n)W_{t-1}^m/\widetilde{W}_{t-1}^m$, so as to correct
for the discrepancy between the resampling weights and the actual weights.

Of particular interest is particle estimate
\[
\ell_{t}^{N}=\frac{1}{N}\sum_{n=1}^{N}w_{t}^{n}=\frac{1}{N}\sum_{n=1}^{N}\frac{W_{t-1}^{A_{t}^{n}}}{\widetilde{W}_{t-1}^{A_{t}^{n}}}G_{t}(X_{t-1}^{A_{t}^{n}},X_{t}^{n})
\]
of normalising constant $\ell_{t}$, and the cumulative product $L_t^N = \prod_{s=0}^t \ell_t^N$, which estimates $L_t=\prod_{s=0}^t \ell_t$.
The latter quantity usually corresponds to the likelihood of the data observed up to time $t$ (for a certain model) and thus plays a central role in parameter estimation methods (e.g. particle Markov chain Monte Carlo, \citealp{PMCMC}).

\begin{thm} \label{thm:APF}
Consider the APF Algorithm (as described above), a given \FK{} model such that Assumptions (G)  and (M)
hold, and assume that functions $\eta_{0}$, $\ldots$, $\eta_{t-2}$ are fixed.
For $\rho=\rho_\multi$, the function $\eta_{t-1}(x_{t-1})=\sqrt{M_{t}(x_{t-1},G_{t}^{2})}$
minimises the
variance of particle estimates $\ell_t^N$ and $L_t^N$.

For $\rho=\rho^\star_\strat$, assuming in addition that $\setX$ is a compact cubifiable set, the quantities
$N\var[\ell_{t}^{N}]$ 
and $N\var[L_t^{N}]$
converge to a limit  which is minimal for $\eta_{t-1}=\eta_{t-1}^\star$, where $\eta_{t-1}^\star(x_{t-1})=\sqrt{V_t(x_{t-1},G_t)}$, among functions $\eta_{t-1}\in\contb(\setX)$ that are positive almost everywhere.
 (In particular, $\eta_{t-1}^\star$ itself is assumed to be positive everywhere.)
\end{thm}

The usual recommendation \citep[e.g.][]{johansen2008note} is to take
$\eta_{t-1}(x_{t-1}) = M_t(x_{t-1},G_t)$ (or some approximation of this
quantity).  Under multinomial resampling, and in the `perfectly
adapted` case (where $G_t$ depends only on $x_{t-1}$), the proposition above
shows that this choice is indeed optimal.  Unfortunately it also shows that the
choice of the auxiliary function in the APF should actually depend on the
resampling scheme. This point deserves further study, which we leave for future
research. We refer to \cite{douc2009optimality} for related results on optimal
auxiliary functions (relative to the asymptotic variance for a given test
function) and \cite{cornebise2008adaptive} for some numerical scheme to
approximate these optimal auxiliary functions within a parametric family.  But
again both papers assume multinomial resampling, and their results and proposed
methodology should be adapted if another resampling scheme is used.



\subsection{Numerical experiments}\label{sec:numerics}

We compare in this section the approximation $(\pi^N_t)_{t=0}^T$ of
$(\pi_t)_{t=0}^T$ generated by Algorithm \ref{alg:Generic-SMC-algorithm} under
the resampling schemes  $\rho_\strat$ (stratified resampling), $\rho^*_\strat$
(ordered stratified resampling) and $\rho_\ssp$ (SSP resampling).

Following \citet{Guarniero2016}, we consider the linear Gaussian state-space
models where $X_{0}\sim \mathcal{N}_{d}(0,I_{d})$, and, for $t=1,\dots,T$,
\begin{align*}
X_{t} & =FX_{t-1}+V_{t},&V_{t}\sim \mathcal{N}_{d}(0,I_{d}),\\
Y_{t} & =X_{t}+W_{t},  &W_{t}\sim \mathcal{N}_{d}(0,I_{d}),
\end{align*}
with $F=(\alpha^{\left|i-j\right|+1})^d_{i,j=1}$,  $\alpha=0.4$, $T=500$ and
$d=5$. We focus on the problem of estimating the log-likelihood of the model,
$\log p(y_{1:T})$, which is estimated from the output of Algorithm
\ref{alg:Generic-SMC-algorithm} by
$\log L^N_{T}=\sum_{t=0}^T \log \ell^N_t$ (see Section \ref{sub:APF}).

We consider two \FK{} models; a `bootstrap' model, where  the Markov kernel  $M_t(x_{t-1}, \dx_t)$
corresponds to the law of $X_t|X_{t-1}=x_{t-1}$, $G_t(x_{t-1},x_t)$ is the probability density
of $Y_t|X_t=x_t$; and  a `guided' model, where $M_t(x_{t-1}, \dx_t)$ is the Gaussian distribution
$\N_d\left( (y_t+Fx_{t-1})/2, I_d/2\right)$, $G_t(x_{t-1}, x_t)$ is the probability
density of $\N_d(Fx_{t-1}, 2I_d)$ at point $y_t$. Both
\FK{} formalisms are such that $\pi_t$ is the filtering
distribution at time $t$ of the model above. The point of the guided formalism is to reduce
the variance of the weights (at each time $t$), and thus to reduce the variance of particle estimates.

%
%

\begin{figure}
\centering
\includegraphics[scale=0.32]{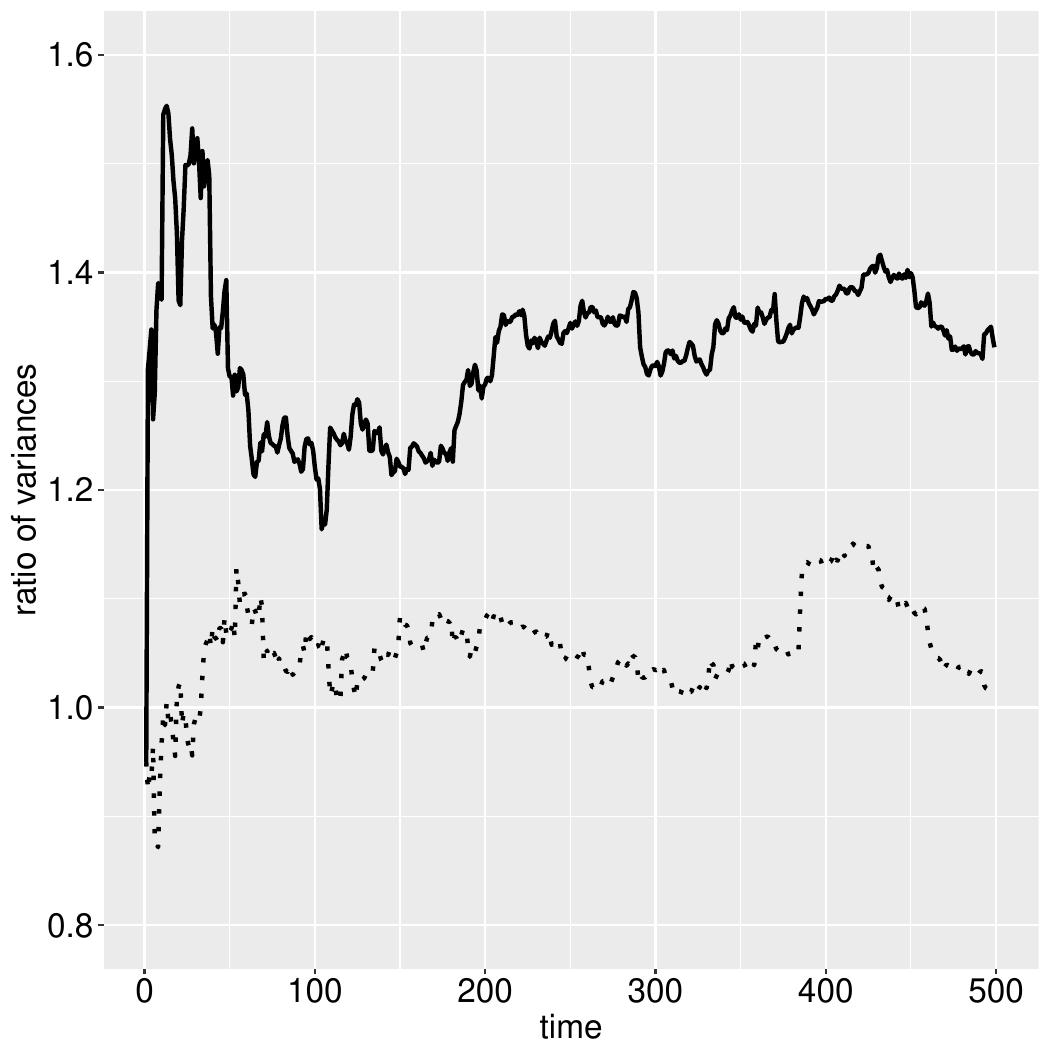}\includegraphics[scale=0.32]{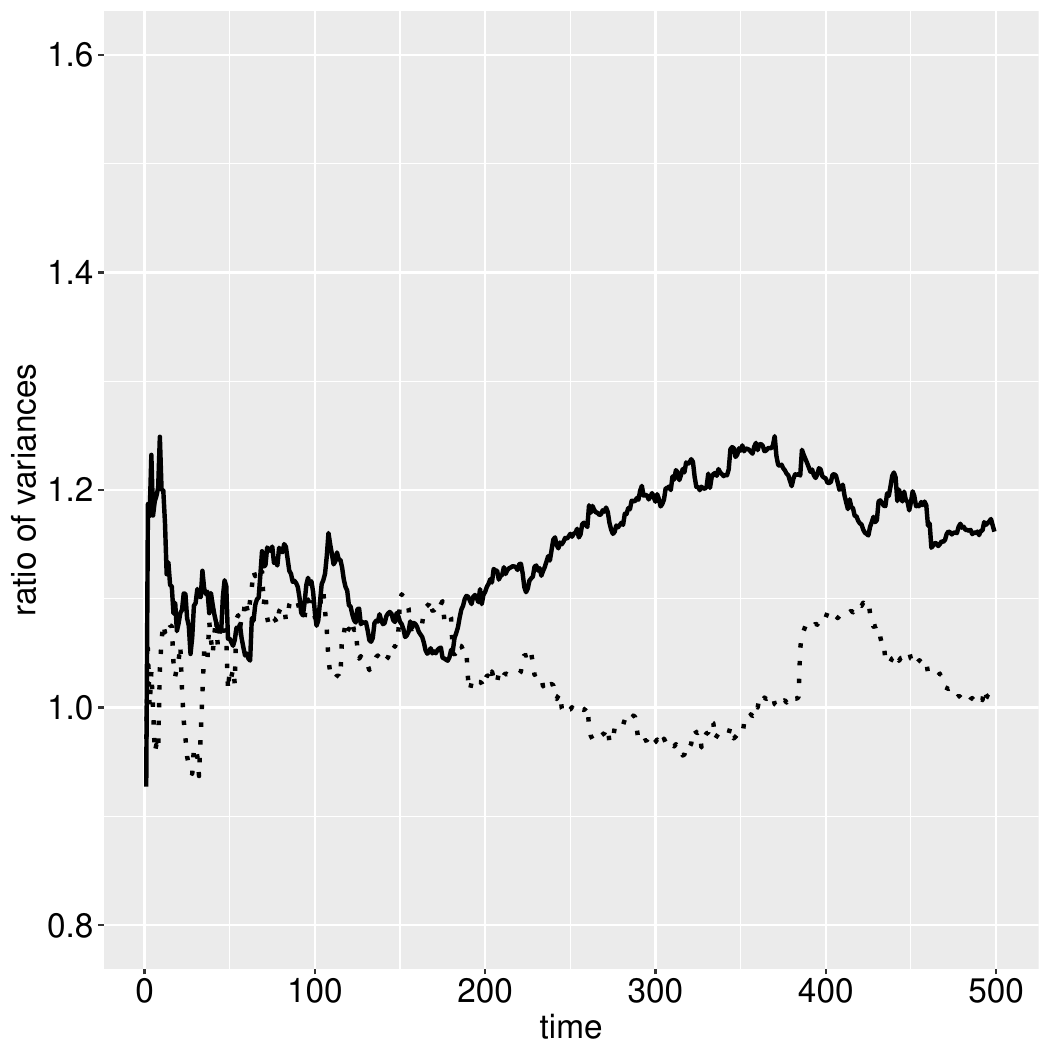}
\caption{Estimation of the log-likelihood function as a function of $t$. The left (resp. right) plot gives  the variance  of SMC based on unordered stratified resampling divided by that of SMC based on Hilbert-ordered stratified resampling (resp. unordered SSP resampling).
Continuous lines are for SMC based on the guided proposal while the dotted line is for the bootstrap particle filter. Results are based on 1\,000 independent runs of the algorithms with $N=2^{13}$ particles.\label{fig:simulations}}
\end{figure}

Figure \ref{fig:simulations} shows the variance of the estimator
$\log L^N_{T}$ obtained under the two above \FK{} formalisms, as a function of
$t\in 1:T$,  and for the resampling schemes $\rho_\strat$, $\rho^*_\strat$ and
$\rho_\ssp$. For each resampling scheme, the results of  Figure
\ref{fig:simulations} are based on 1\,000 independent runs of the two particle algorithms we are considering, with $N=2^{13}$ particles.

As expected from the results of Section \ref{sec:Hilbert}, the variance of
$\log L_t^N$ is smaller with $\rho^*_\strat$ than with $\rho_\strat$;
the relative gains are larger when the guided formalism
is used (where the variances under $\rho_\strat$ are about 40\% higher than
under $\rho^*_\strat$). The results presented in Figure \ref{fig:simulations}
suggest that $\rho_\ssp$ is preferable to $\rho_\strat$. This is particularly
true with the guided formalism where the variances under $\rho_\strat$ are
about 20\% higher than when $\rho_\ssp$ is used. Lastly, the variances under
SSP resampling are larger than under ordered stratified resampling but
$\rho_\ssp$ has the advantage to be faster. Indeed, SSP resampling   requires
$\bigO(N)$ operations against $\bigO(N\log N)$ for $\rho^*_\strat$.

\section{Conclusion}\label{sec:conclusion}



Our results support  the practice in  the SMC literature to abandon multinomial resampling  for stratified resampling by providing  strong theoretical guarantees for this resampling scheme, which has the remarkable property to be both   cheaper  and more accurate than   multinomial resampling. For the same reasons, our results should encourage practitioners to abandon residual resampling for a version of this residual method where the multinomial resampling step is replaced by a stratified resampling step.

The systematic resampling scheme  fails to produce offspring numbers that are negatively associated. As an alternative to it we have introduced the SSP resampling algorithm which   (1) is similar to systematic resampling in term of offspring numbers and (2) verifies the conditions of our general consistency result.  We also built an example suggesting that any general consistency results  for systematic resampling would require to take into account the order of the input particles and have established its validity when they are ordered along the Hilbert curve.


Our practical recommendation is to prefer SSP resampling   to systematic
resampling since both have similar properties while only the former has been
proven to be consistent. Systematic resampling has the advantage to be faster
than SSP resampling  but in most cases this gain is likely to be imperceptible.
Our simulation study suggests that SSP resampling outperforms also stratified
resampling in term of variance but no theoretical result exists to support this
observation.

We have also derived various results showing that the variance of stratified resampling  goes to zero faster than $N^{-1}$ when applied on an input point set ordered along the Hilbert curve, and notably a non-asymptotic bound of size $N^{-1-\frac{1}{d}}$.  Unsurprisingly, when the dimension of the state-space is small and/or when a good proposal distribution is available, our simulation results show that ordering the particle before applying stratified resampling may lead to important variance reduction.  These theoretical results on the variance of Hilbert ordered stratified resamplig are also of particular interest for sequential quasi-Monte Carlo  \citep{MR3351446}, a quasi-Monte Carlo version of SMC, that converges at a faster but currently unknown rate.

\section*{Acknowledgements}

We thank
Patrice Bertail,
Anthony Lee
and
Matthieu Wihelm
for useful comments.
Nicolas Chopin is partly supported by Labex Ecodec (anr-11-labx-0047).

\newpage
\appendix

\section{Convergent sequences of probability measures:  star norm and transformations through the Hilbert curve and its inverse\label{app:hilbert}}
$ $

The following theorem is the main tool for establishing Theorem \ref{thm:Hilbert_Res_Main}.
\begin{thm}\label{thm:Main_Hilbert}
Let $\setX$ be a cubifiable set, $(\pi^N)_{N\geq 1}$ be a sequence in  $\mP(\setX)$, $\pi\in \widetilde{\mP}_b(\setX)$ and $\psi\in\mathcal{D}(\setX)$ be such that $\pi_\psi\in\mP_b((0,1)^d)$. Then, the following assertions are equivalent
\begin{enumerate}
\item[(i)] $\pi^N\wc\pi$;
\item[(ii)] $\lim_{N\rightarrow +\infty}\|\pi^N-\pi\|_\star=0$;
\item[(iii)] $\lim_{N\rightarrow +\infty}\|\pi_{h_{\setX,\psi}}^N-\pi_{h_{\setX,\psi}}\|_\star=0$;
\item[(iv)] $\pi_{h_{\setX,\psi}}^N\wc\pi_{h_{\setX,\psi}}$.
\end{enumerate}
\end{thm}

Implications  $(ii)\Rightarrow(iii)$  and $(iii)\Rightarrow(ii)$ respectively are due to  \citet[][Theorem 3]{MR3351446} and \citet[][Theorem 1]{Schretter2016}.
Implications $(ii)\Rightarrow (i)$ and $(iii)\Rightarrow (iv)$ are direct applications of the Portmanteau lemma  \citep[e.g.][Lemma 2.2, p.6]{MR1652247}.  Implication $(i)\Rightarrow(ii)$ for $d=1$  holds by Poly\`a's theorem (\citealp{Polya1920}; see also \citealp{Bickel1992}, result (A.1)); note that Poly\`a's theorem only requires that $\pi\in\mP(\setX)$ is such that $F_\pi$ is continuous. Implication $(i)\Rightarrow(ii)$ for $d>1$ is new and   proved  following  a similar argument as in \citet[][Theorem 1.2, p.89]{Kuipers1974} while  implication  $(iv)\Rightarrow (iii)$  is a consequence of  Poly\`a's Theorem  and of the continuity of $F_{\pi_{h_{\setX,\psi}}}$, which is established in the next lemma.

\begin{lemma}\label{lemma:continuity}
Let $\setX$ be a cubifiable set, $\pi\in \widetilde{\mP}_b(\setX)$ and   $\psi\in\mathcal{D}(\setX)$ be such that $\pi_\psi\in\mP_b((0,1)^d)$. Then, $\pi_{h_{\setX,\psi}}$ is a continuous probability measure on $(0,1)$.
\end{lemma}

We also note the proofs of implications $(ii)\Rightarrow(iii)$ and $(iii)\Rightarrow(ii)$   in  \citet{MR3351446, gerber2017convergence,Schretter2016}  implicitly assume that the sequence $(\pi^N)_{N\geq 1}$  is such that (with $\setX=(0,1)^d$)
\begin{align*}
\pi^N(\mathcal{H}_d)=0,\quad\text{for all $N$ large enough}
\end{align*}
where  $\mathcal{H}_d$ is the set of points of $[0,1]^d$ that have more than pre-image through $H$.  This point is corrected in the supplementary materials where a complete proof of Theorem \ref{thm:Main_Hilbert} is provided.



\bibliographystyle{apalike}
\bibliography{complete}



\section*{Supplementary material}

# Supplementary materials for "Negative association, ordering, and convergence of resampling methods"

Mathieu Gerber[*]     Nicolas Chopin[†]     Nick Whiteley[*]

## S1 Preliminaries

### S1.1 Additional notation and convention

For a signed measure $\pi$ with respect to $(\mathcal{X}, \mathbb{X})$, we define the extreme norm as

$$\|\pi\|_{\mathrm{E}} = \sup_{[a,b]} \big| \pi([a, b] \cap \mathcal{X}) \big|,$$

and the star norm as

$$\|\pi\|_{\star} = \sup_{(-\infty, b]} \big| \pi((-\infty, b] \cap \mathcal{X}) \big|.$$

In both cases, $a$ and $b$ are vectors in $\mathbb{R}^d$, and the supremums are with respect to multivariate intervals: $[a, b] = \prod_{i=1}^d [a_i, b_i]$, $(-\infty, b] = \prod_{i=1}^d (-\infty, b_i]$.

To avoid tedious repetition of the qualification "$\mathbb{P}$-a.s.", any inequalities involving these quantities are to be understood as holding $\mathbb{P}$-a.s. unless stated otherwise.

When $\mathcal{X}$ is a cubifiable set and $\pi \in \widetilde{\mathcal{P}}_b(\mathcal{X})$, and when dealing with a sequence $(\pi^N)_{n \geq 1}$ in $\mathcal{P}(\mathcal{X})$ that converges to $\pi$ weakly or in the sense of the star/extreme metric, we can assume without loss of generality in our computations that $\mathcal{X} = (0, 1)^d$ and $\pi \in \mathcal{P}_b((0, 1)^d)$. This is indeed true since

$$\|\pi^N - \pi\|_{\star} = \|\pi_{\psi}^N - \pi_{\psi}\|_{\star}, \quad \|\pi^N - \pi\|_{\mathrm{E}} = \|\pi_{\psi}^N - \pi_{\psi}\|_{\mathrm{E}}, \quad \pi^N \overset{\mathrm{w}}{\Rightarrow} \pi \Leftrightarrow \pi_{\psi}^N \overset{\mathrm{w}}{\Rightarrow} \pi_{\psi}$$

while, for a suitable $\psi \in \mathcal{D}(\mathcal{X})$, $\pi_{\psi} \in \mathcal{P}_b((0, 1)^d)$.

Below we abuse notations as follows: $\pi(I) = \pi(I \cap \mathcal{X})$ for $\pi \in \mathcal{P}(\mathcal{X})$ and $I$ any $d$-dimensional interval (possibly not included in $\mathcal{X}$); e.g. for $\mathcal{X} = (0, 1)^d$ and $a \in \mathbb{R}^d$, $\pi([0, a]) = \pi((0, a])$.

---

[*]School of Mathematics, University of Bristol, UK.
[†]CREST-ENSAE, France.

## S1.2 Hilbert space filling curve: Construction and properties

We start by stating some well-known properties of the Hilbert curve (see e.g. Zumbusch, 2003, Chapter 4). The presentation below is inspired by the one in He and Owen (2016).

For $m \geq 1$, let $\mathcal{I}_m^d = \{I_m^d(k)\}_{k=0}^{2^{md}-1}$, with $I_m^d(k) = [k2^{-md}, (k+1)2^{-md}]$, and $\mathcal{S}_m^d = \{S_m^d(k)\}_{k=0}^{2^{md}-1}$ be a collection of closed hyper-cubes of volume $2^{-md}$ that cover $[0,1]^d$. Then, one can define a sequence of mappings $H_m : \mathcal{I}_m^d \to \mathcal{S}_m^d$ such that:

1. $H_m$ is bijective; that is, $H_m(I_m^d(k)) \neq H_m(I_m^d(k'))$ for any $k \neq k'$;

2. The hyper-cubes $H_m(I_m^d(k))$ and $H_m(I_m^d(k+1))$ have one $(d-1)$-dimensional face in common (adjacency property);

3. If we split $I_m^d(k)$ into $2^d$ adjacent intervals $\{I_{m+1}^d(k_i)\}_{i=0}^{2^d-1}$ of length $2^{-(m+1)d}$, then $\cup_{i=0}^{2^d-1} H_{m+1}(I_{m+1}^d(k_i)) = H_m(I_m^d(k))$ (nesting property).

Then, the Hilbert curve is defined as follows. Let $x \in [0,1]$ and note that there exists a sequence $(I_m^d(k_m^x))_{m \geq 1}$ such that (i) $I_{m+1}^d(k_{m+1}^x) \subset I_m^d(k_m^x)$ and (ii) $\{x\} = \cap_{m \geq 1} I_m^d(k_m^x)$. Using the nesting property of $(H_m)_{m \geq 1}$, the set $\cap_{m \geq 1} H_m(I_m^d(k_m^x))$ contains a single point in $[0,1]^d$ and the Hilbert curve $H : [0,1] \to [0,1]^d$ is defined by

$$H(x) \in \cap_{m \geq 1} H_m(I_m^d(k_m^x)), \quad x \in [0,1]^d.$$

Note that Conditions 1-3 listed above do not uniquely define the sequence $(H_m)_{m \geq 1}$ and therefore, although we refer to $H$ as *the* Hilbert curve in this work, there exist in fact several Hilbert curves.

Function $H$ is not bijective. Indeed, if $x \in [0,1]^d$ has at least one dyadic coordinate, then for $m$ large enough there exist at least two distinct indices $k_m$ and $k_m'$ in $0 : 2^{dm} - 1$ such that $x \in S_m^d(k_m) \cap S_m^d(k_m')$; recall that sets in $\mathcal{S}_m^d$ are closed. Since for $m$ large enough $x \in [0,1]^d$ belongs to more than one set in $\mathcal{S}_m^d$ if and only if $x$ has at least one dyadic coordinate, the set $\mathcal{H}_d \subset [0,1]^d$ of points in $[0,1]^d$ that have more than one pre-image through $H$ is such that $\#H^{-1}(x) \leq 2^d$ for all $x \in [0,1]^d$ and such that $\lambda_d(\mathcal{H}_d) = 0$. Lastly, it is easily checked that function $H$ is such that $\lambda_1(A) = \lambda_d(H(A))$ for any measurable set $A \subset [0,1]$ (bi-measure property) and that $\|H(x_1) - H(x_2)\|_\infty \leq C_d |x_1 - x_2|^{1/d}$ for some constant $C_d < +\infty$ and any $x_1, x_2 \in [0,1]$ (i.e. $H$ is Hölder with coefficient $1/d$).

We assume from henceforth that $H$ is such that $H(0) = (0, \ldots, 0)$ and, to simplify the notation, we use the convention that, for any $m \geq 1$, the sets in $\mathcal{S}_m^d$ are labelled so that

$$H(I_m^d(k)) = H_m(I_m^d(k)) = S_m^d(k), \quad \forall k \in 0 : 2^{dm} - 1, \ \forall m \geq 1. \tag{S.1}$$

Thanks to the above properties of the Hilbert curve, there exists a one-to-one Borel measurable function $h : [0,1]^d \to [0,1]$ such that $H(h(x)) = x$ for all $x \in [0,1]^d$, as shown in Proposition 2. For a cubifiable set $\mathcal{X}$ and a $\psi \in \mathcal{D}(\mathcal{X})$ we denote by $h_{\mathcal{X},\psi}$ the mapping $x \mapsto h \circ \psi(x)$. By construction, the mapping $h_{\mathcal{X},\psi} : \mathcal{X} \to (0,1)$ is one-to-one and Borel measurable.

### S1.3 Some preliminary results

The following lemma is a direct extension of Niederreiter (1992, Lemma 2.5, p.15).

**Lemma S1.** *Let $\mathcal{X}$ be a cubifiable set, $\epsilon > 0$ and $z, \tilde{z} \in \mathcal{Z}^N$ be such that*

$$z = (x^n, W^n)_{n=1}^N, \quad \tilde{z} = (\tilde{x}^n, W^n)_{n=1}^N, \quad \max_{n \in 1:N} \|x^n - \tilde{x}^n\|_\infty \leq \epsilon.$$

*Let $\pi^N = \sum_{n=1}^N W^n \delta(x^n)$ and $\tilde{\pi}^N = \sum_{n=1}^N W^n \delta(\tilde{x}^n)$. Then, for any $\pi \in \widetilde{\mathcal{P}}_b(\mathcal{X})$, there exists a constant $c_\pi < +\infty$ (which depends only on $\pi$) such that*

$$\left| \|\tilde{\pi}^N - \pi\|_\star - \|\pi^N - \pi\|_\star \right| \leq c_\pi \, \epsilon$$

*and*

$$\left| \|\tilde{\pi}^N_{h_{\mathcal{X},\psi}} - \pi_{h_{\mathcal{X},\psi}}\|_\star - \|\pi^N_{h_{\mathcal{X},\psi}} - \pi_{h_{\mathcal{X},\psi}}\|_\star \right| \leq c_\pi \max_{n \in 1:N} \|h_{\mathcal{X},\psi}(x^n_{\pi^N}) - h_{\mathcal{X},\psi}(\tilde{x}^n)\|_\infty$$

*with $\psi \in \mathcal{D}(\mathcal{X})$ such that $\pi_\psi \in \mathcal{P}_b((0,1)^d)$.*

*Proof of Lemma S1.* Without loss of generality we assume that $\mathcal{X} = (0,1)^d$ and take $h_{\mathcal{X},\psi} = h$.

Let $B = [0, b] \in [0, 1)^d$, $B^+ = [0, b + \epsilon] \cap [0, 1)^d$ and $B^- = [0, b - \epsilon]$. If $\epsilon > b_i$ for at least one $i \in 1 : d$, $B^- = \varnothing$. Then,

$$\pi^N(B^-) \leq \tilde{\pi}^N(B) \leq \pi^N(B^+). \tag{S.2}$$

By the definition of the star norm, we have

$$\left| \pi^N(B^+) - \pi(B^+) \right| \leq \|\pi^N - \pi\|_\star, \ \left| \pi^N(B^-) - \pi(B^-) \right| \leq \|\pi^N - \pi\|_\star. \tag{S.3}$$

Combining (S.2) and (S.3) yields:

$$\begin{cases} -\left(\pi(B) - \pi(B^-)\right) - \|\pi^N - \pi\|_\star \leq \tilde{\pi}^N(B) - \pi(B) \\ \tilde{\pi}^N(B) - \pi(B) \leq (\pi(B^+) - \pi(B)) + \|\pi^N - \pi\|_\star. \end{cases} \tag{S.4}$$

Then, as $\pi$ admits a bounded density $p_\pi$, we have,

$$\begin{aligned} \pi(B) - \pi(B^-) &\leq \|p_\pi\|_\infty \lambda_d(B \setminus B^-) \leq \|p_\pi\|_\infty d\epsilon, \\ \pi(B^+) - \pi(B) &\leq \|p_\pi\|_\infty \lambda_d(B^+ \setminus B) \leq \|p_\pi\|_\infty d\epsilon. \end{aligned} \tag{S.5}$$

Therefore, combining (S.4) and (S.5), we obtain,

$$-\|p_\pi\|_\infty \, d\epsilon - \|\pi^N - \pi\|_\star \leq \tilde{\pi}^N(B) - \pi(B) \leq \|\pi^N - \pi\|_\star + \|p_\pi\|_\infty \, d\epsilon$$

and thus

$$\|\tilde{\pi}^N - \pi\|_\star \leq \|\pi^N - \pi\|_\star + \|p_\pi\|_\infty \, d\epsilon.$$

To complete the proof of the first part of the lemma it suffices to repeat the above computations while swapping the role of $\pi^N$ and $\tilde{\pi}^N$.

The second part of the lemma follows from similar computations, where $\epsilon$ is replaced by

$$\epsilon' := \max_{n \in 1:N} \|h(x^n) - h(\tilde{x}^n)\|_\infty$$

and where instead of (S.5) we have, by the definition of $\pi_h$ and by the bi-measure property of the Hilbert curve (see Section S1.2),

$$\pi_h(B) - \pi_h(B^-) = \pi\big(H(B \setminus B^-)\big) \le \|p_\pi\|_\infty \lambda_1(B \setminus B^-) \le \|p_\pi\|_\infty \epsilon'$$
$$\pi(B^+) - \pi(B) = \pi\big(H(B^+ \setminus B)\big) \le \|p_\pi\|_\infty \lambda_1(B^+ \setminus B) \le \|p_\pi\|_\infty \epsilon'.$$

$\square$

The next lemma follows from the computations of Gerber and Chopin (2015, Theorem 3).

**Lemma S2.** *Let* $\mu \in \mathcal{P}((0,1)^d)$ *and* $\pi \in \mathcal{P}_b((0,1)^d)$. *Then,*

$$\|\mu - \pi\|_\star = \sup_{0 \le a \le 1} \big|\mu([0,a)) - \pi([0,a))\big|, \quad \|\mu - \pi\|_E = \sup_{0 \le a < b \le 1} \big|\mu([a,b)) - \pi([a,b))\big|$$

*and*

$$\|\mu_h - \pi_h\|_\star = \sup_{0 \le a \le 1} \big|\mu_h([0,a)) - \pi_h([0,a))\big|, \quad \|\mu_h - \pi_h\|_E = \sup_{0 \le a < b \le 1} \big|\mu_h([a,b)) - \pi_h([a,b))\big|.$$

*Proof of Lemma S2.* Below we only prove the second equality, the other ones being proved in a similar way.

Let $p$ be the density of $\pi$ with respect to $\lambda_d$. Let $\epsilon > 0$, $a \in [0,1]$, and $\delta_{a,\epsilon} \in [0, \epsilon/\|p\|_\infty]$, be small enough so that $\mu_h([0,a]) = \mu_h([0, a + \delta_{a,\epsilon}))$ and $a + \delta_{a,\epsilon} \le 1$. (If $a = 1$ or $\mu$ is continuous, take $\delta_{a,\epsilon} = 0$.) Then,

$$|\mu_h([0,a)) - \pi_h([0,a))| \le |\mu_h([0, a + \delta_{a,\epsilon})) - \pi_h([0, a + \delta_{a,\epsilon}))| + \pi_h([a, a + \delta_{a,\epsilon})) \quad (S.6)$$

and

$$|\mu_h([0,a]) - \pi_h([0,a])| \ge |\mu_h([0, a + \delta_{a,\epsilon})) - \pi_h([0, a + \delta_{a,\epsilon}))| - \pi_h([a, a + \delta_{a,\epsilon})). \quad (S.7)$$

By the bi-measure property of the Hilbert curve (see Section S1.2), the set $H((a, a + \delta_{a,\epsilon}))$ has Lebesgue measure $\delta_{a,\epsilon}$. Thus,

$$\pi_h([a, a + \delta_{a,\epsilon})) = \pi\big(H([a, a + \delta_{a,\epsilon}))\big) \le \|p\|_\infty \delta_{a,\epsilon} \le \epsilon.$$

Replacing $\pi_h((a, a + \delta_{a,\epsilon}))$ by $\epsilon$ in (S.6) and (S.7), and taking the supremum other $a$ yields

$$-\epsilon \le \|\mu_h - \pi_h\|_\star - \sup_{0 \le a \le 1} |\mu_h([0,a)) - \pi_h([0,a))| \le \epsilon$$

implying that

$$\|\mu_h - \pi_h\|_\star = \sup_{0 \le a \le 1} |\mu_h([0,a)) - \pi_h([0,a))|.$$

$\square$

The next result notably follows from the computations of Gerber and Chopin (2015, Theorem 7).

**Lemma S3.** *Let $\mathcal{X}$ be a cubifiable set, $\pi \in \widetilde{\mathcal{P}}_b(\mathcal{X})$ be such that $\pi(x) = p(x)\lambda_d(\mathrm{d}x)$ for a strictly positive density $p : \mathcal{X} \to \mathbb{R}_+^*$ and $\psi \in \mathcal{D}(\mathcal{X})$ be such that $\pi_\psi \in \mathcal{P}_b((0,1)^d)$. Then,*

1. *$F_{\pi_{h_{\mathcal{X},\psi}}} : (0,1) \to [0,1]$ is strictly increasing.*

2. *For any sequence $(z^N)_{N \geq 1}$ such that, for all $N \geq 1$, $z^N \in \mathcal{Z}^N$ and such that*
$$\pi^N \overset{w}{\Rightarrow} \pi, \quad \lim_{N \to +\infty} \big( \max_{n \in 1:N} W^{n,N} \big) = 0,$$
   *we have $\lim_{n \to +\infty} \|F_{\pi_{h_{\mathcal{X},\psi}}^N}^- - F_{\pi_{h_{\mathcal{X},\psi}}}^-\|_\infty = 0$.*

3. *For any sequence $(F_N)_{N \geq 1}$ of continuous CDF such that $\lim_{N \to +\infty} \|F_N - F_{\pi_{h_{\mathcal{X},\psi}}}\|_\infty = 0$ we have $\lim_{n \to +\infty} \|F_N^- - F_{\pi_{h_{\mathcal{X},\psi}}}^-\|_\infty = 0$.*

*Proof of Lemma S3.* Without loss of generality we assume that $\mathcal{X} = (0,1)^d$ and take $h_{\mathcal{X},\psi} = h$. Let $(z^N)_{N \geq 1}$ and $\pi$ be as in the statement of the theorem.

We first show the first part of the lemma. Let $0 \leq a < b \leq 1$, and let $k$, $m$ be integers such that $I_m^d(k) \subset [a,b]$ where $I_m^d(k) = [k2^{-md}, (k+1)2^{-md})$ (as defined in Section S1.2). Then

$$F_{\pi_h}(b) - F_{\pi_h}(a) = \pi_h([a,b]) \geq \pi_h(I_m^d(k)) = \pi(S_m^d(k)) > 0$$

where $S_m^d(k) = H(I_m^d(k))$ is a hyper-cube of volume $2^{-m}$, again see Section S1.2. The last inequality comes from the fact that $\pi$ admits a positive density. Thus $F_{\pi_h}$ is increasing.

To establish the second part of the lemma we first show that, for all $u \in (0,1)$.

$$\lim_{N \to +\infty} |F_{\pi_h}^-(u) - F_{\pi_h^N}^-(u)| = 0. \tag{S.8}$$

This result is derived in the computations of Gerber and Chopin (2015, Theorem 7) but for sake of completness it is proved below.

Let $\epsilon > 0$ and $u \in (0,1)$. Because $F_{\pi_h}$ is continuous (Lemma 2) and strictly increasing, $F_{\pi_h}^{-1}$ is continuous and thus there exists a $\delta_{u,\epsilon} > 0$ such that,

$$|u' - u| \leq \delta_{u,\epsilon}, \implies |F_{\pi_h}^-(u') - F_{\pi_h}^-(u)| \leq \epsilon. \tag{S.9}$$

By assumption, for any $\delta_0 > 0$, there exists a $N_{\delta_0}$ such that, for all $N \geq N_{\delta_0}$,

$$\|F_{\pi_h^N} - F_{\pi_h}\|_\infty \leq \delta_0. \tag{S.10}$$

Let $x_N = F_{\pi_h^N}^-(u)$ and $u_N = F_{\pi_h}(x_N)$. Then, by (S.10),

$$|F_{\pi_h^N}(x_N) - F_{\pi_h}(x_N)| \leq \delta_0, \quad \forall N \geq N_{\delta_0}.$$

Let $r_N(u) = F_{\pi_h^N}(F_{\pi_h^N}^-(u)) - u$ so that

$$|F_{\pi_h^N}(x_N) - F_{\pi_h}(x_N)| = |u + r_N(u) - u_N| \leq \delta_0, \quad \forall N \geq N_{\delta_0}$$

Now note that $|r_N(u)| \leq \max_{n \in 1:N} W^{n,N}$ and thus, by assumption, for all $\delta' > 0$, there exists a $N_{\delta'}$ such that, a.s., $|r_N(u)| \leq \delta'$ for all $N \geq N_{\delta'}$. Let $\delta = \delta_0 + \delta'$ and set $N_\delta := N_{\delta_0} \vee N_{\delta'}$. Then, for $N \geq N_\delta$, we have $|u - u_N| \leq \delta$. By taking $\delta_0$ and $\delta'$ such that $\delta = \delta_{u,\epsilon}$, (S.9) implies that

$$|F_{\pi_h}^-(u) - F_{\pi_h}^-(u_N)| \leq \epsilon, \quad \forall N \geq N_{\delta_{u,\epsilon}}.$$

In addition, $F_{\pi_h}^-(u_N) = x_N = F_{\pi_h^N}^-(u)$ and therefore

$$|F_{\pi_h}^-(u) - F_{\pi_h^N}^-(u)| \leq \epsilon, \quad \forall N \geq N_{\delta_{u,\epsilon}}.$$

This shows (S.8).

Then, to show the second part of the theorem remark that, since $F_{\pi_h}^-$ is continuous on $(0,1)$ and such that

$$\lim_{u \to 0} F_{\pi_h}^-(u) = 0, \quad \lim_{u \to 1} F_{\pi_h}^-(u) = 1,$$

the mapping $F_{\pi_h}^-$ can be extended to a continuous function $\tilde{F}_{\pi_h}^- : [0,1] \to [0,1]$ (Mytrofanov and Ravsky, 2012, Lemma 2) which is thus uniformly continuous on $[0,1]$. Consequently, for any $\epsilon > 0$ there exists a $\delta_\epsilon > 0$ such that,

$$|u' - u| \leq \delta_\epsilon, \implies |\tilde{F}_{\pi_h}^-(u') - \tilde{F}_{\pi_h}^-(u)| \leq \epsilon$$

and thus, by replacing $\delta_{u,\epsilon}$ by $\delta_\epsilon$ in the above computations, it follows that

$$|\tilde{F}_{\pi_h}^-(u) - F_{\pi_h^N}^-(u)| \leq \epsilon, \quad \forall N \geq N_{\delta_\epsilon}.$$

Since $N_{\delta_\epsilon}$ is independent of $u$,

$$\|\tilde{F}_{\pi_h}^- - F_{\pi_h^N}^-\|_\infty \leq \epsilon, \quad \forall N \geq N_{\delta_\epsilon}$$

and the proof is completed upon noting that

$$\|F_{\pi_h}^- - F_{\pi_h^N}^-\|_\infty \leq \|\tilde{F}_{\pi_h}^- - F_{\pi_h^N}^-\|_\infty, \quad \forall N \geq 1.$$

The last part of the lemma is obvious from the computations carried out to show the third part.

$\square$

# S2 Proofs for Section 2: Proof of Lemma 1

*Proof of Lemma 1.* We prove the result for the case $\mathcal{X} = \mathbb{R}^d$, its extension to any cubifiable sets being trivial.

Let $\pi \in \mathcal{P}_b(\mathbb{R}^d)$ be such that there exists a constant $C_\pi < +\infty$ such that, for any $I \subseteq 1 : d$ and $x_{\backslash I} \in \mathbb{R}^{d-I}$, we have $\sup_{x_I \in \mathbb{R}^{|I|}} p_\pi(x) \prod_{i \in I} |x_i|^{1+\delta} \leq C_\pi$, and let $\psi \in \mathcal{D}(\mathbb{R}^d)$. Then, using the change of variable formula,

$$p_{\pi_\psi}(u) = p_\pi(\psi^{-1}(u)) \prod_{i=1}^d \frac{\mathrm{d}\psi_i(u_i)}{\mathrm{d}u_i}, \quad u \in (0,1)^d.$$

Hence, because $p_\pi$ is continuous and $\psi$ is a $C^1$-diffeomorphism, $p_{\pi_\psi}$ is continuous on $(0,1)^d$. We now show that for a suitable choice of $\psi \in \mathcal{D}(\mathbb{R}^d)$ the density $p_{\pi_\psi}$ is bounded.

To construct $\psi$, let $\alpha > 0$ and $\tilde{\psi} : \mathbb{R} \to (0,1)$ be such that

$$\tilde{\psi}^{-1}(u) = \frac{2u-1}{u^\alpha(1-u)^\alpha}, \quad u \in (0,1).$$

It is easily checked that $\lim_{u \to 0} \psi^{-1}(u) = -\infty$ and $\lim_{u \to 1} \psi^{-1}(u) = +\infty$. In addition,

$$\frac{\mathrm{d}\psi^{-1}(u)}{\mathrm{d}u} = \frac{2}{u^\alpha(1-u)^\alpha} + \frac{\alpha(2u-1)^2}{u^{\alpha+1}(1-u)^{\alpha+1}} > 0, \quad \forall u \in (0,1).$$

and thus $\tilde{\psi} \in \mathcal{D}(\mathbb{R})$.

Let $\psi \in \mathcal{D}(\mathbb{R}^d)$ be defined by $\psi(x) = (\tilde{\psi}(x_1), \ldots, \tilde{\psi}(x_d))$, $x \in \mathbb{R}^d$. Then, for $u \in (0,1)^d$, we have

$$
\begin{aligned}
&p_{\pi_\psi}(u) \\
&= p_\pi\left(\frac{2u_1-1}{u_1^\alpha(1-u_1)^\alpha}, \ldots, \frac{2u_d-1}{u_d^\alpha(1-u_d)^\alpha}\right) \prod_{i=1}^d \left(\frac{2}{u_i^\alpha(1-u_i)^\alpha} + \alpha \frac{(2u_i-1)^2}{u_i^{\alpha+1}(1-u_i)^{\alpha+1}}\right).
\end{aligned}
$$

Let $\underline{u} = \frac{1}{2} - \frac{1}{2}\left(\sqrt{1 - \frac{4\alpha}{4\alpha+2}}\right)$ and $\bar{u} = \frac{1}{2} + \frac{1}{2}\left(\sqrt{1 - \frac{4\alpha}{4\alpha+2}}\right)$ so that

$$\frac{2}{u_1^\alpha(1-u_1)^\alpha} \leq \alpha \frac{(2u_1-1)^2}{u_1^{\alpha+1}(1-u_1)^{\alpha+1}}, \quad \forall u_1 \in (0, \underline{u}] \cup [\bar{u}, 1)$$

while

$$\alpha \frac{(2u_1-1)^2}{u_1^{\alpha+1}(1-u_1)^{\alpha+1}} \leq \frac{2}{u_1^\alpha(1-u_1)^\alpha}, \quad \forall u_1 \in [\underline{u}, \bar{u}].$$

Next, let $\mathcal{B}_\alpha = \tilde{\psi}^{-1}([\underline{u}, \bar{u}])$, and, for $u \in (0,1)^d$, let $I_u = \{i \in 1 : d : u_i \notin [\underline{u}, \bar{u}]\}$ and $x_i^{(u)} = \tilde{\psi}^{-1}(u_i)$, $i \in 1 : d$. Then, for any $u \in (0,1)^d$ and $\alpha \geq 1$, and with the convention

that empty products equal one,

$$
\begin{aligned}
p_{\pi_\psi}(u) &\leq (2\alpha)^d p_\pi(x_1^{(u)}, \ldots, x_d^{(u)}) \prod_{i \notin I_u} \frac{2}{u_i^\alpha(1-u_i)^\alpha} \prod_{i \in I_u} \frac{\alpha(2u_i-1)^2}{u_i^{\alpha+1}(1-u_i)^{\alpha+1}} \\
&\leq \Big( \max_{u \in [\underline{u}, \bar{u}]} \frac{4\alpha}{u^\alpha(1-u)^\alpha} \Big)^d \max_{z_{\backslash I_u} \in \mathcal{B}_\alpha^{d-|I_u|}} p_\pi\Big(x_{I_u}^{(u)}, z_{\backslash I_u}\Big) \prod_{i \in I_u} |2u_i-1|^{\frac{\alpha-1}{\alpha}} |x_i^{(u)}|^{\frac{1+\alpha}{\alpha}} \\
&= \Big( \max_{u \in [\underline{u}, \bar{u}]} \frac{4\alpha}{u^\alpha(1-u)^\alpha} \Big)^d p_\pi\Big(x_{I_u}^{(u)}, \tilde{x}_{\backslash I}^{(u)}\Big) \prod_{i \in I_u} |2u_i-1|^{\frac{\alpha-1}{\alpha}} |x_i^{(u)}|^{\frac{1+\alpha}{\alpha}} \\
&\leq \Big( \max_{u \in [\underline{u}, \bar{u}]} \frac{4\alpha}{u^\alpha(1-u)^\alpha} \Big)^d p_\pi\Big(x_{I_u}^{(u)}, \tilde{x}_{\backslash I_u}^{(u)}\Big) \prod_{i \in I_u} |x_i^{(u)}|^{\frac{1+\alpha}{\alpha}}.
\end{aligned}
$$

for a $\tilde{x}_{\backslash I_u}^{(u)} \in \mathcal{B}_\alpha^{d-|I|}$ and where the equality holds because $\tilde{p}_\psi$ is continuous and $B_\alpha$ is compact. Then, the result follows by noting that $\alpha \geq 1$ is arbitrary and that, as $\alpha \to +\infty$, $(1+\alpha)/\alpha \to 1$. $\qquad\square$

# S3 Proofs for Section 3

## S3.1 Proof of Theorem 1

Before proving the result we recall the following maximal inequality.

**Theorem S1.** *(Shao, 2000, Theorem 3) Let $(Z^n)_{n=1}^N$ be a sequence of NA random variables with zero means and finite second moments. Let $B_N = \sum_{n=1}^N \mathbb{E}[(Z^n)^2]$. Then, for all $\epsilon > 0$ and $a > 0$,*

$$
\begin{aligned}
\mathbb{P}\Big( &\max_{m \in 1:N} \Big| \sum_{n=1}^m Z^n \Big| \geq \epsilon \Big) \\
&\leq 2\mathbb{P}\Big( \max_{n \in 1:N} |Z^n| > a \Big) + 4\exp\Big( -\frac{\epsilon^2}{8B_N} \Big) + 4\Big( \frac{B_N}{4(\epsilon a + B_N)} \Big)^{\frac{\epsilon}{12a}}.
\end{aligned}
$$

*Proof of Theorem 1.* Let $\rho$ be an unbiased resampling scheme that satisfies $(H_1)$ and $(H_2)$, $\pi \in \widetilde{\mathcal{P}}_b(\mathcal{X})$, $\psi \in \mathcal{D}(\mathcal{X})$ be such that $\pi_\psi \in \mathcal{P}_b((0,1)^d)$ and $(\zeta^N)_{N \geq 1}$ be such that $\pi^N \xRightarrow{\text{w}} \pi$, $\mathbb{P}$-a.s.

By Theorem 3, for the required $\widetilde{\mathcal{P}}_b(\mathcal{X})$-consistency of $\rho$, it is necessary and sufficient that

$$
\lim_{N \to +\infty} \|\rho(\zeta^N)_{h_{\mathcal{X},\psi}} - \pi^N_{h_{\mathcal{X},\psi}}\|_\star = 0. \quad \mathbb{P}-a.s.
$$

For a given $N \geq 1$ and $z = (x^n, w^n)_{n=1}^N$ let $\sigma^*(z, \cdot)$ be a permutation of $1:N$ such that $h_{\mathcal{X},\psi}(x^{\sigma^*(z,1)}) \leq \cdots \leq h_{\mathcal{X},\psi}(x^{\sigma^*(z,N)})$, i.e., $\sigma^*$ sorts the points $x^n$ using the Hilbert curve.

Notice then that using the definition of $(\Delta_{\rho,z}^n)_{n=1}^N$ given in (2), we have:

$$\|\rho(\zeta^N)_{h_{\mathcal{X},\psi}} - \pi_{h_{\mathcal{X},\psi}}^N\|_\star = \frac{1}{N} \max_{m \in 1:N} \Big| \sum_{n=1}^m \Delta_{\rho,\zeta^N}^{\sigma^*(\zeta^N,n)} \Big|,$$

so our goal in the following is to show that

$$\lim_{N \to +\infty} \frac{1}{N} \max_{m \in 1:N} \Big| \sum_{n=1}^m \Delta_{\rho,\zeta^N}^{\sigma^*(\zeta^N,n)} \Big| = 0, \quad \mathbb{P} - a.s. \tag{S.11}$$

Under $(H_1)$, for any $N \geq 1$ and $z \in \mathcal{Z}^N$, the random variables $\big(\#^n(U,z)\big)_{n=1}^N$ are NA, using the definition of NA random variables and (2), the random variables $(\Delta_{\rho,z}^n)_{n=1}^N$ are NA too, as are also $(\Delta_{\rho,z}^{\sigma^*(z,n)})_{n=1}^N$. In addition, because $\rho$ is assumed to be unbiased, $\mathbb{E}[\Delta_{\rho,z}^{\sigma^*(z,n)}] = 0$ for all $n \in 1:N$ and thus $(\Delta_{\rho,z}^{\sigma^*(z,n)})_{n=1}^N$ satisfy the assumptions of Theorem S1.

Let $t > 0$ and $z \in \mathcal{Z}^N$. Then, using Theorem S1 with $a = r_N$ and $\epsilon = tN$, we have

$$\mathbb{P}\Big( \max_{m \in 1:N} \Big| \sum_{n=1}^m \Delta_{\rho,z}^{\sigma^*(z,n)} \Big| \geq tN \Big) \leq 2\mathbb{P}\Big( \max_{n \in 1:N} |\Delta_{\rho,z}^n| > r_N \Big)$$
$$+ 4\exp\Big( -\frac{(Nt)^2}{8\sum_{n=1}^N \mathbb{E}\big[(\Delta_{\rho,z}^n)^2\big]} \Big) + 4\Big( \frac{\sum_{n=1}^N \mathbb{E}\big[(\Delta_{\rho,z}^n)^2\big]}{4tNr_N + 4\sum_{n=1}^N \mathbb{E}\big[(\Delta_{\rho,z}^n)^2\big]} \Big)^{\frac{tN}{12r_N}}.$$

Under $(H_2)$, $\sum_{n=1}^N \mathbb{E}\big[(\Delta_{\rho,z}^n)^2\big] \leq r_N N$ for $N$ large enough and thus, since for any $b \geq 0$, the mapping $x \mapsto x/(b + 4x)$ is non-decreasing, we have

$$\Big( \frac{\sum_{n=1}^N \mathbb{E}\big[(\Delta_{\rho,z}^n)^2\big]}{4tNr_N + 4\sum_{n=1}^N \mathbb{E}\big[(\Delta_{\rho,z}^n)^2\big]} \Big)^{\frac{tN}{12r_N}} \leq \Big( \frac{r_N N}{4tNr_N + 4r_N N} \Big)^{\frac{tN}{12r_N}} \leq \Big( \frac{1}{4} \Big)^{\frac{tN}{12r_N}}.$$

The condition $\sum_{n=1}^N \mathbb{E}\big[(\Delta_{\rho,z}^n)^2\big] \leq r_N N$ also implies that

$$\exp\Big( -\frac{(Nt)^2}{8\sum_{n=1}^N \mathbb{E}\big[(\Delta_{\rho,z}^n)^2\big]} \Big) \leq \exp\Big( -\frac{Nt^2}{8r_N} \Big)$$

and thus

$$\mathbb{P}\Big( \max_{m \in 1:N} \Big| \sum_{n=1}^m \Delta_{\rho,z}^{\sigma^*(z,n)} \Big| \geq N t \Big) \leq 2\mathbb{P}\Big( \max_{n \in 1:N} |\Delta_{\rho,z}^n| > r_N \Big)$$
$$+ 4\exp\Big( -\frac{Nt^2}{8r_N} \Big) + 4\Big( \frac{1}{4} \Big)^{\frac{tN}{12r_N}}.$$

Since $r_N = o(N/\log N)$, we have $\sum_{N=1}^\infty \beta^{N/r_N} < +\infty$ for any $\beta \in (0,1)$; take e.g. $\beta = (1/4)^{\frac{t}{12}}$ and $\beta = \exp(-t^2/8)$ so that

$$\sum_{N=1}^\infty \sup_{z \in \mathcal{Z}^N} \mathbb{P}\Big( \max_{n \in 1:N} |\Delta_{\rho,z}^n| > r_N \Big) < +\infty, \quad \sum_{N=1}^\infty \exp\Big( -\frac{Nt^2}{8r_N} \Big) < +\infty,$$

and

$$\sum_{N=1}^{\infty} \left(\frac{1}{4}\right)^{\frac{tN}{12r_N}} < +\infty.$$

Using the tower property and the fact that $U$ is independent of $\zeta^N$,

$$\mathbb{P}\Big(\max_{m\in 1:N} \Big|\sum_{n=1}^{m} \Delta_{\rho,\zeta^N}^{\sigma^*(\zeta^N,n)}\Big| \geq N\,t\Big) = \mathbb{E}\left[\mathbb{P}_{\zeta^N}\Big(\max_{m\in 1:N}\Big|\sum_{n=1}^{m} \Delta_{\rho,\zeta^N}^{\sigma^*(\zeta^N,n)}\Big| \geq N\,t\Big)\right]$$

$$\leq \sup_{z\in\mathcal{Z}^N} \mathbb{P}\Big(\max_{m\in 1:N}\Big|\sum_{n=1}^{m} \Delta_{\rho,z}^{\sigma^*(z,n)}\Big| \geq N\,t\Big),$$

so we have proved:

$$\sum_{N=1}^{\infty} \mathbb{P}\Big(\max_{m\in 1:N}\Big|\sum_{n=1}^{m} \Delta_{\rho,\zeta^N}^{\sigma^*(\zeta^N,n)}\Big| \geq N\,t\Big) < +\infty.$$

Therefore (S.11) holds by the Borel-Cantelli lemma and the proof is complete.

$\square$

## S3.2 Proof of Proposition 1

*Proof of Proposition 1.* Let $\mathcal{X} = (0,1)$ be a cubifiable set. To construct $\pi \in \widetilde{\mathcal{P}}_b(\mathcal{X})$, let $p = 1/2$, $p' = 3$, $k = 3$, $a_1 = 2$, $a_2 = 5$, $b_1 = a_1 2^{-k}$, $b_2 = a_2 2^{-k}$ and $\epsilon = 2^{-k}$. Then,

$$p(1 + (a_1 - a_2 - 1)2^{-k}) + p'(a_2 - a_1 - 1)2^{-k} = 1 \tag{S.12}$$

and the function $p : (0,1) \to \mathbb{R}_+$ defined by

$$p(x) = \begin{cases} p, & x \in (0, b_1] \\ p + \frac{x - b_1}{\epsilon}(p' - p), & x \in (b_1, b_1 + \epsilon) \\ p', & x \in [b_1 + \epsilon, b_2 - \epsilon] \\ p' - \frac{x - (b_2 - \epsilon)}{\epsilon}(p' - p), & x \in (b_2 - \epsilon, b_2) \\ p, & x \in [b_2, 1) \end{cases}$$

is a continuous and bounded probability density on $\mathcal{X}$ (w.r.t. to $\lambda_1$). Thus, $\pi(\mathrm{d}x) := p(x)\lambda_1(\mathrm{d}x)$ belongs to $\widetilde{\mathcal{P}}_b(\mathcal{X})$, as required.

We now construct a sequence $(\zeta^N)_{N\geq 1}$ such that $\pi^N \overset{\mathrm{w}}{\Rightarrow} \pi$, $\mathbb{P}$-a.s. Let $\big((\tilde{X}^{n,N})_{n=1}^N\big)_{N\geq 1}$ be a sequence of point sets in $\mathcal{X}$ such that, for all $m \geq 2$,

$$(\tilde{X}^{1,2^m}, \ldots, \tilde{X}^{2^m,2^m}) = \big(2^{-m}, 2^{1-m}, \ldots, 1 - 2^{-m}, v\big)$$

for some fixed non-dyadic number $v \in (b_2, 1)$. (This ensures that the $2^m$ points of the point set are all distinct.) For values of $N$ which are not powers of 2 we take for $(\tilde{X}^{n,N})_{n=1}^N$ a set of i.i.d. uniform random numbers in $\mathcal{X}$.

Next, for $N \geq 1$, let for $n = 1, \ldots, N$ $\tilde{W}^{n,N} = p(\tilde{X}^{n,N})/\sum_{m=1}^{N} p(\tilde{X}^{m,N})$ and define $\zeta^N = \left(\tilde{X}^{\sigma_N(n),N}, \tilde{W}^{\sigma_N(n),N}\right)_{n=1}^{N}$ where the sequence of permutations $(\sigma_N)_{N \geq 1}$ is defined below. Then, it is easily checked that $(\zeta^N)_{N \geq 1}$ is such that $\pi^N \overset{w}{\Rightarrow} \pi$, $\mathbb{P}$-a.s., as required.

We now construct a sequence of permutations $(\sigma_N)_{N \geq 1}$ for which $\mathbb{P}(\rho_{\text{syst}}(\pi^N) \overset{w}{\Rightarrow} \pi) < 1$. To follow the notation used throughout the paper we define $X^{n,N} = \tilde{X}^{\sigma_N(n),N}$ and $W^{n,N} = \tilde{W}^{\sigma_N(n),N}$ for $n \in 1 : N$ so that $\zeta^N = \left(X^{n,N}, W^{n,N}\right)_{n=1}^{N}$.

Let $m \geq k$ and, with the shorthand $N_m = 2^m$, remark that, with $\mathbb{P}$-probability one (as $\zeta^{N_m}$ is non-random for any $m \geq 1$)

$$\sum_{n=1}^{N_m} \mathbb{1}\left(X^{n,N_m} \in (0, b_1]\right) = a_1 2^{m-k}, \quad \sum_{n=1}^{N_m} \mathbb{1}\left(X^{n,N_m} \in [b_2, 1)\right) = 2^m - a_2\, 2^{m-k} + 1,$$

$$\sum_{n=1}^{N_m} \mathbb{1}\left(X^{n,N_m} \in (b_1, b_1 + \epsilon)\right) = \sum_{n=1}^{N_m} \mathbb{1}\left(X^{n,N_m} \in (b_2 - \epsilon, b_2)\right) = 2^{m-k} - 1,$$

$$\sum_{n=1}^{N_m} \mathbb{1}\left(X^{n,N_m} \in [b_1 + \epsilon, b_2 - \epsilon]\right) = (a_2 - a_1 - 2)2^{m-k} + 1$$

$$\sum_{n=1}^{N_m} p(X^{n,N_m})\mathbb{1}\left(X^{n,N_m} \in (b_1, b_1 + \epsilon)\right) = \sum_{n=1}^{N_m} p(X^{n,N_m})\mathbb{1}\left(X^{n,N_m} \in (b_2 - \epsilon, b_2)\right)$$

$$= \left(2^{m-k} - 1\right)\frac{p' - p}{2}.$$

and thus, taking $a = a_2 - a_1$, we have

$$\frac{1}{N_m} \sum_{n=1}^{N_m} p(X^{n,N_m})$$
$$= p\left(1 - a2^{-k} + 2^{-m}\right) + p'\left((a-2)2^{-k} + 2^{-m}\right) + (p' - p)\left(2^{-k} - 2^{-m}\right)$$
$$= p\left(1 - (a+1)2^{-k} + 2^{-m+1}\right) + p'(a-1)2^{-k} = p\,2^{-m+1} + 1$$

using (S.12). Since $p = 0.5$, for $m \geq k$ and $n$ such that $X^{n,N_m} \in (0, b_1] \cup [b_2, 1)$,

$$N_m W^{n,N_m} = \frac{N_m}{2(N_m + 1)} = \frac{1}{2} - \frac{1}{2(N_m + 1)}. \tag{S.13}$$

Let $P_m$ denote the number of points $X^{n,N_m}$ in $(0, b_1]$; $P_m = a_1 2^{m-k} = N_m/4$. Note that $P_m \leq 2^m - a_2 2^{m-k} + 1$, i.e. it is possible to pair each point in $(0, b_1]$ with a different point in $[b_2, 1)$.

We take $\sigma_{N_m}$ to be a permutation that alternates between points in $(0, b_1)$ and points in $[b_2, 1]$; $\sigma_{N_m}(1 : N_m) = \left(1, N_m, 2, N_m - 1, 3, \ldots, P_m, N_m - P_m + 1, \ldots\right)$; the remaining components are arbitrary.

We now show that for this sequence $(\zeta^N)_{N \geq 1}$ and probability measure $\pi \in \widetilde{\mathcal{P}}_b(\mathcal{X})$ we have $\mathbb{P}(\rho_{\text{syst}}(\pi^N) \overset{w}{\Rightarrow} \pi) \leq 3/4$.

Given (S.13) and given how systematic resampling operates, all the points in $(0, b_1]$ (resp. in $[b_2, 1]$) will have exactly one (resp. 0) off-spring as soon as:

$$u_1 < \frac{1}{2} - \frac{1}{2N_m + 1} - \frac{P_m - 1}{N_m + 1} = \frac{1}{4}$$

where $u_1$ is the first component of $u$.

Then, with $\mathbb{P}$-probability at least $1/4$, we have, for any $m \geq k$ and using (S.13),

$$\|\rho_{\text{syst}}(\zeta^{N_m}) - \pi^N\|_\star = \frac{1}{N_m} \max_{i \in 1:N_m} \left| \sum_{n=1}^{i} \Delta_{\rho_{\text{syst}}, \zeta^{N_m}}^{\sigma^\star(\zeta^{N_m}, n)} \right|$$

$$\geq \frac{1}{N_m} \left| \sum_{n=1}^{P_m} \Delta_{\rho_{\text{syst}}, \zeta^{N_m}}^{\sigma^\star(\zeta^{N_m}, n)} \right|$$

$$= \frac{P_m}{N_m} \left| \frac{1}{2} + \frac{1}{2(1 + N_m)} \right|$$

$$\geq \frac{1}{8}.$$

where $\sigma^\star$ is a permutation that orders the points. This shows that

$$\mathbb{P}\Big( \lim_{N \to +\infty} \|\rho_{\text{syst}}(\pi^N) - \pi\|_\star = 0 \Big) \leq 3/4$$

and thus, by Theorem 3, $\mathbb{P}(\rho_{\text{syst}}(\pi^N) \overset{\text{w}}{\Rightarrow} \pi) \leq 3/4$. The proof is complete. $\qquad \square$

## S3.3 Proof of Corollary 1

*Proof of Corollary 1.* Condition $(H_1)$ holds for multinomial resampling as shown by **?**.

We show below that $(H_2)$ holds for $r_N = \max(\sqrt{3N \log(N)/2}, 1) = o(N/\log N)$.

Let $N \geq 2$ and $z \in \mathcal{Z}^N$. For any $n \in 1 : N$, $\Delta_{\rho_{\text{multi}}, z}^n = \sum_{i=1}^{N}(B_n^i - w^n)$ where $B_n^1, \ldots, B_n^N$ are i.i.d. random variables in $\{0, 1\}$ such that $\mathbb{E}[B_n^i] = w^n$. Therefore, by Hoeffding's inequality,

$$\mathbb{P}\Big( \big| \Delta_{\rho_{\text{multi}}, z}^n \big| > \epsilon N \Big) \leq 2e^{-2\epsilon^2 N}, \quad \forall \epsilon > 0, \; \forall n \in 1 : N.$$

Applying this result with $\epsilon = \sqrt{3 \log N / (2N)}$, we have

$$\mathbb{P}\Big( \max_{n \in 1:N} \big| \Delta_{\rho_{\text{multi}}, z}^n \big| > r_N \Big) \leq \sum_{n=1}^{N} \mathbb{P}\Big( \big| \Delta_{\rho_{\text{multi}}, z}^n \big| > r_N \Big) \leq 2N e^{-3 \log N} = \frac{2}{N^2}.$$

To conclude the proof it remains to show that $\sum_{n=1}^{N} \mathbb{E}\big[ (\Delta_{\rho_{\text{multi}}, z}^n)^2 \big] \leq r_N N$. To this end, remark that

$$\mathbb{E}\big[ (\Delta_{\rho_{\text{multi}}, z}^n)^2 \big] = \text{Var}\Big( \sum_{i=1}^{N} B_n^i \Big) = N w^n (1 - w^n), \quad \forall n \in 1 : N$$

and thus $\sum_{n=1}^{N} \mathbb{E}\big[ (\Delta_{\rho_{\text{multi}}, z}^n)^2 \big] \leq N \leq r_N N$ as required.

$\qquad \square$

## S3.4 Proof of Corollary 2

*Proof of Corollary 2.* Let $N \geq 1$ and $z \in \mathcal{Z}^N$. Then, as $|\Delta_{\rho_{\text{strat}}, z}| \leq 2$, $\mathbb{P}$-a.s., it follows that

$$\sum_{n=1}^{N} \mathbb{E}\big[(\Delta_{\rho_{\sigma,\text{strat}}, z}^n)^2\big] \leq 4\,N, \quad \mathbb{P}\Big(\max_{n \in 1:N} |\Delta_{\rho_{\text{strat}}, z}^n| > 4\Big) = 0$$

and thus $(H_2)$ holds for sequence $r_N = 4$.

To show $(H_1)$ we define, for $n, m \in 1 : N$,

$$A_n = \big\{ u \in [0,1] : F_N^-(u) = n \big\}, \; V^n = \frac{n - 1 + U_n}{N}$$

and $p_m^n := \mathbb{P}(V^n \in A_m)$.

The collection of sets $(A_n)_{n=1}^N$ form a partition of $[0,1]$. Consequently, we can see $(A_n)_{n=1}^N$ as bins and the collection of independent random variables $(V^m)_{m=1}^N$ as balls in a "Ball and Bins" problem (see, e.g. Dubhashi and Ranjan, 1998), where for $n, m \in 1 : N$, the probability that ball $n$ falls into bin $m$ is given by $p_m^n \in [0,1]$. The collection of random variables $(\#^n(U, z))_{n=1}^N$ can therefore be interpreted as occupancy numbers for the "Ball and Bins" problem we just described; that is, $\#^n(U, z)$ is the number of balls that fall into bin $n$. By Dubhashi and Ranjan (1998, Theorem 13), occupancy numbers in "Ball and Bins" problems are NA and thus $(\#^n(U, z))_{n=1}^N$ satisfies $(H_1)$. □

## S3.5 Proof of Corollary 3

*Proof of Corollary 3.* Let $N \geq 1$ and $z \in \mathcal{Z}^N$. Then, following a similar argument as in the proof of Corollary 2, it is easily checked that $|\Delta_{\rho_{\text{ssp}}, z}| \leq 1$, $\mathbb{P}$-a.s. so that $(H_2)$ is verified for sequence $r_N = 1$. Lastly, $(H_1)$ is verified as well because $\rho_{\text{ssp}}$ is based on what is called a linear SSP process in Kramer et al. (2011) and thus, by Dubhashi et al. (2007, Theorem 5.1), the collection of random variables $(\#^n(U, z))_{n=1}^N$ is NA. □

## S3.6 Proofs for Section 4

### S3.6.1 Proof of Theorem 2

*Proof of Theorem 2.* The first part of the theorem is a particular case of Theorem 4 while the second part is direct consequence of Theorem 6 and of the computations used in the proof of Lemma 1. □

### S3.6.2 Proof of Proposition 2

*Proof of Proposition 2.* Remind first that, for any $x \in [0,1]^d$, the set $H^{-1}(x) \subset [0,1]$ contains at most $2^d$ elements. Then, let $h : [0,1]^d \to [0,1]$ be the mapping defined by

$$h(x) = \min H^{-1}(x), \quad x \in [0,1]^d.$$

By construction $H(h(x)) = x$ for all $x \in [0,1]^d$ and $h$ is one-to-one. Thus, to establish the proposition it remains to show that $h$ is a Borel measurable function.

To see this, remark that the mapping $h$ is such that, for all $m \geq 1$,

$$h^{-1}(I_m^d(0)) = S_m^d(0), \quad h^{-1}(I_m^d(k)) = S_m^d(k) \setminus \cup_{i=0}^{k-1} S_m^d(i), \quad k \in 1 : 2^{md} - 1 \quad \text{(S.14)}$$

where, for $k \geq 1$, the set $S_m^d(k) \setminus \cup_{i=0}^{k-1} S_m^d(i)$ is obtained by removing the edges that $S_m^d(k)$ has in common with the closed hyper-cubes $S_m^d(i)$, $i = 0, \ldots, k-1$. To show that (S.14) indeed holds, let $m \geq 1$ and note that, using the definition of $h$ and (S.1), we have

$$\begin{aligned}
h^{-1}(I_m^d(0)) &= \{x \in [0,1]^d : h(x) \in I_m^d(0)\} \\
&= \{x \in [0,1]^d : H^{-1}(x) \cap I_m^d(0) \neq \varnothing\} \\
&= S_m^d(0)
\end{aligned}$$

while, for $k \in 1 : 2^{md} - 1$,

$$\begin{aligned}
&h^{-1}(I_m^d(k)) \\
&= \{x \in [0,1]^d : h(x) \in I_m^d(k)\} \\
&= \{x \in [0,1]^d : H^{-1}(x) \cap I_m^d(k) \neq \varnothing\} \setminus \{x \in [0,1]^d : H^{-1}(x) \cap \left(\cup_{p=0}^{k-1} I_m^d(p)\right) \neq \varnothing\} \\
&= S_m^d(k) \setminus \left(\cup_{p=0}^{k-1} S_m^d(p)\right).
\end{aligned}$$

Next, let $\mathcal{B}([0,1])$ be the Borel $\sigma$-algebra on $[0,1]$ and $I \in \mathcal{B}([0,1])$. Then, because dyadic numbers are dense in $[0,1]$, there exists a sequence of closed dyadic intervals $(I_{m_n}^d(k_n))_{n \geq 1}$ such that $I = \cup_{n \geq 1} I_{m_n}^d(k_n)$, and thus

$$h^{-1}(I) = \cup_{n \geq 1} h^{-1}(I_{m_n}^d(k_n)).$$

By (S.14), the set $h^{-1}(I_{m_n}^d(k_n)) \subset [0,1]^d$ is an hypercube (which may be either open, or closed, or neither closed nor open) and is therefore a Borel set of $[0,1]^d$. This completes the proof. $\qquad \square$

### S3.6.3 Proof of Theorem 3

*Proof of Theorem 3.* Since $\mathcal{X}$ is cubifiable and we consider probability measures in $\widetilde{\mathcal{P}}_b(\mathcal{X})$, we can assume without loss of generality that $\mathcal{X} = (0,1)^d$ and take $h_{\mathcal{X},\psi} = h$.

Let $\pi \in \widetilde{\mathcal{P}}_b(\mathcal{X})$ and $(\pi^N)_{N \geq 1}$ be a sequence such that $\pi^N \in \mathcal{P}_f^N(\mathcal{X})$ and $\pi^N \overset{\mathrm{w}}{\Rightarrow} \pi$. Note that, by Theorem 9, $\pi^N \overset{\mathrm{w}}{\Rightarrow} \pi$ implies that

$$\lim_{N \to \infty} \|\pi_h^N - \pi_h\|_\star = 0. \quad \text{(S.15)}$$

To establish the "if" part simply note that, by (S.15) and under the sufficient condition provided in the statement of the theorem,

$$\lim_{n \to +\infty} \|\rho(\zeta^N)_h - \pi_h\|_\star \leq \lim_{n \to +\infty} \|\rho(\zeta^N)_h^N - \pi_h^N\|_\star + \lim_{n \to +\infty} \|\pi_h^N - \pi_h\|_\star = 0, \quad \mathbb{P} - a.s.$$

so that the result follows from Theorem 9.

To establish the "only if" part assume that $\rho(\zeta^N) \overset{w}{\Rightarrow} \pi$, $\mathbb{P}$-a.s. By Theorem 9, $\lim_{N\to\infty} \|\rho(\zeta^N)_h - \pi_h\|_\star = 0$, $\mathbb{P}$-a.s. and therefore, by (S.15) and the triangle inequality,

$$\lim_{N\to+\infty} \|\rho(\zeta^N)_h - \pi_h^N\|_\star \leq \lim_{N\to+\infty} \|\rho(\zeta^N)_h - \pi_h\|_\star + \lim_{N\to+\infty} \|\pi_h^N - \pi_h\|_\star, \quad \mathbb{P} - a.s.$$

This completes the proof. $\qquad\square$

### S3.6.4 Proof of Theorem 4

*Proof of Theorem 4.* Since $\mathcal{X}$ is cubifiable and $\pi \in \widetilde{\mathcal{P}}_b^*(\mathcal{X})$, we can assume without loss of generality that $\mathcal{X} = (0,1)^d$ and thus $h = h_{\mathcal{X}}$. Let $\varphi \in \mathcal{C}_b(\mathcal{X})$ and denote $\varphi_H = \varphi \circ H \in \mathcal{C}_b((0,1))$. Then, for any $N \geq 1$, we have:

$$N\mathrm{Var}_{\zeta^N}\left[\rho_{\mathrm{strat}}^*(\zeta^N)(\varphi)\right]$$

$$= \pi^N(\varphi^2) - N\sum_{n=1}^N \Big(\int_{\frac{n-1}{N}}^{\frac{n}{N}} \varphi_H \circ F_{\pi_h^N}^-(u)\mathrm{d}u\Big)^2$$

$$= \pi^N(\varphi^2) - N\sum_{n=1}^N \Big(\int_{\frac{n-1}{N}}^{\frac{n}{N}} \varphi_H \circ F_{\pi_h}^-(u)\mathrm{d}u + \int_{\frac{n-1}{N}}^{\frac{n}{N}} \big(\varphi_H \circ F_{\pi_h^N}^- - \varphi_H \circ F_{\pi_h}^-\big)(u)\mathrm{d}u\Big)^2$$

$$= \pi^N(\varphi^2) - N\sum_{n=1}^N \Big(\int_{\frac{n-1}{N}}^{\frac{n}{N}} \varphi_H \circ F_{\pi_h}^-(u)\mathrm{d}u\Big)^2$$

$$- N\sum_{n=1}^N \Big(\int_{\frac{n-1}{N}}^{\frac{n}{N}} \big(\varphi_H \circ F_{\pi_h^N}^- - \varphi_H \circ F_{\pi_h}^-\big)(u)\mathrm{d}u\Big)^2$$

$$- 2N\sum_{n=1}^N \Big(\int_{\frac{n-1}{N}}^{\frac{n}{N}} \varphi_H \circ F_{\pi_h}^-(u)\mathrm{d}u\Big)\Big(\int_{\frac{n-1}{N}}^{\frac{n}{N}} \big(\varphi_H \circ F_{\pi_h^N}^- - \varphi_H \circ F_{\pi_h}^-\big)(u)\mathrm{d}u\Big). \quad \text{(S.16)}$$

For the first term, under the assumptions of the theorem

$$\lim_{N\to+\infty} \pi^N(\varphi^2) = \pi(\varphi^2), \quad \mathbb{P} - a.s. \quad \text{(S.17)}$$

For the second term, since $F_{\pi_h}$ is continuous (Lemma 2) and strictly increasing on $[0,1]$ (Lemme S3), $F_{\pi_h}^-$ is a continuous function on $[0,1]$. Hence, the function $\varphi_H \circ F_{\pi_h}^-$ belongs to $\mathcal{C}_b([0,1])$ and is Riemann integrable. Consequently,

$$N\sum_{n=1}^N \Big(\int_{\frac{n-1}{N}}^{\frac{n}{N}} \varphi_H \circ F_{\pi_h}^-(u)\mathrm{d}u\Big)^2 = \frac{1}{N}\sum_{n=1}^N \{\varphi_H \circ F_{\pi_h}^-(u_n)\}^2 \to \pi(\varphi^2) \quad \text{(S.18)}$$

for some $u_i \in [n - 1/N, n/N]$ (mean value theorem).

For the fourth term,

$$N\Big|\sum_{n=1}^N \Big(\int_{\frac{n-1}{N}}^{\frac{n}{N}} \varphi_H \circ F_{\pi_h}^-(u)\mathrm{d}u\Big)\Big(\int_{\frac{n-1}{N}}^{\frac{n}{N}} \big(\varphi_H \circ F_{\pi_h^N}^- - \varphi_H \circ F_{\pi_h}^-\big)(u)\mathrm{d}u\Big)\Big|$$

$$\leq \|\varphi\|_\infty \int_0^1 |\varphi_H \circ F_{\pi_h^N}^-(u) - \varphi_H \circ F_{\pi_h}^-(u)|\mathrm{d}u. \quad \text{(S.19)}$$

By Lemma S3, on an event of $\mathbb{P}$-probability 1, $\lim_{N \to +\infty} |F_{\pi_h^N}^-(u) - F_{\pi_h}^-(u)| = 0$ for all $u \in [0,1]$. Therefore, using the fact that $\varphi_H$ is continuous and bounded, we have, by the dominated convergence theorem,

$$\lim_{N \to +\infty} \int_0^1 |\varphi_H \circ F_{\pi_h^N}^-(u) - \varphi_H \circ F_{\pi_h}^-(u)| \mathrm{d}u = 0, \quad \mathbb{P} - a.s. \tag{S.20}$$

and, similarly for the third term:

$$\lim_{N \to +\infty} N \sum_{n=1}^{N} \Big( \int_{\frac{n-1}{N}}^{\frac{n}{N}} \big( \varphi_H \circ F_{\pi_h^N}^- - \varphi_H \circ F_{\pi_h}^- \big)(u) \mathrm{d}u \Big)^2$$
$$\leq 2\|\varphi\|_\infty \lim_{N \to +\infty} \int_0^1 |\varphi_H \circ F_{\pi_h^N}^-(u) - \varphi_H \circ F_{\pi_h}^-(u)| \mathrm{d}u \tag{S.21}$$
$$= 0.$$

Putting (S.16)-(S.21) together shows that $N \mathrm{Var}\left[\rho_{\mathrm{strat}}^*(\pi^N)(\varphi)\right] \to 0$, $\mathbb{P}$-a.s. as required. $\qquad\square$

*Proof of Corollary 4.* Let $N \geq 1$, $z \in \mathcal{Z}^N$, and $u_{1:N} \in [0,1]^N$. Then, for any $a \in [0,1]$,

$$\left| \frac{1}{N} \sum_{n=1}^{N} \mathbb{1}\big(F_{\pi_{h_{\mathcal{X}}}^N}^-(u_n) \leq a\big) - \pi_{h_{\mathcal{X}}}^N([0,a]) \right| = \left| \frac{1}{N} \sum_{n=1}^{N} \mathbb{1}\big(u_n \leq F_{\pi_{h_{\mathcal{X}}}^N}(a)\big) - F_{\pi_{h_{\mathcal{X}}}^N}(a) \right|$$
$$\leq D_N^\star(u_{1:N}).$$

Thus, the condition (10) implies the condition (8) of Theorem 3. For the second part of the corollary, we simply use the well-known fact (Niederreiter, 1992, Theorem 2.6, p.15) that

$$D_N^\star(u_{1:N}) = \frac{1}{2N} + \max_{1 \leq n \leq N} \left| u_n - \frac{n-1/2}{N} \right| \tag{S.22}$$

which implies that for the instances of $\phi_N^n$ corresponding to stratified resampling and systematic resampling given in the statement of the corollary, $D_N^\star(\phi_N^n(U)) \leq 1/N$, $\mathbb{P}$-a.s. $\qquad\square$

### S3.6.5 Proof of Theorem 5

We first recall the following result due to Aistleitner and Dick (2015) that will play a key role in the poof of Theorems 5 and 6.

**Theorem S2.** *(Aistleitner and Dick, 2015, Theorem 1) Let $\varphi : [0,1]^d \to \mathbb{R}$ be a measurable function, $\pi \in \mathcal{P}([0,1]^d)$ and $(x_n)_{n=1}^N$ be a set of $N \geq 1$ points in $[0,1]^d$. Then,*

$$\left| \frac{1}{N} \sum_{n=1}^{N} \varphi(x_n) - \int_{[0,1]^d} \varphi(x) \pi(\mathrm{d}x) \right| \leq V(\varphi) \left\| N^{-1} \sum_{n=1}^{N} \delta(x_n) - \pi \right\|_\star$$

*where $V(\varphi)$ is the variation of $\varphi$ in the sense of Hardy and Krause.*

*Proof of Theorem 5.* Since $\mathcal{X}$ is cubifiable and $\pi \in \widetilde{\mathcal{P}}_b^*(\mathcal{X})$, we can assume without loss of generality that $\mathcal{X} = (0,1)^d$ and thus $h_{\mathcal{X}} = h$.

Let $N \geq 1$, $z \in \mathcal{Z}^N$ and, for $n \in 1 : N$,

$$V^n = \frac{n-1+U_n}{N}, \quad \hat{X}^n = H \circ F_{\pi_h^N}^-(V^n), \quad \bar{x}^n = H\Big(\mathbb{E}_{\zeta^N}\big[F_{\pi_h^N}^-(V^n)\big]\Big).$$

Let $\varphi_I : [0,1] \to [0,1]$ be such that $\varphi_I(x) = x$, $\forall x \in [0,1]$, and remark that the variation of $\varphi_I$ is $V(\varphi_I) = |\varphi_I(1) - \varphi_I(0)| = 1$. Therefore, by Theorem S2,

$$\mathrm{Var}\left[\frac{1}{N}\sum_{n=1}^N F_{\pi_h^N}^-(V^n)\right] \leq V(\varphi_I)^2 \mathbb{E}\big[\|\rho_{\mathrm{strat}}^*(\zeta^N)_h - \pi_h^N\|_\star^2\big] \leq \mathbb{E}\big[D_N^\star\big((V^n)_{n=1}^N\big)^2\big] \tag{S.23}$$
$$\leq \frac{1}{N^2}$$

where the second inequality comes from equation (10) in the proof of Corollary 4 and the last one is due to (S.22) and the definition of $(V^n)_{n=1}^N$.

Next, let $\varphi : \mathcal{X} \to \mathbb{R}$ and $C_{\varphi,\psi_{\mathcal{X}}} \in (0,+\infty)$ be as in the statement of the theorem. Then, denoting by $\tilde{C}_d \in (0,+\infty)$ the Hölder constant of the Hilbert curve $H$ for the $\|\cdot\|_2$ norm, we have

$$\mathrm{Var}[\rho_{\mathrm{strat}}^*(z)(\varphi)] = \frac{1}{N^2}\sum_{n=1}^N \mathrm{Var}\Big[\varphi(\hat{X}^n) - \varphi(\bar{x}^n)\Big]$$
$$\leq \frac{1}{N^2}\sum_{n=1}^N \mathbb{E}\Big[\big(\varphi(\hat{X}^n) - \varphi(\bar{x}^n)\big)^2\Big]$$
$$\leq \frac{C_{\varphi,\psi_{\mathcal{X}}}^2}{N^2}\sum_{n=1}^N \mathbb{E}\Big[\big\|\hat{X}^n - \bar{x}^n\big\|_2^{2\gamma}\Big]$$
$$\leq \frac{C_{\varphi,\psi_{\mathcal{X}}}^2 \tilde{C}_d^{2\gamma}}{N^2}\sum_{n=1}^N \mathbb{E}\Big[\big(F_{\pi_h^N}^-(V^n) - \mathbb{E}\big[F_{\pi_h^N}^-(V^n)\big]\big)^{\frac{2\gamma}{d}}\Big] \tag{S.24}$$
$$\leq \frac{C_{\varphi,\psi_{\mathcal{X}}}^2 \tilde{C}_d^{2\gamma}}{N^2}\sum_{n=1}^N \mathbb{E}\Big[\big(F_{\pi_h^N}^-(V^n) - \mathbb{E}\big[F_{\pi_h^N}^-(V^n)\big]\big)^2\Big]^{\frac{\gamma}{d}} \tag{S.25}$$
$$\leq \frac{C_{\varphi,\psi_{\mathcal{X}}}^2 \tilde{C}_d^{2\gamma}}{N}\Big(\frac{1}{N}\sum_{n=1}^N \mathbb{E}\Big[\big(F_{\pi_h^N}^-(V^n) - \mathbb{E}\big[F_{\pi_h^N}^-(V^n)\big]\big)^2\Big]\Big)^{\frac{\gamma}{d}} \tag{S.26}$$
$$= \frac{C_{\varphi,\psi_{\mathcal{X}}}^2 \tilde{C}_d^{2\gamma}}{N}\Big(\frac{1}{N}\sum_{n=1}^N \mathrm{Var}\big[F_{\pi_h^N}^-(V^n)\big]\Big)^{\frac{\gamma}{d}}$$
$$\leq C_{\varphi,\psi_{\mathcal{X}}}^2 \tilde{C}_d^{2\gamma} N^{-1-\frac{\gamma}{d}} \tag{S.27}$$

where (S.24) is due to the Hölder continuity of the Hilbert curve, (S.25) and (S.26) are due to Jensen's inequality and the fact that, for any $\alpha \in (0,1)$, the function $x^\alpha$ is concave on $\mathbb{R}_+$ while (S.27) comes from (S.23). The result follows from the fact that $\tilde{C}_d \leq 2\sqrt{d+3}$ (see e.g. the proof of Zumbusch, 2003, Lemma 4.3, pp 97-99). $\qquad\square$

### S3.6.6 Proof of Theorem 6

*Proof of Theorem 6.* Using the same notation and computations as in the proof of Theorem 5 we note that, to establish the result, it is enough to show below that

$$\mathrm{Var}_{\zeta^N}\left[\frac{1}{N}\sum_{n=1}^{N}F_{\pi_h^N}^{-}(V^n)\right] = o(N^{-2}), \quad \mathbb{P}-a.s.$$

Let $N \geq 1$ and define

$$X_h^n = h(X^n), \quad a_n = F_{\pi_h^N}(X_h^{n-1}), \quad b_n = F_{\pi_h^N}(X_h^n), \quad n \in 1:N$$

with the convention $F_{\pi_h^N}(X_h^n) = 0$ when $n = 0$.

Next, let $\alpha > 0$ and $F_{\pi_h^N,\alpha} : [0, X_h^N] \to [0, 1]$ be the mapping defined by

$$F_{\pi_h^N,\alpha}(z) = a_n + (b_n - a_n)\left(\frac{z - X_h^{n-1}}{X_h^n - X_h^{n-1}}\right)^{\frac{1}{\alpha}}, \quad z \in [X_h^{n-1}, X_h^n], \ n \in 1:N$$

with the convention $X_h^n = 0$ when $n = 0$. Let $F_{\pi_h^N,\alpha}^{-} : [0, 1] \to [0, X_h^N]$ be the inverse of $F_{\pi_h^N,\alpha}$, which is defined by

$$F_{\pi_h^N,\alpha}^{-}(u) = X_h^{n-1} + \left(X_h^n - X_h^{n-1}\right)\left(\frac{u - a_n}{b_n - a_n}\right)^{\alpha}, \quad u \in [a_n, b_n], \ n \in 1:N.$$

For any $\alpha > 0$, the function $F_{\pi_h^N,\alpha}$ is continuous and strictly increasing on $[0, X_h^N]$ and such that $F_{\pi_h^N,\alpha}(X_h^n) = F_{\pi_h^N}(X_h^n)$ for all $n \in 1:N$. This last property implies that

$$\|F_{\pi_h^N,\alpha} - F_{\pi_h^N}\|_\infty \leq \max_{n\in 1:N} W^{n,N}. \tag{S.28}$$

As preliminary computations, remark that

$$F_{\pi_h^N}^{-}(u) - F_{\pi_h^N,\alpha}^{-}(u) = (X_h^n - X_h^{n-1})\left[1 - \left(\frac{u - a_n}{b_n - a_n}\right)^{\alpha}\right], \quad u \in (a_n, b_n], \ n \in 1:N$$

and thus

$$\int_{a_n}^{b_n}\left(F_{\pi_h^N}^{-}(u) - F_{\pi_h^N,\alpha}^{-}(u)\right)^2 \mathrm{d}u = (X_h^n - X_h^{n-1})^2(b_n - a_n)\frac{2\alpha^2}{(\alpha + 1)(2\alpha + 1)}. \tag{S.29}$$

Lastly, let $\alpha_N > 0$ be such that

$$\frac{2\alpha_N^2}{(\alpha_N + 1)(2\alpha_N + 1)} = \frac{1}{N^2}$$

and, to simplify the notation, we use the shorthand $\widetilde{F}_{\pi_h^N} = F_{\pi_h^N,\alpha_N}$ and $\widetilde{F}_{\pi_h^N}^{-} = F_{\pi_h^N,\alpha_N}^{-}$ in what follows.

By construction $\widetilde{F}_{\pi_h^N}$ is continuous and strictly increasing on $[0, X_h^N]$ and thus $\widetilde{F}_{\pi_h^N}^-$ is continuous on $[0, 1]$. In addition, by (S.28), Lemma S3 and the assumptions of the theorem,

$$\lim_{N \to +\infty} \|\widetilde{F}_{\pi_h^N}^- - F_{\pi_h}^-\|_\infty = 0, \quad \mathbb{P} - a.s. \tag{S.30}$$

Let $g_N : [0, 1] \to [0, 1]$ be defined by

$$g_N(u) = \widetilde{F}_{\pi_h^N}^-(u) - F_{\pi_h}^-(u), \quad u \in [0, 1].$$

Then,

$$\mathrm{Var}_{\zeta^N}\left[\frac{1}{N}\sum_{n=1}^N F_{\pi_h^N}^-(V^n)\right]$$
$$\leq \left(\mathrm{Var}_{\zeta^N}\left[\frac{1}{N}\sum_{n=1}^N F_{\pi_h}^-(V^n)\right]^{\frac{1}{2}} + \mathrm{Var}_{\zeta^N}\left[\frac{1}{N}\sum_{n=1}^N g_N(V^n)\right]^{\frac{1}{2}}\right.$$
$$\left. + \mathrm{Var}_{\zeta^N}\left[\frac{1}{N}\sum_{n=1}^N \left(\widetilde{F}_{\pi_h^N}^-(V^n) - F_{\pi_h^N}^-(V^n)\right)\right]^{\frac{1}{2}}\right)^2$$

where, by assumption, the first variance is of order $o(N^{-2})$, $\mathbb{P}$-a.s. In addition, for the last variance, we have, using the properties of the random variables $(V^n)_{n=1}^N$ and (S.29),

$$\mathrm{Var}_{\zeta^N}\left[\frac{1}{N}\sum_{n=1}^N \left(\widetilde{F}_{\pi_h^N}^-(V^n) - F_{\pi_h^N}^-(V^n)\right)\right] = \frac{1}{N^2}\sum_{n=1}^N \mathrm{Var}_{\zeta^N}\left[\widetilde{F}_{\pi_h^N}^-(V^n) - F_{\pi_h^N}^-(V^n)\right]$$
$$\leq \frac{1}{N^2}\sum_{n=1}^N \mathbb{E}_{\zeta^N}\left[\widetilde{F}_{\pi_h^N}^-(V^n) - F_{\pi_h^N}^-(V^n)\right]^2$$
$$= \frac{1}{N}\int_0^1 \left(\widetilde{F}_{\pi_h^N}^-(u) - F_{\pi_h^N}^-(u)\right)^2 \mathrm{d}u$$
$$= \frac{1}{N}\sum_{n=1}^N \int_{a_n}^{b_n} \left(F_{\pi_h^N}^-(u) - \widetilde{F}_{\pi_h^N}^-(u)\right)^2 \mathrm{d}u$$
$$\leq \frac{1}{N}\frac{2\alpha_N^2}{(\alpha_N+1)(2\alpha_N+1)}$$
$$= \frac{1}{N^3}.$$

Hence, to prove the theorem it remains to show that

$$\mathrm{Var}_{\zeta^N}\left[\frac{1}{N}\sum_{n=1}^N g_N(V^n)\right] = o(N^{-2}), \quad \mathbb{P} - a.s. \tag{S.31}$$

To establish this result note that, using the properties of $(V^n)_{n=1}^N$ and the mean value theorem (which can be used because $g_N$ is continuous, see Lemmas 2 and S3),

$$\text{Var}_{\zeta^N}\Big[\sum_{n=1}^N g_N(V^n)\Big] = N\Big(\int_0^1 g_N(u)^2\mathrm{d}u - \frac{1}{N}\sum_{n=1}^N g_N(v_n)^2\Big)$$

for some $v_n \in [(n-1)/N, n/N]$.

By (S.22), we have $D_N^*((v_n)_{n=1}^N) \leq N^{-1}$ and thus, by Theorem S2,

$$\text{Var}_{\zeta^N}\Big[\sum_{n=1}^N g_N(V^n)\Big] \leq N\, D_N^*((v_n)_{n=1}^N)\, V(g_N^2) \leq V(g_N^2) \tag{S.32}$$

with $V(g_N^2)$ the variation of the function $g_N^2$ on $[0,1]$.

To control this quantity recall that $V(g_N^2) = \sup_{P \in \mathrm{P}} S_{g_N^2}(P)$ where $\mathrm{P}$ is the set of all partitions of $[0,1]$ and where, for a partition $P = (z_i)_{i=0}^{M_P} \in \mathrm{P}$ of size $M_p$,

$$S_{g_N^2}(P) := \sum_{i=1}^{M_P} |g_N(z_i)^2 - g_N(z_{i-1})^2|.$$

Next, remark that for any $P = (z_i)_{i=0}^{M_P} \in \mathrm{P}$ we have

$$
\begin{aligned}
S_{g_N^2}(P) &= \sum_{i=1}^{M_P} |g_N(z_i)^2 - g_N(z_{i-1})^2| \\
&\leq \sum_{i=1}^{M_P} |g_N(z_i)||g_N(z^i) - g_N(z_{i-1})| + \sum_{i=1}^{M_P} |g_N(z_{i-1})||g_N(z_i) - g_N(z_{i-1})| \\
&\leq 2\|g_N\|_\infty \sum_{i=1}^{M_P} |g_N(z_i) - g_N(z_{i-1})|
\end{aligned}
$$

so that $V(g_N^2) \leq 2\|g_N\|_\infty V(g_N) \leq 4\|g_N\|_\infty$ where the last inequity uses the fact that

$$
\begin{aligned}
V(g_N) = V(\widetilde{F}_{\pi_h^N}^- - F_{\pi_h}^-) &\leq V(\widetilde{F}_{\pi_h^N}^-) + V(F_{\pi_h}^-) \\
&= |\widetilde{F}_{\pi_h^N}^-(1) - \widetilde{F}_{\pi_h^N}^-(0)| + |F_{\pi_h}^-(1) - F_{\pi_h}^-(0)| \\
&\leq 2.
\end{aligned}
$$

Using (S.30),

$$\lim_{N\to+\infty}\|g_N\|_\infty = \lim_{n\to+\infty}\|\widetilde{F}_{\pi_h^N}^- - F_{\pi_h}^-\|_\infty = 0, \quad \mathbb{P}-a.s.$$

and thus, together with (S.32), it follows that

$$\lim_{N\to+\infty}\text{Var}_{\zeta^N}\Big[\sum_{n=1}^N g_N(V^n)\Big] = o(1), \quad \mathbb{P}-a.s.$$

showing (S.31). This completes the proof of the first part of the theorem.

We now prove the second part of the theorem. Recall that, by the bi-measure property of Hilbert curve, for any $0 < a < b < 1$ we have

$$F_{\pi_h}(b) - F_{\pi_h}(a) = \pi_h\big((a,b]\big) = \pi\big(H((a,b])\big),$$

where $\lambda_d(H(a,b]) = (b-a)$. Therefore, because by assumption there exists a constant $c < +\infty$ such that $c^{-1}\lambda_d(A) \le \pi(A) \le c\lambda_d(A)$ for all measurable sets $A \subseteq \mathcal{X}$, if follows that

$$\frac{1}{c}|b-a| \le \big|F_{\pi_h}(b) - F_{\pi_h}(a)\big| \le c\,|b-a|, \quad \forall a,b \in (0,1).$$

Therefore, $F_{\pi_h}$ is bi-Lipschitz on $(0,1)$ and thus $F_{\pi_h}^{-}$ is Lipschitz on $(0,1)$. Using this last property of $F_{\pi_h}$ it is readily checked that the rate in (11) is $\mathcal{O}(N^{-3})$ and the result follows.

$\square$

## S3.7 Proofs of Section 5

### S3.7.1 Proof of Theorem 7

We prove below this more general result.

**Theorem S3.** *For Algorithm 2, assuming that $\mathcal{X}$ is a cubifiable set, $\mathcal{P}_0 = \mathcal{P}_b^*(\mathcal{X})$, $\rho \in \{\rho_{\mathrm{multi}}, \rho_{\mathrm{res/multi}}, \rho_{\mathrm{strat}}^*\}$ and that the Feynman-Kac model fulfils assumptions (G) and (M), for any test function $\varphi \in \mathcal{C}_b(\mathcal{X})$, we have that (for any $t \ge 0$)*

$$N^{1/2}\left\{\frac{1}{N}\sum_{n=1}^{N}\varphi(X_t^n) - (\pi_{t-1}M_t)(\varphi)\right\} \overset{w}{\Rightarrow} \mathcal{N}_d\left(0, \widetilde{\mathcal{V}}_t[\varphi]\right) \qquad (\text{S.33})$$

$$N^{1/2}\left\{\sum_{n=1}^{N} W_t^n \varphi(X_t^n) - \pi_t(\varphi)\right\} \overset{w}{\Rightarrow} \mathcal{N}_d(0, \mathcal{V}_t[\varphi]) \qquad (\text{S.34})$$

$$N^{1/2}\left\{\frac{1}{N}\sum_{n=1}^{N}\varphi(X_t^{A_{t+1}^n}) - \pi_t(\varphi)\right\} \overset{w}{\Rightarrow} \mathcal{N}_d\left(0, \widehat{\mathcal{V}}_t[\varphi]\right), \quad t \ge 0 \qquad (\text{S.35})$$

*where the $\mathcal{V}_t(\varphi)$ are defined recursively as follows: $\widetilde{\mathcal{V}}_0[\varphi] = V_0(\varphi)$,*

$$\mathcal{V}_t[\varphi] = \frac{1}{\ell_t^2}\widetilde{\mathcal{V}}_t[G_t\{\varphi - \pi_t(\varphi)\}]$$

$$\widehat{\mathcal{V}}_t[\varphi] = \mathcal{V}_t[\varphi] + R_t(\rho,\varphi)$$

$$\widetilde{\mathcal{V}}_{t+1}[\varphi] = \widehat{\mathcal{V}}_t[M_{t+1}(\varphi)] + \pi_t[V_{t+1}(\varphi)]$$

*and*

$$0 = R_t(\rho_{\mathrm{strat}^*},\varphi) \le R_t(\rho_{\mathrm{res/multi}},\varphi) \le R_t(\rho_{\mathrm{multi}},\varphi).$$

*Proof of Theorem S3.* There is nothing to prove for multinomial and residual resampling, and, for $\rho = \rho^{\star}_{\text{strat}}$, it is enough to prove that (S.34) $\Rightarrow$ (S.35) for all $t \geq 0$, as (S.35) $\Rightarrow$ (S.33) $\Rightarrow$ (S.34) have already been established in e.g. Chopin (2004). Note in addition that Assumptions (M) and (V) ensure that the operators $\mathcal{V}$, $\widehat{\mathcal{V}}$ and $\widetilde{\mathcal{V}}$ map $\mathcal{C}_b(\mathcal{X})$ into itself.

Assuming (S.34),

$$N^{1/2} \left\{ \sum_{n=1}^{N} W_t^n \varphi(X_t^n) - \pi_t(\varphi) \right\} \overset{\text{w}}{\Rightarrow} \mathcal{N}_d \left( 0, \mathcal{V}_t[\varphi] \right)$$

we have

$$N^{1/2} \left\{ \frac{1}{N} \sum_{n=1}^{N} \varphi(X_t^{A_{t+1}^n}) - \pi_t(\varphi) \right\} = N^{1/2} \left\{ \frac{1}{N} \sum_{n=1}^{N} \varphi(X_t^{A_{t+1}^n}) - \sum_{n=1}^{N} W_t^n \varphi(X_t^n) \right\} +$$
$$N^{1/2} \left\{ \sum_{n=1}^{N} W_t^n \varphi(X_t^n) - \pi_t(\varphi) \right\}.$$

and the result is proven provided we can apply Theorem 4 to the first term: then this term converges to 0 in $L^2$, and thus in probability (by Chebyshev's inequality), and we can apply Slutsky's theorem to conclude.

To apply Theorem 4, we need to establish that, $\mathbb{P} - a.s.$, $\max_n W_t^n \to 0$ as $N \to +\infty$. By assumption (G), there exists a $\gamma_t < +\infty$ such that

$$W_t^n = \frac{G_t(X_t^n)}{\sum_{m=1}^{N} G_t(X_t^m)} \leq \frac{\gamma_t}{N} \left( \frac{1}{N} \sum_{m=1}^{N} G_t(X_{t-1}^{A_t^m}, X_t^m) \right)^{-1}$$

and we know that

$$\left\{ \frac{1}{N} \sum_{m=1}^{N} G_t(X_{t-1}^{A_t^m}, X_t^m) - (\pi_{t-1} M_t)(G_t) \right\} \to 0 \quad \mathbb{P} - a.s.$$

where $(\pi_{t-1} M_t)(G_t) = \ell_t > 0$. Thus $\mathbb{P} \left( \max_{n \in 1:N} W_t^n \to 0 \right) = 1$. $\qquad\square$

### S3.7.2 Proof of Theorem 8

*Proof of Theorem 8.* Let $\mathcal{F}_{t-1} = \sigma(X_0^{1:N}, \ldots, U_{t-1}, X_{t-1}^{1:N})$, then, for any unbiased scheme:

$$\mathbb{E}\left[\ell_t^n | \mathcal{F}_{t-1}\right] = \sum_{n=1}^{N} W_{t-1}^n M_t(X_{t-1}^n, G_t)$$

which does not depend on $\eta_{t-1}$. Thus we wish to minimise the expectation of $\text{Var}\left[\ell_t^n | \mathcal{F}_{t-1}\right]$. (The same remark applies to $L_t^N$, as $\mathbb{E}\left[L_t^N | \mathcal{F}_{t-1}\right] = L_{t-1}^N \mathbb{E}\left[\ell_t^N | \mathcal{F}_{t-1}\right]$. For simplicity, we

work with $\ell_t^N$ from now on.) Under multinomial resampling, the $(A_t^n, X_t^n)$'s are IID conditional on $\mathcal{F}_{t-1}$, thus (for any $n$)

$$\mathrm{Var}\left[\ell_t^N|\mathcal{F}_{t-1}\right] = N^{-1}\mathrm{Var}\left[w_t^n|\mathcal{F}_{t-1}\right]$$

and this quantity is minimal when

$$\mathbb{E}\left(\left(w_t^n\right)^2|\mathcal{F}_{t-1}\right) = \sum_{n=1}^{N} \frac{\left(W_{t-1}^n\right)^2}{\widetilde{W}_{t-1}^n} M_t(X_{t-1}^n, G_t^2)$$

is minimal. Using constrained optimisation (the constraint being $\sum_{n=1}^{N} \widetilde{W}_{t-1}^n = 1$), it is easy to see that this quantity is minimal (with respect to the $N$ twisted weights) when

$$\widetilde{W}_{t-1}^n \propto W_{t-1}^n \sqrt{M_t(X_{t-1}^n, G_t^2)}$$

and thus taking $\eta_{t-1}(x_{t-1}) = M_t(x_{t-1}, G_t^2)$ gives a lower bound for the corresponding expectation.

We now take $\rho = \rho_{\mathrm{strat}}^\star$; the $(A_t^n, X_t^n)$'s are not conditionally IID any more, but

$$\mathrm{Var}\left[\ell_t^N|\mathcal{F}_{t-1}\right] = \mathrm{Var}\left\{\mathbb{E}\left[\ell_t^N|\hat{\mathcal{F}}_{t-1}\right]|\mathcal{F}_{t-1}\right\} + \mathbb{E}\left\{\mathrm{Var}\left[\ell_t^N|\hat{\mathcal{F}}_{t-1}\right]|\mathcal{F}_{t-1}\right\}$$

where $\hat{\mathcal{F}}_{t-1} = \sigma(X_{t-1}^{1:N}, A_t^{1:N})$. For the first term:

$$\mathbb{E}\left[\ell_t^N|\hat{\mathcal{F}}_{t-1}\right] = \frac{1}{N}\sum_{n=1}^{N} \frac{W_{t-1}^{A_t^n}}{\widetilde{W}_{t-1}^{A_t^n}} M_t(X_{t-1}^{A_t^n}, G_t) = \frac{\sum_{n=1}^{N} \widetilde{w}_{t-1}^n}{\sum_{n=1}^{N} w_{t-1}^n} \times \frac{1}{N}\sum_{n=1}^{N} \psi(X_{t-1}^{A_t^n})$$

where $\psi(x_{t-1}) = M_t(x_{t-1}, G_t)/\eta_{t-1}(x_{t-1})$. Since $\mathcal{X}$ is compact and $\eta_{t-1}$ takes values in $\mathbb{R}_{>0}$, $\psi \in \mathcal{C}_b(\mathcal{X})$, and we can apply Theorem 4:

$$\mathrm{MSE}\left[\frac{1}{N}\sum_{n=1}^{N} \psi(X_{t-1}^{A_t^n})\right] = o(N^{-1})$$

whith a constant that depends only on $\psi$. Hence, the first term is at a $o(N^{-1/2})$ $L^2-$distance of quantity

$$\frac{\sum_{n=1}^{N} \widetilde{w}_{t-1}^n}{\sum_{n=1}^{N} w_{t-1}^n} \times \sum_{n=1}^{N} \widetilde{W}_{t-1}^n \psi(X_{t-1}^n) = \frac{\sum_{n=1}^{N} \widetilde{w}_{t-1}^n \psi(X_{t-1}^n)}{\sum_{n=1}^{N} w_{t-1}^n} = \frac{\sum_{n=1}^{N} w_{t-1}^n M_t(X_{t-1}^n, G_t)}{\sum_{n=1}^{N} w_{t-1}^n}$$

which (a) does not depend on $\eta_{t-1}$; (b) converges at rate $\mathcal{O}_P(N^{-1/2})$. Hence the part that depends on $\eta_{t-1}$ becomes negligible when $N \to +\infty$.

Now for the second term:

$$\mathrm{Var}\left[\ell_t^N|\hat{\mathcal{F}}_{t-1}\right] = \frac{1}{N^2}\sum_{n=1}^{N} \left(\frac{W_{t-1}^{A_t^n}}{\widetilde{W}_{t-1}^{A_t^n}}\right)^2 \mathrm{Var}_{M_t}(X_{t-1}^{A_t^n}, G_t)$$

the expectation of which is

$$\mathbb{E}\left\{ \mathrm{Var}\left[\ell_t^N|\tilde{\mathcal{F}}_{t-1}\right]|\mathcal{F}_{t-1}\right\} = \frac{1}{N}\sum_{n=1}^{N}\frac{(W_{t-1}^n)^2}{\widetilde{W}_{t-1}^n}\mathrm{Var}_{M_t}(X_{t-1}^n, G_t)$$

which is mininal (under the constraint that $\sum_{n=1}^N \widetilde{W}_{t-1}^n = 1$) when $\widetilde{W}_{t-1}^n \propto W_{t-1}^n \sqrt{\mathrm{Var}_{M_t}(X_{t-1}^n, G_t)}$, hence the following function minimises the second term:

$$\eta_{t-1}(x_{t-1}) = \sqrt{\mathrm{Var}_{M_t}(X_{t-1}^n, G_t)}.$$

$\square$

# S4 Proofs of Appendix A

## S4.1 Proof of Lemma 2

*Proof of Lemma 2.* Without loss of generality we assume that $\mathcal{X} = (0,1)^d$ and take $h_{\mathcal{X},\psi} = h$. Let $(z^N)_{N\geq 1}$ and $\pi$ be as in the statement of the lemma and take $\epsilon > 0$, $a \in [0,1]$, and $\gamma > 0$ small enough so that the ball $B$ of centre $H(a)$ and radius $\gamma$ is such that $\pi(B) \leq \epsilon$. Since $H$ is continuous, there exists $\delta > 0$ such that $|b - a| \leq \delta$ implies that $\|H(b) - H(a)\| \leq \gamma$, and thus $H(b) \in B$. For any such $b$,

$$|F_{\pi_h}(b) - F_{\pi_h}(a)| \leq \pi(B) \leq \epsilon$$

and the result follows. $\square$

## S4.2 Proof of Theorem 9

We prove Theorem 9 by a succession of lemmas. Lemma S4 shows the implication $(ii) \Rightarrow (iii)$, Lemma S5 shows the implication $(iii) \Rightarrow (ii)$ while Lemma S6 shows the remaining equivalences.

**Lemma S4.** *Let $\mathcal{X}$ be a cubifiable set, $\pi \in \widetilde{\mathcal{P}}_b(\mathcal{X})$, $\psi \in \mathcal{D}(\mathcal{X})$ be such that $\pi_\psi \in \mathcal{P}_b((0,1)^d)$ and $(\pi^N)_{N\geq 1}$ be a sequence in $\mathcal{P}(\mathcal{X})$ such that $\lim_{N\to+\infty}\|\pi^N - \pi\|_\star = 0$. Then,*

$$\lim_{N\to+\infty}\|\pi_{h_{\mathcal{X},\psi}}^N - \pi_{h_{\mathcal{X},\psi}}\|_\star = 0.$$

*Proof of Lemma S4.* Without loss of generality we assume that $\mathcal{X} = (0,1)^d$ and take $h_{\mathcal{X},\psi} = h$.

We first assume that

$$\pi_h^N(I_m^d(k)) = \pi^N(S_m^d(k)), \quad \text{for all } k \in 0 : 2^{md}-1 \text{ and } m \geq 1 \text{ large enough.} \quad \text{(S.36)}$$

In this case, the result follows from Gerber and Chopin (2015, Theorem 3) but for sake of completeness the whole argument is presented below.

Let $I = [0, b]$, $b \in (0, 1)$, and $m \in \mathbb{N}$ (which may depend on $N$) and assume first that $b \geq 2^{-dm}$, so that $I_m^d(0) \subseteq I$. Take $\tilde{I} = [0, k^* 2^{-dm}]$, where $k^* \leq (2^{dm} - 1)$ is the largest integer such that $k^* 2^{-dm} \leq b$. Then

$$
\begin{aligned}
\left| \pi_h^N(I) - \pi_h(I) \right| &\leq \left| F_{\pi_h^N}(k^* 2^{-dm}) - F_{\pi_h}(k^* 2^{-dm}) \right| \\
&\quad + \left| \pi_h^N(I) - F_{\pi_h^N}(k^* 2^{-dm}) - \left\{ \pi_h(I) - F_{\pi_h}(k^* 2^{-dm}) \right\} \right| \\
&= \left| \pi^N(J) - \pi(J) \right| + \left| \pi_h^N \big( (k^* 2^{-dm}, b] \big) - \pi_h \big( (k^* 2^{-dm}, b] \big) \right|
\end{aligned}
\tag{S.37}
$$

with $J = H(\tilde{I})$. Note that the last equality holds by the definition of $\pi_h$ and by (S.36).

Next, since $\tilde{I}$ is the union of $k^*$ intervals in $\mathcal{I}_m^d$, $J$ is the union of $k^* \leq 2^{md}$ closed hypercubes in $\mathcal{S}_m^d$, and thus,

$$
\left| \pi^N(J) - \pi(J) \right| \leq k^* \| \pi^N - \pi \|_{\mathrm{E}} \leq 2^{dm} \| \pi^N - \pi \|_{\mathrm{E}}.
$$

For the second term of (S.37), we have under (S.36)

$$
\begin{aligned}
\left| \pi_h^N \big( (k 2^{-dm}, b] \big) - \pi_h \big( (k 2^{-dm}, b] \big) \right| &\leq \pi_h^N \big( I_m^d(k) \big) + \pi_h \big( I_m^d(k) \big) \\
&\leq \pi^N \big( S_m^d(k) \big) + \pi \big( S_m^d(k) \big) \\
&\leq 2\pi \big( S_m^d(k) \big) + \| \pi^N - \pi \|_{\mathrm{E}} \\
&= \mathcal{O} \left( 2^{-dm} \vee \| \pi^N - \pi \|_{\mathrm{E}} \right)
\end{aligned}
$$

where the last inequality comes from the fact that $\pi$ has a bounded bounded density with respect to $\lambda_d$.

In case $b < 2^{-dm}$, similar computations show that

$$
\left| \pi_h^N(I) - \pi_h(I) \right| \leq \pi_h^N(I_m^d(0)) + \pi_h(I_m^d(0)) = \mathcal{O} \left( 2^{-dm} \vee \| \pi^N - \pi \|_{\mathrm{E}} \right).
$$

To conclude the proof under (S.36), remark that

$$
\begin{aligned}
\| \pi^N - \pi \|_{\mathrm{E}} &= \sup_{0 \leq a < b \leq 1} \left| \pi^N([a, b)) - \pi([a, b)) \right| \\
&\leq 2^d \sup_{0 < b \leq 1} \left| \pi^N([0, b)) - \pi([0, b)) \right| \\
&= 2^d \| \pi^N - \pi \|_\star
\end{aligned}
$$

where the two equalities are due to Lemma S2, first part, and the inequality to Niederreiter (1992, Proposition 2.4, p.15). Hence, under the assumptions of the lemma, $\| \pi^N - \pi \|_{\mathrm{E}} = o(1)$ and thus, choosing $m$ so that $2^{-dm} = \mathcal{O}(\| \pi^N - \pi \|_{\mathrm{E}}^{1/2})$ gives

$$
\| \pi_h^N - \pi_h \|_\star = \mathcal{O} \left( \| \pi^N - \pi \|_{\mathrm{E}}^{1/2} \right) = o(1).
$$

This shows the results under (S.36).

Assume now that (S.36) does not hold. To facilitate the presentation we assume below that $\pi^N(\mathrm{d}x) = \sum_{n=1}^N W^{n,N} \delta(x^{n,N})$ for a $z^N = (x^{n,N}, W^{n,N})_{n=1}^N \in \mathcal{Z}^N$; that is, that $(\pi^N)_{N \geq 1}$ is a sequence in $\mathcal{P}_f(\mathcal{X})$. Then, because $x \in \mathcal{H}_d$ if and only if $x$ has at least one dyadic coordinate, for any $\epsilon > 0$ there exists a sequence $(\tilde{z}^N)_{N \geq 1}$ such that (S.36) holds and such that, for all $N \geq 1$,

$$\tilde{z}^N = (\tilde{x}^{n,N}, W^{n,N}), \quad \max_{n \in 1:N} \|x^{n,N} - \tilde{x}^{n,N}\|_\infty \leq \epsilon.$$

Then, by Lemma S1, first part,

$$\left| \|\tilde{\pi}^N - \pi\|_\star - \|\pi^N - \pi\|_\star \right| \leq c_\pi \epsilon$$

for a constant $c_\pi < +\infty$ which depends only on $\pi$.

Under the assumptions of the theorem we therefore have $\lim_{N \to +\infty} \|\tilde{\pi}^N - \pi\|_\star = 0$ and thus, from above,

$$\lim_{N \to +\infty} \|\tilde{\pi}_h^N - \pi_h\|_\star = 0. \tag{S.38}$$

To conclude the proof we choose the sequence $(\tilde{z}^N)_{N \geq 1}$ such that, for all $N \geq 1$,

$$\max_{n \in 1:N} |h(x^{n,N}) - h(\tilde{x}^{n,N})| \leq (\epsilon/C_d)^d$$

with $C_d < +\infty$ the Hölder constant of the Hilbert curve for the $\|\cdot\|_\infty$ norm. Such a sequence indeed exists because, since $H(h(x)) = x$ for all in $x \in [0,1]^d$ and by the Hölder property of the Hilbert curve,

$$\begin{aligned}
\|x^{n,N} - \tilde{x}^{n,N}\|_\infty &= \|H(h(x^{n,N})) - H(h(\tilde{x}^{n,N}))\|_\infty \\
&\leq C_d |h(x^{n,N}) - h(\tilde{x}^{n,N})|^{1/d} \leq \epsilon.
\end{aligned} \tag{S.39}$$

Then, by Lemma S1, second part,

$$\left| \|\tilde{\pi}_h^N - \pi_h\|_\star - \|\pi_h^N - \pi_h\|_\star \right| \leq c_\pi (\epsilon/C_d)^d$$

with $c_\pi + \infty$ depending only on $\pi$. Hence, since $\epsilon > 0$ is arbitrary and using (S.38),

$$\lim_{N \to +\infty} \|\pi_h^N - \pi_h\|_\star = 0.$$

This completes the proof of the lemma.

$\square$

**Lemma S5.** *Let $\mathcal{X}$ be a cubifiable set, $\pi \in \widetilde{\mathcal{P}}_b(\mathcal{X})$, $\psi \in \mathcal{D}(\mathcal{X})$ be such that $\pi_\psi \in \mathcal{P}_b((0,1)^d)$ and $(\pi^N)_{N \geq 1}$ be a sequence in $\mathcal{P}(\mathcal{X})$ such that $\lim_{N \to +\infty} \|\pi_{h_{\mathcal{X},\psi}}^N - \pi_{h_{\mathcal{X},\psi}}\|_\star = 0$. Then,*

$$\lim_{N \to +\infty} \|\pi^N - \pi\|_\star = 0.$$

*Proof.* Without loss of generality we assume that $\mathcal{X} = (0,1)^d$ and take $h_{\mathcal{X},\psi} = h$.

We first assume (S.36). In this case, the result follows from similar computations as in Schretter et al. (2016, Theorem 1) but for sake of completeness the whole argument is reproduced below.

Let $m \geq 0$ be an arbitrary integer and $a \in [0,1)^d$ be such that $S_m^d(0) \subseteq B := [0,a]$. Let $\mathcal{S}_m^B = \{W \in \mathcal{S}_m^d : W \subseteq B\}$, $\tilde{B} = \cup \mathcal{S}_m^B$ and $\mathcal{D}_m^B = \{W \in \mathcal{S}_m^d : (B \setminus \tilde{B}) \cap W \neq \varnothing\}$. Then, let $\tilde{\mathcal{D}}_m^B$ be the set of $\#\mathcal{D}_m^B$ disjoint subsets of $[0,1]^d$ such that

1. $\forall \tilde{W} \in \tilde{\mathcal{D}}_m^B, \exists W \in \mathcal{D}_m^B \mid \tilde{W} \subseteq W$,

2. $\cup \tilde{\mathcal{D}}_m^B = \mathcal{D}_m^B$,

3. $\tilde{B} \cap \{\cup \tilde{\mathcal{D}}_m^B\} = \varnothing$.

Note that $\tilde{\mathcal{D}}_m^B$ is obtained by removing boundaries of the elements in $\mathcal{D}_m^B$ such that the above conditions 2 and 3 are satisfied. Then, we have

$$|\pi^N(B) - \pi(B)| \leq |\pi^N(\tilde{B}) - \pi(\tilde{B})| + \sum_{\tilde{W} \in \tilde{\mathcal{D}}_m^B} |\pi^N(\tilde{W} \cap B) - \pi(\tilde{W} \cap B)| \qquad (S.40)$$

where, under (S.36),

$$|\pi^N(\tilde{B}) - \pi(\tilde{B})| = |\pi_h^N(h(\tilde{B})) - \pi_h(h(\tilde{B}))| \leq 2^{md} \|\pi_h^N - \pi_h\|_{\mathrm{E}}.$$

For the second term of (S.40), take $\tilde{W} \in \tilde{\mathcal{D}}_m^B$ and note that $\tilde{W} \subseteq S_m^d(k)$ for a $k \in \{0, \ldots, 2^{dm} - 1\}$. Then, with $p$ the bounded density of $\pi$ with respect to the Lebesgue measure,

$$\begin{aligned}
|\pi^N(\tilde{W} \cap B) - \pi(\tilde{W} \cap B)| &\leq \pi^N(S_m^d(k))) + \pi(S_m^d(k)) \\
&\leq \pi_h^N(I_m^d(k)) + \|p\|_\infty \lambda_d(S_m^d(k)) \\
&\leq 2\|p\|_\infty \lambda_d(S_m^d(k)) + \|\pi_h^N - \pi_h\|_{\mathrm{E}} \\
&= 2\|p\|_\infty 2^{-dm} + \|\pi_h^N - \pi_h\|_{\mathrm{E}}.
\end{aligned}$$

Thus,

$$\sum_{\tilde{W} \in \tilde{\mathcal{D}}_m^B} |\pi^N(\tilde{W} \cap B) - \pi(\tilde{W} \cap B)| \leq 2\|p\|_\infty d\, 2^{-m} + d\, 2^{m(d-1)} \|\pi_h^N - \pi_h\|_{\mathrm{E}}$$

since $\#\tilde{\mathcal{D}}_m^B = \#\mathcal{D}_m^B \leq d\, 2^{m(d-1)}$ (Schretter et al., 2016).

Hence, for all $a \in [0,1)^d$ such that $S_m^d(0) \subseteq [0,a]$ we have

$$|\pi^N([0,a]) - \pi([0,a])| \leq 2\|p\|_\infty d\, 2^{-m} + \|\pi_h^N - \pi_h\|_{\mathrm{E}} (1 + 5d\, 2^{md}).$$

Finally, if $a \in [0,1)^d$ is such that $S_m^d(0) \nsubseteq [0,a]$, we proceed exactly as above, but now $\tilde{B}$ is empty and therefore the first term in (S.40) disappears.

To conclude the proof under (S.36) remark that

$$\|\pi_h^N - \pi_h\|_{\mathrm{E}} = \sup_{0 \le a < b \le 1} \left| \pi_h^N([a, b)) - \pi_h([a, b)) \right|$$

$$\le 2 \sup_{0 < b \le 1} \left| \pi_h^N([0, b)) - \pi_h([0, b)) \right|$$

$$= 2\|\pi_h^N - \pi_h\|_\star$$

where the two equalities are due to Lemma S2, second part, and the inequality to Niederreiter (1992, Proposition 2.4, p.15). Hence, $\|\pi_h^N - \pi_h\|_{\mathrm{E}} = o(1)$ under the assumptions of the lemma. We then choose $m$ such that $2^{-m} \sim 2^{md} \|\pi_h^N - \pi_h\|_{\mathrm{E}}$, which implies

$$\lim_{N \to +\infty} \|\pi^N - \pi\|_\star = 0$$

as required

If (S.36) does not hold the result follows using a similar argument as in the proof of lemma S4. To facilitate the presentation we assume below that $\pi^N(dx) = \sum_{n=1}^N W^{n,N}\delta(x^{n,N})$ for a $z^N = (x^{n,N}, W^{n,N})_{n=1}^N \in \mathcal{Z}^N$; that is, that $(\pi^N)_{N\ge1}$ is a sequence in $\mathcal{P}_f(\mathcal{X})$. Let $\epsilon > 0$ and choose a sequence $(\tilde{z}^N)_{N\ge1}$ such that (S.36) holds and such that, for all $N \ge 1$,

$$\tilde{z}^N = (\tilde{x}^{n,N}, W^{n,N}), \quad \max_{n \in 1:N} \|x^{n,N} - \tilde{x}^{n,N}\|_\infty \le \epsilon$$

and

$$\max_{n \in 1:N} |h(x^{n,N}) - h(\tilde{x}^{n,N})| \le (\epsilon/C_d)^d$$

for a constant $C_d < \infty$; note that such a sequence $(\tilde{z}^N)_{N\ge1}$ exists by (S.39). Then, by Lemme S1 and under the assumptions of the lemma, $\lim_{N\to+\infty} \|\tilde{\pi}_h^N - \pi_h\|_\star = 0$ and thus, from above, $\lim_{N\to+\infty} \|\tilde{\pi}^N - \pi\|_\star = 0$. Using again Lemma S1 we conclude that $\lim_{N\to+\infty} \|\pi^N - \pi\|_\star = 0$ and the proof is complete. $\qquad\square$

**Lemma S6.** *Let $\mathcal{X}$ a cubifiable set, $\pi \in \widetilde{\mathcal{P}}_b(\mathcal{X})$, $\psi \in \mathcal{D}(\mathcal{X})$ be such that $\pi_\psi \in \mathcal{P}_b((0,1)^d)$ and $(\pi^N)_{N\ge1}$ a sequence in $\mathcal{P}(\mathcal{X})$. Then,*

$$\pi^N \overset{w}{\Rightarrow} \pi \Leftrightarrow \lim_{N\to+\infty} \|\pi^N - \pi\|_\star = 0$$

*and*

$$\pi_{h_{\mathcal{X},\psi}}^N \overset{w}{\Rightarrow} \pi_{h_{\mathcal{X},\psi}} \Leftrightarrow \lim_{N\to+\infty} \|\pi_{h_{\mathcal{X},\psi}}^N - \pi_{h_{\mathcal{X},\psi}}\|_{\mathrm{E}} = 0$$

In order to prove Lemma S6 we need Lemma S7 below, which is a straightforward consequence of e.g. van der Vaart (1998, Lemma 2.2, p.6).

**Lemma S7.** *Let $\pi \in \mathcal{P}(\mathcal{X})$ and $(\pi^N)_{N\ge1}$ be a sequence in $\mathcal{P}(\mathcal{X})$. Then, the following statements are equivalent:*

$$\lim_{N\to+\infty} |\pi^N(\varphi) - \pi(\varphi)| = 0, \qquad\qquad \forall\varphi \in \mathcal{C}_b(\mathcal{X}) \tag{a}$$

$$\lim_{N\to+\infty} |F_{\pi^N}(a) - F_\pi(a)| = 0, \qquad\qquad \forall a \in \mathcal{X} \tag{b}$$

*In addition, if $\pi \in \mathcal{P}_b(\mathcal{X})$, statements (a) and (b) are equivalent to*

$$\lim_{N \to +\infty} |\pi^N(B) - \pi(B)| = 0, \qquad \text{for all Borel set } B \in \mathbb{X}. \qquad \text{(c)}$$

*Proof of Lemma S6.* Without loss of generality we assume that $\mathcal{X} = (0,1)^d$ and take $h_{\mathcal{X},\psi} = h$. Let $\pi$ and $(\pi^N)_{N \geq 1}$ be as in the statement of the lemma and $p : \mathcal{X} \to \mathbb{R}_+$ be the continuous and bounded density of $\pi$.

The implications "$\Leftarrow$" are direct consequences of (b)$\Rightarrow$(a) in Lemma S7.

We first show the implication "$\Rightarrow$" for the first part of the lemma. To this end, we follow a similar argument as in Kuipers and Niederreiter (1974, Theorem 1.2, p.89). Note first that $\pi^N \overset{\text{w}}{\Rightarrow} \pi$ implies (c) in Lemma S7, hence $\pi^N \overset{\text{w}}{\Rightarrow} \pi$ implies that

$$|\pi^N([a,b]) - \pi([a,b])| \to 0, \quad \forall [a,b] \subset (0,1)^d. \qquad \text{(S.41)}$$

For a fixed $\epsilon > 0$, let $m_\epsilon \geq 2$ be the smallest positive integer such that

$$\left( \frac{2\|p\|_\infty}{m_\epsilon}\left(2 + \frac{1}{m_\epsilon}\right)\right) \vee \left( \frac{1}{m_\epsilon} + \|p\|_\infty \left(\frac{2}{m_\epsilon}\right)^d \left(1 + \frac{1}{m_\epsilon}\right)\right) \leq \epsilon \qquad \text{(S.42)}$$

and let $B_k = \prod_{i=1}^d [k_i/m_\epsilon, (k_i+1)/m_\epsilon]$ with $k_i \in 0 : (m_\epsilon - 1)$ for all $i \in 1 : d$. Then, using (S.41), there exists a $N_\epsilon \geq 1$ such that, for all $N \geq N_\epsilon$ and $k \in 0 : (m_\epsilon - 1)$, we have

$$\pi(B_k)\left(1 - \frac{1}{m_\epsilon}\right) \leq \pi^N(B_k) \leq \pi(B_k)\left(1 + \frac{1}{m_\epsilon}\right). \qquad \text{(S.43)}$$

Let $J = [a,b] \subset (0,1)^d$. There exist hypercubes $J_1$ and $J_2$, which are (possibly empty) finite unions of hypercubes $B_k$, such that

$$J_1 \subseteq J \subseteq J_2, \quad \lambda_d(J - J_1) \leq (2/m_\epsilon)^d, \quad \lambda_d(J_2 - J) \leq (2/m_\epsilon)^d.$$

This implies that $\pi(J) - \pi(J_1) \leq \|p\|_\infty (2/m_\epsilon)^d$ and $\pi(J_2) - \pi(J) \leq \|p\|_\infty (2/m_\epsilon)^d$.

Assume first that that $J$ is such that $J_1 \neq \varnothing$. Then, we have for all $N \geq N_\epsilon$

$$\pi(J_1)\left(1 - \frac{1}{m_\epsilon}\right) \leq \pi^N(J_1) \leq \pi^N(J) \leq \pi^N(J_2) \leq \pi(J_2)\left(1 + \frac{1}{m_\epsilon}\right)$$

thus

$$\left(\pi(J) - \|p\|_\infty \left(\frac{2}{m_\epsilon}\right)^d\right)\left(1 - \frac{1}{m_\epsilon}\right) \leq \pi^N(J) \leq \left(\pi(J) + \|p\|_\infty \left(\frac{2}{m_\epsilon}\right)^d\right)\left(1 + \frac{1}{m_\epsilon}\right)$$

and since $\pi(J) \leq 1$

$$-\frac{1}{m_\epsilon} - \|p\|_\infty \left(\frac{2}{m_\epsilon}\right)^d\left(1 + \frac{1}{m_\epsilon}\right) \leq \pi^N(J) - \pi(J) \leq \frac{1}{m_\epsilon} + \|p\|_\infty \left(\frac{2}{m_\epsilon}\right)^d\left(1 + \frac{1}{m_\epsilon}\right).$$
$$\text{(S.44)}$$

Consider now the case $J_1 = \varnothing$ and define $J_2$ as above. Then, $\lambda_d(J_2) \leq 2/m_\epsilon$ and thus, for all $N \geq N_\epsilon$,

$$|\pi^N(J) - \pi(J)| \leq \pi^N(J_2) + \pi(J_2) \leq \pi(J_2)\Big(2 + \frac{1}{m_\epsilon}\Big) \leq \frac{2\|p\|_\infty}{m_\epsilon}\Big(2 + \frac{1}{m_\epsilon}\Big). \qquad (S.45)$$

Therefore, combining (S.42), (S.44) and (S.45), we have that, for all $N \geq N_\epsilon$ and using (S.43),

$$\sup_{[a,b]\subset(0,1)} |\pi^N\big([a,b]\big) - \pi\big([a,b]\big)| \leq \epsilon,$$

which concludes the proof of the first part of the lemma.

The implication "$\Rightarrow$" in the second part of the lemma is due to the continuity of $F_{\pi_h}$ (Lemma 2) and to the Polyà's Theorem (Pólya, 1920; see also Bickel and Millar, 1992, result (A.1)). Alternatively, we can establish this implication following the same computation as per above. To do so, take $d = 1$ and $m_\epsilon = 2^{\tilde{m}_\epsilon}$ for some $\tilde{m}_\epsilon \geq 1$. Then, as $\pi_h$ is a continuous probability measure under the assumptions of the lemma (Lemma 2), $\pi_h(B_k) = \pi(S_{\tilde{m}_\epsilon}(k)) \leq \|p\|_\infty/m_\epsilon$ for all $k$ and, by part (c) of Lemma S7, when $\pi_h^N \overset{w}{\Rightarrow} \pi_h$ we have

$$|\pi_h^N\big([a,b]\big) - \pi_h\big([a,b]\big)| \to 0, \quad \forall [a,b] \subset (0,1).$$

which replaces (S.41). $\qquad \square$

# References

Aistleitner, C. and Dick, J. (2015). Functions of bounded variation, signed measures, and a general Koksma-Hlawja inequality. *Acta Arith.*, 167(2):143–171.

Bickel, P. and Millar, P. (1992). Uniform convergence of probability measures on classes of functions. *Statistica Sinica*, pages 1–15.

Chopin, N. (2004). Central limit theorem for sequential Monte Carlo methods and its application to Bayesian inference. *Ann. Statist.*, 32(6):2385–2411.

Dubhashi, D., Jonasson, J., and Ranjan, D. (2007). Positive influence and negative dependence. *Combinatorics, Probability and Computing*, 16(01):29–41.

Dubhashi, D. and Ranjan, D. (1998). Balls and bins: a study in negative dependence. *Random Structures Algorithms*, 13(2):99–124.

Gerber, M. and Chopin, N. (2015). Sequential quasi Monte Carlo. *J. R. Stat. Soc. Ser. B. Stat. Methodol.*, 77(3):509–579.

He, Z. and Owen, A. B. (2016). Extensible grids: uniform sampling on a space filling curve. *J. R. Stat. Soc. Ser. B. Stat. Methodol.*, 78(4):917–931.

Kramer, J. B., Cutler, J., and Radcliffe, A. (2011). Negative dependence and Srinivasan's sampling process. *Combinatorics, Probability and Computing*, 20(03):347–361.

Kuipers, L. and Niederreiter, H. (1974). *Uniform distribution of sequences*. Wiley-Interscience.

Mytrofanov, M. A. and Ravsky, A. V. (2012). Approximation of continuous functions on Fréchet spaces. *Journal of Mathematical Sciences*, 185(6):792–799.

Niederreiter, H. (1992). *Random number generation and quasi-Monte Carlo methods*, volume 63 of *CBMS-NSF Regional Conference Series in Applied Mathematics*. Society for Industrial and Applied Mathematics (SIAM), Philadelphia, PA.

Pólya, G. (1920). Über den zentralen grenzwertsatz der wahrscheinlichkeitsrechnung und das momentenproblem. *Mathematische Zeitschrift*, 8(3):171–181.

Schretter, C., He, Z., Gerber, M., Chopin, N., and Niederreiter, H. (2016). Van der Corput and golden ratio sequences along the Hilbert space-filling curve. In *Monte Carlo and Quasi-Monte Carlo Methods*, pages 531–544. Springer.

Shao, Q.-M. (2000). A comparison theorem on moment inequalities between negatively associated and independent random variables. *J. Theoret. Probab.*, 13(2):343–356.

van der Vaart, A. W. (1998). *Asymptotic statistics*, volume 3 of *Cambridge Series in Statistical and Probabilistic Mathematics*. Cambridge University Press, Cambridge.

Zumbusch, G. (2003). *Parallel multilevel methods*. Springer.



\end{document}